%% file: main.tex
\PassOptionsToPackage{nameinlink,capitalise}{cleveref}
\PassOptionsToPackage{naturalnames}{hyperref}
\documentclass[a4paper,UKenglish,autoref, cleveref, thm-restate,numberwithinsect]{lipics-v2021} 

\title{Deterministic regular functions of infinite words} 

\titlerunning{Deterministic regular functions of infinite words} 

\author{Olivier Carton}{Université Paris Cité, CNRS, IRIF, F-75013, Paris, France}%
{carton@irif.fr}{https://orcid.org/0000-0002-2728-6534}
{}
\author{Gaëtan Douéneau-Tabot}{Université Paris Cité, CNRS, IRIF, F-75013, Paris, France
\and Direction générale de l'armement - Ingénierie des projets, Paris, France
}{doueneau@irif.fr}{}{}
\author{Emmanuel Filiot}{Université libre de Bruxelles \& F.R.S.-FNRS, Brussels, Belgium}{efiliot@ulb.ac.be}%
{https://orcid.org/0000-0002-2520-5630}{}
\author{Sarah Winter}{Université libre de Bruxelles \& F.R.S.-FNRS, Brussels, Belgium}{sarah.winter@ulb.ac.be}%
{https://orcid.org/0000-0002-3499-1995}{}

\authorrunning{O. Carton, G. Douéneau-Tabot, E. Filiot and S. Winter} 

\Copyright{\phantom{a}} 

\ccsdesc[100]{Theory of computation~Formal languages and automata theory~Automata extensions~Transducers} 

\keywords{infinite words, streaming string transducers, two-way transducers,
monadic second-order logic, look-aheads, factorization forests} 

\category{} 

\relatedversion{} 

\acknowledgements{}

\nolinenumbers 

\pdfoutput=1 

\hideLIPIcs  

\EventEditors{}
\EventNoEds{1}
\EventLongTitle{}
\EventShortTitle{}
\EventAcronym{}
\EventYear{}
\EventDate{}
\EventLocation{}
\EventLogo{}
\SeriesVolume{}
\ArticleNo{}

\usepackage[dvipsnames]{xcolor}
\usepackage{todonotes}
\usepackage{xspace}
\usepackage{subcaption}

\usetikzlibrary{arrows,automata,positioning,patterns,fit}

\usepackage{tipa}
\usepackage{url}

\usepackage{stmaryrd}
\SetSymbolFont{stmry}{bold}{U}{stmry}{m}{n}
\usepackage{anyfontsize}

\crefname{claim}{Claim}{Claims}
\crefname{item}{Item}{Items}

\usepackage{amssymb}
\usepackage{amsfonts}
\usepackage{mathtools}
\usepackage{amsthm}
\usepackage{thmtools}
\usepackage{hyperref}

\usepackage{forest}
\useforestlibrary{edges}

\usepackage{caption}

\renewcommand{\epsilon}{\varepsilon} 

\newcommand{\saf}{\text{sf}}
\newcommand{\mv}{\text{mv}}
\newcommand{\la}{\text{la}}
\newcommand{\fst}{\text{fst}}

\newcommand{\dom}{\text{dom}}
\newcommand{\tDT}{{\text{$2$-$\operatorname{\normalfont\textsf{\small dT}}$}{}}}
\newcommand{\oDT}{{\text{$1$-$\operatorname{\normalfont\textsf{\small dT}}$}{}}}
\newcommand{\tDTLA}{{\text{$2$-$\operatorname{\normalfont\textsf{\small dT}}^{\operatorname{\normalfont \textsf{FLA}}}$}{}}}

\newcommand{\tDTLAS}{{\text{$2$-$\operatorname{\normalfont\textsf{\small dT}}^{\operatorname{\normalfont \textsf{FLA}}}_{\sf S}$}{}}}
\newcommand{\tDTLARS}{{\text{$2$-$\operatorname{\normalfont\textsf{\small dT}}^{\operatorname{\normalfont \textsf{FLAR}}}_{\sf S}$}{}}}
\newcommand{\gv}{\textsf{g}}
\newcommand{\DSST}{{$\operatorname{\normalfont\textsf{\small BSST}}${}}}
\newcommand{\BSST}{\DSST}
\newcommand{\SST}{{$\operatorname{\normalfont\textsf{\small SST}}${}}}
\newcommand{\trans}{\mathcal{T}}
\newcommand{\aut}{\mathcal{A}}
\newcommand{{\lmark}}{\vdash}
\newcommand{\lmove}{\triangleleft}
\newcommand{\rmove}{\triangleright}
\newcommand{\parfonc}{\rightharpoonup}
\newcommand{\fonc}{\rightarrow}
\newcommand{\defined}{\coloneqq}
\newcommand{\Regs}{\mathfrak{R}}
\newcommand{\reg}{\mathfrak{r}}
\newcommand{\pref}{\sqsubseteq}
\newcommand{\movi}{\varepsilon}
\newcommand{\cro}[1]{ {{\llbracket} #1{\rrbracket}}}
\newcommand{\regg}{\mathfrak{s}}

\newcommand{\out}{\mathfrak{out}}
\newcommand{\Dom}{\operatorname{\normalfont\textsf{\small Dom}}}
\newcommand{\Nat}{\mathbb{N}}
\newcommand{\tnorm}[1]{\text{\normalfont {#1}}}
\newcommand{\subst}[2]{\mathcal{S}_{#1}^{#2}}
\newcommand{\replace}{ \operatorname{\normalfont\textsf{replace}}}
\newcommand{\double}{ \operatorname{\normalfont\textsf{double}}}
\newcommand{\conddouble}{ \operatorname{\normalfont\textsf{copy}}}
\newcommand{\MSO}{\textnormal{\textsf{MSO}}\xspace}
\newcommand{\GMSO}{\textnormal{\textsf{MSO}$_g$}\xspace}
\newcommand{\MSOT}{\textnormal{\textsf{MSOT}}\xspace}
\newcommand{\GMSOT}{\ensuremath{\textnormal{\textsf{MSOT}}_g}\xspace}
\newcommand{\GMSOTO}{\ensuremath{\textnormal{\textsf{MSOT}}_g[\leq]}\xspace}
\newcommand{\guard}{\textsf{G}}
\newcommand{\first}{\textsf{first}}
\newcommand\sem[1]{\llbracket #1 \rrbracket}

\newcommand{\hash}[1]{|_{#1}}
\newcommand{\Hash}[1]{\Delta_{#1}}
\newcommand{\mapcr}{\operatorname{ \normalfont\textsf{map-copy-reverse}}}

\newcommand{\mirror}[1]{\widetilde{#1}}
\newcommand{\finin}[1]{{#1}^{\infty}}
\newcommand{\FRS}[1]{\operatorname{\normalfont \textsf{FRS}}_{#1}}
\newcommand{\oright}[1]{\overrightarrow{#1}}
\newcommand{\oleft}[1]{\overleftarrow{#1}}

\newcommand{\Class}{\mathcal{C}}

\newcommand{\mapc}{\operatorname{\normalfont \textsf{map-copy}}}
\newcommand{\mapr}{\operatorname{\normalfont \textsf{map-reverse}}}
\newcommand{\runleft}[1]{\oleft{\operatorname{\normalfont \textsf{run}}}_{#1}}
\newcommand{\tryleft}{\oleft{\operatorname{\normalfont \textsf{try-run}}}}
\newcommand{\tryright}{\oright{\operatorname{\normalfont \textsf{try-run}}}}
\newcommand{\runright}[1]{\oright{\operatorname{\normalfont \textsf{run}}}_{#1}}
\newcommand{\oleftright}[1]{\overline{#1}}
\newcommand{\spli}{\operatorname{\normalfont \textsf{split}}}
\newcommand{\frs}{\mathfrak{s}}
\newcommand{\ide}{\operatorname{\normalfont \textsf{id}}}
\newcommand{\norm}{\operatorname{\normalfont \textsf{normalize}}}
\newcommand{\then}[1]{[#1]}

\newcommand{\manu}[2][]{\todo[inline,color=green!50!blue!30,#1]{\small M. #2}}

\begin{document}

\maketitle

\begin{abstract}
    Regular functions of infinite words are (partial) functions realized by
    deterministic two-way transducers with \emph{infinite}
    look-ahead. Equivalently, Alur et.\ al.\ have shown that they
    correspond to functions realized by deterministic Muller
    streaming string transducers, and to functions defined by
    MSO-transductions. Regular functions are however not computable in general (for a classical
    extension of Turing computability to infinite inputs), and  we consider in this
    paper the class of \emph{deterministic regular functions} of
    infinite words, realized by deterministic two-way transducers
    \emph{without} look-ahead. We prove that it is a well-behaved
    class of functions: they are computable, closed under composition,
    characterized by the \emph{guarded} fragment of MSO-transductions,
    by deterministic B\"uchi streaming string transducers, by
    deterministic two-way transducers with \emph{finite} look-ahead, 
    and by finite compositions of sequential functions and one fixed basic
    function called $\mapcr$.
\end{abstract}

\section{Introduction}

Transducers extend automata with output mechanisms, turning finite
state machines from language acceptors to computational models for
functions. Inspired by a seminal work by Engelfriet and
Hoogeboom~\cite{engelfriet2001mso}, the last decade has seen an increasing interest in
characterizing the class of functions defined by deterministic two-way
transducers over finite words (\tDT), now called the class of
\emph{regular functions of finite words}. This class
admits several (effective) characterizations: it corresponds to the functions definable by \MSO-transductions~\cite{engelfriet2001mso}, by an \MSO-based
logic on origin graphs~\cite{dartois2018logics}, by an extension of regular expressions
called combinator expressions~\cite{alur2014regular,baudru2018two,dave2018regular},
and computed by \emph{copyless streaming string transducers} (\SST)
(a deterministic one-way model which uses
registers to store and update partial output words~\cite{alur2010expressiveness}).
Moreover, the class of regular functions over finite words is closed under composition~\cite{chytil1977serial},
and it has decidable equivalence problem~\cite{gurari1982equivalence}.

\begin{example} \label{ex:mapcr-fini}
Let $\Sigma$ be an alphabet, the function
 $\mapcr : (\Sigma\uplus \{\hash{}\})^* \fonc (\Sigma\uplus \{\hash{}\})^* $
 takes any word of the form $u_1\hash{}\dots\hash{}u_n$ where each $u_i$ is $\hash{}$-free,
and outputs $u_1\hash{}\mirror{u_1}\hash{}\dots \hash{} u_n\hash{}\mirror{u_n}$, where
$\mirror{u_i}$ is the mirror image of $u_i$. The function
$\mapcr$ is regular.
\end{example}

Regular functions can also be characterized as the compositions of
sequential functions (functions computed by deterministic \emph{one-way}
finite transducers) and $\mapcr$~\cite{bojanczyk2020single}.

\subparagraph{Regular functions of infinite words} The class of
regular functions has been extended to infinite
words in~\cite{alur2012regular}, and defined as the 
class of functions definable by \MSO-transductions over infinite words.
Equivalently, they have
been shown to be the functions realized by deterministic two-way
transducers with regular look-ahead, and by streaming string transducers with
a Muller selection condition (the register holding the final output word
is selected depending on the set of states seen infinitely often).
As for finite words, regular functions of infinite words are
closed under composition, and have decidable equivalence
problem~\cite{alur2012regular}.

\begin{example} 
\label{ex:intro1} 
    Let $\Sigma = \{a,b,c\}$ be an alphabet, and consider the function
    $\double:\Sigma^\omega\rightarrow \Sigma^\omega$
    which behaves like the identity function except that
    any occurrence of $a$ is replaced by $aa$ if there exists a $b$
    in the future of that occurence.
    For example, $(ab)^\omega$ is mapped to $(aab)^\omega$ and
    $acaab(ac)^\omega$ is mapped to $aacaaaab(ac)^\omega$. The function
    $\double$ is regular, as it can be realized by a one-way
    transducer which, when reading an $a$,  uses regular look-aheads to determine whether
    there exists a $b$ or not in the future, and produces either $a$
    or $aa$ accordingly.
\end{example}

\begin{example}
\label{ex:intro1bis}
    Let $\Sigma' = \{a,b,1,2\}$ and consider the function $\conddouble$ which maps:
    \begin{itemize}
    \item $u_1\sigma_1u_2\dots \sigma_nu
    \mapsto u_1^{\sigma_1}\sigma_1\dots u_n^{\sigma_n}\sigma_n u$
    where $u_1u_2\dots
    u_nu\in \{a,b\}^\omega$ and $\sigma_1,\dots,\sigma_n\in\{1,2\}$;
    \item $u_1\sigma_1\dots u_i\sigma_i \dots
    \mapsto u_1^{\sigma_1}\sigma_1\dots u_i^{\sigma_i}\sigma_i\dots$
    (if there are infinitely many $\sigma_i \in \{1,2\}$).
    \end{itemize}
For example, $\conddouble(ab2a1b^\omega) = abab2a1b^\omega$
and $\conddouble((a2)^\omega) =
(aa2)^\omega$.  The function $\conddouble$ is regular, for instance
realized by a deterministic two-way transducer which, using
two-wayness, makes one or two passes on the blocks $u_i$, depending on
whether they are followed by $\sigma_i=2$. On the first pass, it always
outputs what it reads, so that if no separator in $\{1,2\}$ is ever
read again (which means it is reading the infinite suffix $u$), then
it outputs $u$. 
\end{example}

Despite the robustness of the class of regular functions of infinite
words, witnessed by its various characterizations and algorithmic
properties, they suffer from a severe downside when it comes to
computability. Indeed, there are regular functions of infinite words
which are \emph{not} computable. At this point, we make clear what is
meant by computability, since the input is infinite. We refer the
reader to~\cite{dave2020synthesis,DBLP:journals/lmcs/DaveFKL22} (and the references therein) for a formal definition of
computability, and rather give intuitions here. A function $f$ of infinite
words is computable if there is a Turing machine with an infinite
read-only tape which contains some infinite input word $u$ in the
domain of the function, a bidirectional working tape, and a write-only
left-to-right output tape, such that by reading longer and longer input
prefixes, the machine writes longer and longer prefixes of $f(u)$ on the
output tape. Informally, it is an algorithm which takes the input as a
stream and is able to produce the output as a stream, so that
infinitely often, at least one output symbol is produced. For
instance, the function $\double$ above is not computable. On reading
prefixes of the form $ac^n$ for increasing values of $n$, it can
safely output one $a$ symbol, but not more. Indeed, if it outputs one
more $a$, then it is a wrong output for continuation $c^\omega$, and
if it outputs a $c$, then it is a wrong output for continuation
$b^\omega$, as $\double(ac^nc^\omega) = ac^\omega$ and $\double(ac^nb^\omega)
= aac^nb^\omega$. Its implementation by a two-way transducer indeed
requires an infinite look-ahead to check the absence of a $b$ in the
future. On the other hand, $\conddouble$ is realized by a deterministic two-way
transducer with \emph{no} look-ahead, so it is computable. 
So, deterministic two-way transducers with (infinite) look-ahead, and there equivalent model Muller
streaming string transducers, \emph{cannot} be considered as models of
computation for infinite word functions. This was observed in~\cite{DBLP:journals/lmcs/DaveFKL22}, where it
is shown that the problem of deciding whether a given regular function of
infinite words is computable is \textsc{Pspace-c}. On the other hand, 
deterministic two-way transducers \emph{without look-ahead} are a
proper model of computation for functions of infinite words.

\subparagraph{Deterministic regular functions of infinite words}
Motivated by the latter observation, the class of functions computed
by deterministic two-way transducers \emph{without}
look-ahead, coined
the class of \emph{deterministic regular functions}, was introduced
in~\cite{carton2022continuous}, where it is shown that they are also equivalently
computed by B\"uchi
\SST{} (\BSST). In \BSST, there is one special designated register $\out$ in
which to write the output word, which is required to be updated with
at least one new symbol infinitely often. For example, $\conddouble$ can be implemented by a single-state \BSST{} with two
registers $\out$ and $\reg$, updated as follows. On reading
$\sigma \in\{a,b\}$, it performs the updates $\out\mapsto \out.\sigma$
and $\reg\mapsto \reg.\sigma$, on reading~$1$, it does
$\out\mapsto \out.1$ and $\reg\mapsto \epsilon$, and on reading $2$,
it does $\out\mapsto \out.\reg.2$ and $\reg\mapsto \epsilon$.

Several important questions remain on the class of deterministic
regular functions, such as whether it is closed under composition,
whether it can be logically characterized by a natural fragment of
\MSO-transductions, and whether they can be obtained as finite compositions
of ``simple'' functions. In this paper, we provide positive answers to
these questions.

\subparagraph{Contributions} Concerning the class of
deterministic regular functions, our main results are:
\begin{itemize}
  \item its effective closure under composition;
  \item its characterization by means of finite compositions of
    sequential functions and  an extension of $\mapcr$ to infinite
    words;
  \item a logical characterization by a natural syntactic fragment of \MSO-transductions,
    the guarded fragment, called $\GMSOT$. 
\end{itemize}

An \MSO-transduction is defined as an \MSO-interpretation,
where the predicates of the output word structure, namely the successor
and label relations, are defined by \MSO formulas with
two and one free first-order variables respectively, interpreted over
a fixed number of copies of the input. The guarded fragment is defined by a classical restriction  (see
e.g.~\cite{DBLP:journals/jsyml/Gradel99} and references therein)
on the \MSO formulas composing the
\MSO-transduction. They have to be prefixed by an existential
quantifier $\exists \gv$, where $\gv$ is a word position, and all
quantifiers of the formula are guarded by the guard $x\leq \gv$ (and
$\forall x\in X, x\leq \gv$ for any set variable $X$). So, guarded
\MSO formulas on infinite words, only speak about finite
prefixes. Consider again the function $\conddouble$. Two copies of the
input are needed to account for potential duplication of the blocks,
but the presence or not of a successor edge between two nodes of the
output word structure, only depends on local properties, which are
definable by guarded \MSO formulas. E.g., such a property may
be ``if position $x+1$ is labeled $2$, then there is a successor
between the 1st copy of  $x$ and the 2nd copy of first position
of the block to which $x$ belongs''. 

In general, guarded \MSO formulas can test non-local properties, which
is the main source of technical difficulties in the paper. It is
illustrated by the next example.
\begin{example}
\label{ex:replace}
The function $\replace:  \{0,a,b\}^\omega \parfonc \{a,b\}^\omega$ of
  domain 
  $\Dom(\replace) = \{u \in  \{0,a,b\}^\omega: |u|_{a} {=} \infty \tnorm{ or } |u|_{b} {=}
  \infty\}$ and mapping $ 0^{n_1} {\sigma_1} 0^{n_2} {\sigma_2} \cdots
  \mapsto {\sigma_1}^{n_1+1} {\sigma_2}^{n_2+1} \cdots$ if $\sigma_i \in
  \{a,b\}$ and $n_i \in \Nat$, is deterministic regular. Replacing a zero at
  position $x$ by $a$ or $b$ depends on the next non-zero symbol in
  the future of $x$, which can be arbitrarily faraway, but occurs in a
  finite prefix if $u \in \Dom(\replace)$. This property is expressible with a guarded
  \MSO formula, which  defines the position holding this non-zero symbol
  as a guard. 
\end{example}

\subparagraph{Proof techniques and additional results} We now give an overview of the proof techniques used to show the
logical characterization, along with some other interesting
and useful results. We prove that deterministic two-way transducers
(\tDT) are expressively equivalent to \GMSOT. The conversion of \tDT{}
into \GMSOT is standard and follows the same line as~\cite{engelfriet2001mso}. The
converse is more involved and requires new techniques. First, we
convert \GMSO-transductions into deterministic two-way transducers
with \emph{finite look-ahead} (\tDTLA), which account for non-local, but finite, properties, as illustrated before. \tDTLA{} are equipped with regular languages of finite words
on their transitions, which act as finite look-aheads in the
following sense: when the reading head is at some position $i$ of an
infinite word $u$, in some state $q$, a transition from $q$ with
look-ahead $L$ is enabled if there exists a position $j\geq i$, called
witness, such
that the infix $u[i{:}j]$ starting at position $i$ and ending at
position $j$, belongs to $L$. If no transition is enabled at state
$q$, the computation fails. To ensure determinism, if several 
transitions are enabled, only the transition with minimal
(i.e.\ smallest) witness $j$ is
triggered, and a disjointness requirement on the look-aheads make sure
that this $j$ is unique. The condition to consider only the transition
with minimal witness $j$
is crucial to ensure that \tDTLA{} define only computable
functions. Indeed, a \tDTLA{} can be executed as follows: all finite
look-aheads, supposed for instance to be finitely represented by DFA,
are executed in parallel. By the minimality requirement for
$j$ and the disjointness of  look-aheads, \emph{as soon as} a prefix is accepted by one look-ahead DFA, the
corresponding transition is triggered.

Adding look-aheads to two-way transducers in order to capture \MSO-transductions is standard on finite words~\cite{engelfriet2001mso,courcelle2012graph}, for example because the ``moves''
of the \MSO-transduction depends on non-local properties. Look-aheads
are then directly removed by using the closure under composition
of deterministic two-way transducers~\cite{chytil1977serial}. Closure under
composition of deterministic two-way transducers on infinite words is,
to the best of our knowledge, unknown, and instead we give a direct
proof of finite look-ahead removal. It is
our main technical result: any \tDTLA{} is effectively equivalent to
some \tDT. To prove this result, classical techniques, such as
Hopcroft-Ullman construction~\cite{aho1969general} or the tree outline
construction~\cite{dartois2017reversible} do not apply, as they heavily rely on the fact
that words are finite. In our setting, we instead use a new
technique, based on summarizing the computations of the look-aheads
into trees which we prove to be bounded. As a side result of finite
look-ahead removal, we prove that \tDT{} (and so deterministic
regular functions) are closed under composition. Classically, closure
under composition of \MSO-transductions is direct, by formula
substitutions~\cite{courcelle2012graph}. This technique however does not apply here, as
the guarded \MSO formulas are not syntactically  closed under formula
substitution, making the correspondence between \GMSOT and \tDT{}
crucial to obtain closure under composition of \GMSO-transductions.

\subparagraph{Structure of the paper} In \cref{sec:detreg}, we
introduce the class of deterministic regular functions. In
\cref{sec:decomp}, we prove its closure under composition and
the decomposition result. In \cref{sec:MSO}, we introduce
guarded \MSO-transductions and state the logical
characterization. Since its proof is based on a compilation into
deterministic two-way transducers with finite look-ahead, we prove in
\cref{sec:twoway} how to remove those look-aheads. Finally, we
prove the logical characterization is \cref{sec:proofmain}. All
transformations are effective in the paper. Some proofs are only
sketched or simply omitted, but the proof details can be
found in Appendix.

\section{Deterministic regular functions}\label{sec:detreg}

In this section, we introduce the class of \emph{deterministic
regular functions of infinite words} and recall that it can be described
by two computation models: \emph{deterministic two-way transducers}
and \emph{deterministic Büchi streaming string transducers}.

\subparagraph{Notations}
Letters $\Sigma,\Gamma$ denote alphabets, i.e.\ finite sets of letters. The
set $\Sigma^*$ (resp.\ $\Sigma^+$, $\Sigma^\omega$) denotes the set of
finite words (resp.\  non-empty finite words, infinite words) over the
alphabet $\Sigma$. Let $\finin{\Sigma} \defined \Sigma^* \cup
\Sigma^\omega$.  If $u \in \finin{\Sigma}$, we let $|u| \in
\Nat \cup \{\infty\}$ be its length, $|u|_\sigma \in \Nat \cup \{\infty\}$
be the number of occurrences of $\sigma \in \Sigma$ and $u[i] \in \Sigma$
be the $i$-th letter of $u$ for $1 \le i \le |u|$.  If $1 \le i \le j \le
|u|$, $u[i{:}j]$ stands for $u[i] \cdots u[j]$.  We write $u[i{:}]$ for
$u[i{:}|u|]$.  If $j > |u|$ we let $u[i{:}j] \defined u[i{:}|u|]$.  If $j <
i$ we let $u[i{:}j] \defined \movi$. In this paper, functions are
\emph{by default} partial (i.e.\ possibly with non-total
domain). A (partial) function $f$ from $S$ to $T$ is denoted
$f: S \parfonc T$, and its domain is denoted $\Dom(f)\subseteq S$. A
total function from $S$ to $T$ is denoted $f: S \fonc T$.

\subparagraph{Two-way transducers} Let us recall
the syntax of two-way transducers. We consider here that the machines
work on infinite words, and have a Büchi acceptance
condition.

\begin{definition}[Two-way transducer] \label{def:two-way}
  A \emph{deterministic two-way transducer} \emph{(\tDT)} denoted
  $\trans = (\Sigma,\Gamma,Q,q_0,F,\delta,\lambda)$ consists of:
  \begin{itemize}
  \item an input alphabet $\Sigma$ and an output alphabet $\Gamma$;
  \item a finite set of states $Q$
    with an initial state $q_0 \in Q$ and
    a set of final states $F \subseteq Q$;
  \item a transition function $\delta: Q \times (\Sigma \uplus
    \{{\lmark}\}) \parfonc Q \times \{\lmove, \rmove\}$;
  \item an output function $\lambda: Q \times (\Sigma \uplus \{{\lmark}\})
    \parfonc \Gamma^*$ with same domain as $\delta$.
  \end{itemize}
\end{definition}

A \emph{configuration} of $\trans$ over
$u \in \finin{(\Sigma \cup \{\lmark\})}$ is a tuple $(q,i)$ where $q \in Q$ is the current state
and $1 \le i \le |u|$ is the current position of the reading head.  The \emph{transition
relation} $\rightarrow$ is defined as follows.  Given a configuration
$(q,i)$, let $(q',\star):= \delta(q,u[i])$.  Then $(q, i) \rightarrow (q',
i')$ whenever either $\star = \lmove$ and $i' = i-1$ (move left), or $\star
= \rmove$ and $i' = i+1$ (move right).  A \emph{run} over $u$ is a
(finite or infinite) sequence of consecutive configurations $(q_1,i_1)
\rightarrow (q_2,i_2) \rightarrow \cdots$.

Now, we define the infinite output
produced by $\trans$ when given the infinite word
$u \in \Sigma^\omega$ as input.
First, we let $u[0] \defined{} \lmark$,
i.e.\ we force the symbol ${\lmark}$ to be used
to mark the beginning of the input. 
An \emph{accepting} run is an infinite run that starts in $(q_0,0)$, visits
infinitely often configurations of the form $(q,i)$ with $q \in F$ and such
that $i_n \rightarrow \infty$ when $n \rightarrow \infty$ (without this
last condition, the transducer may enter an infinite loop without reading
its whole input).
The partial function $f: \Sigma^\omega \parfonc \Gamma^\omega$ computed by
$\trans$ is defined as follows. Let $u \in \Sigma^\omega$ be such that
there exists a (unique) accepting run $(q^u_0,i^u_0) \rightarrow
(q^u_1,i^u_1) \rightarrow \cdots$ labelled by ${\lmark} u$. Let $v \defined
\prod_{j=1}^{\infty} \lambda(q^u_j, ({\lmark} u)[i^u_j]) \in \Gamma^* \cup
\Gamma^\omega$ be the concatenation of the outputs produced along this run.
If $v \in \Gamma^\omega$, we define $f(u) \defined v$.  Otherwise $f(u)$ is
undefined.

\begin{definition}
  The class of \emph{deterministic regular functions of infinite words} is the class of
  (partial) functions computed by deterministic two-way transducers.
\end{definition}

We have explained in \cref{ex:intro1bis} how to compute
the function $\conddouble$ using a \tDT{}
(without look-aheads). Observe that the function  $\replace$
from \cref{ex:replace} can be computed in a similar fashion.
Hence both functions are deterministic regular.



\begin{example} \label{ex:mapcr}
        Let us extend the function $\mapcr$ of \cref{ex:mapcr-fini}
        to infinite words. Let $\Sigma$ be an alphabet,
        we define $\mapcr : (\Sigma\uplus \{\hash{}\})^\omega \fonc (\Sigma\uplus \{\hash{}\})^\omega$
        as follows:
  \begin{itemize}
  \item $\mapcr(u_1 \hash{} u_2 \hash{} \cdots) \defined u_1 \hash{} \mirror{u_1}
    \hash{} u_2  \hash{} \mirror{u_2} \hash{} \cdots$ with $u_i \in \Sigma^*$ for all $i
    \ge 0$;
  \item $\mapcr(u_1 \hash{} \cdots \hash{} u_n \hash{} u) \defined u_1 \hash{}
    \mirror{u_1} \hash{} \cdots \hash{} u_n \hash{} \mirror{u_n} \hash{} u$ for
    $u_i \in \Sigma^*$ and $u \in \Sigma^\omega$.
  \end{itemize}
  This function is deterministic regular since we can build a \tDT{} that processes
  twice each $\hash{}$-free factor (or only once for the last
  infinite one if it exists).
\end{example}

\subparagraph{B\"uchi Streaming String Transducers}
Now, we describe a model of a one-way machine
with registers which captures deterministic regular functions
of infinite words. Over finite words, it is
well-known that deterministic two-way transducers are equivalent to
\emph{copyless streaming string transducers} \cite{alur2010expressiveness}.
A similar equivalence holds for the class
of \emph{regular functions of infinite words}, which can
equivalently be described by
\emph{deterministic two-way transducers with regular look-aheads}
or \emph{copyless streaming string transducers with Muller
  conditions}~\cite{alur2012regular}. However, Muller
conditions enable to check regular properties of the infinite input,
and thus describe functions which are not (Turing)
computable~\cite{alur2012regular}.
Now, let us recall
the model of \emph{B\"uchi deterministic streaming string
  transducer} (\DSST), introduced by Carton and Douéneau-Tabot in
\cite{carton2022continuous}, that captures exactly the class
of deterministic regular functions.

Formally, a B\"uchi deterministic streaming string
  transducer consists of a one-way deterministic automaton with
a finite set $\Regs$ of registers that store words from $\Gamma^*$. We use
a distinguished register $\out$ to store the output produced when reading
an infinite word.  The registers are modified when reading the input using
\emph{substitutions}, i.e.\ mappings $\Regs \rightarrow (\Gamma \uplus
\Regs)^*$.  We denote by $\subst{\Regs}{\Gamma}$ the set of these
substitutions.  They can be extended morphically from $(\Gamma \uplus
\Regs)^*$ to $ (\Gamma \uplus \Regs)^*$ by preserving the elements of
$\Gamma$.

\begin{example}[Substitutions] \label{ex:subst} Let
  $\Regs = \{\reg,\regg\}$ and $\Gamma = \{b\}$.  Consider $\tau_1:= \reg
  \mapsto b, \regg \mapsto b\reg \regg b$ and $\tau_2:= \reg \mapsto \reg b,
  \regg \mapsto \reg \regg$, then $ \tau_1 \circ \tau_2 (\reg) =
  \tau_1(\reg b) = bb$ and $\tau_1 \circ \tau_2 (\regg) = \tau_1(\reg
  \regg) = bb\reg \regg b$.
\end{example}

\begin{definition}
  A \emph{B\"uchi deterministic streaming string transducer} \emph{(\DSST)}
  denoted by $\trans = (\Sigma,\Gamma, Q,F,q_0, \delta, \Regs, \out, \lambda)$
  consists of:
  \begin{itemize}
  \item a finite input (resp.\ output) alphabet $\Sigma$ (resp.\ $\Gamma$);
  \item a finite set of states $Q$ with $q_0 \in Q$ initial and $F
    \subseteq Q$ final;
  \item a transition function $\delta: Q \times \Sigma \parfonc Q$;
  \item a finite set of registers $\Regs$ with a distinguished output
    register $\out \in \Regs$;
  \item an update function $\lambda: Q \times \Sigma \parfonc
    \subst{\Regs}{\Gamma}$ such that for all $(q,\sigma) \in \Dom(\lambda)
    = \Dom(\delta)$:
    \begin{itemize}
    \item $\lambda(q,\sigma)(\out) = \out \cdots$;
    \item there is no other occurrence of $\out$ among the
      $\lambda(q,\sigma)(\reg)$ for $\reg \in \Regs$.
    \end{itemize}
  \end{itemize}
\end{definition}

This machine defines a partial function $f: \Sigma^\omega \parfonc
\Gamma^\omega$ as follows.  For $i \ge 0$ let $q^u_i:=
\delta(q_0,u[1{:}i])$ (when defined).  For $i \ge 1$, we let $\lambda^u_i
\defined \lambda(q^u_{i-1},u[i])$ (when defined) and $\lambda^u_0(\reg) =
\movi$ for all $\reg \in \Regs$.  For $i \ge 0$, let
$\cro{\cdot}^u_i \defined \lambda^u_0 \circ \cdots \circ \lambda^u_i$.  By
construction $\cro{\out}^u_i$ is a prefix of $\cro{\out}^u_{i+1}$ (when
defined). If $\cro{\out}^u_i$ is defined for all $i \ge 0$, $q^u_i$ is a
state of $F$ infinitely often, and $| \cro{\out}^u_i| \rightarrow +
\infty$, then we let $f(u) \defined \bigvee_i \cro{\out}^u_i$ (the symbol
$\vee$ is used to denote the unique $v\in \Gamma^\omega$ such that
$\cro{\out}^u_i $ is a prefix of $v$ for all $i \ge 0$).  Otherwise $f(u)$ is
undefined.

\begin{example} \label{ex:replace-sst}
  The function $\replace$ from \cref{ex:replace} can be computed by
  a \DSST{}. For all $i \ge 1$, it crosses the block $0^{n_i}$ and computes
  $1^{n_i}$ and $2^{n_i}$ in two registers.  Once it sees ${\sigma_i}$ it
  adds in $\out$ the register storing ${\sigma_i}^{n_i}$.
\end{example}



\begin{definition}[Copyless, bounded copy]
  We say that a substitution $\tau \in \subst{\Regs}{B}$ is
  \emph{copyless} (resp.\ \emph{$K$-bounded}) if for all $\reg \in \Regs$,
  $\reg$ occurs at most once in $\{ \tau(\regg): \regg \in \Regs\}$
  (resp.\ for all $\reg, \regg \in \Regs$, $\reg$ occurs at most $K$ times
  in $\tau(\regg)$).
  We say that a \DSST{} $\trans = (\Sigma,\Gamma, Q,q_0, \delta, \Regs, \out,
  \lambda)$ is \emph{copyless} (resp.\ \emph{$K$-bounded}) if for all $u \in
  \Sigma^\omega$ and $i \le j$ such that $\lambda^u_i \circ \cdots \circ
  \lambda^u_j$ is defined, this substitution is copyless (resp.\ 
  $K$-bounded).
\end{definition}

\begin{remark}
  The composition of two copyless substitutions is copyless, hence a
  \DSST{} is copyless as soon as $\lambda(q,\sigma)$ is copyless for all $q
  \in Q$ and $\sigma \in \Sigma$.  However, $K$-boundedness is not
  necessarily preserved under composition.
\end{remark}

Observe that the \DSST{} described in \cref{ex:replace-sst}
is copyless. Now, we recall the result of Carton and Dou\'eneau-Tabot
that proves equivalence between two-way transducers,
copyless, and bounded copy B\"uchi deterministic streaming string transducers.

\begin{theorem}[{\cite[Theorem 3.7]{carton2022continuous}}]
\label{theo:2dt-dsst} The following machines compute
the same class of partial functions over infinite words:
\begin{enumerate}
\item deterministic two-way  transducers \emph{(\tDT{})};
\item $K$-bounded deterministic B\"uchi
streaming string transducers \emph{(}$K$-bounded \DSST\emph{)};
\item copyless deterministic B\"uchi
streaming string transducers \emph{(}copyless \DSST\emph{)}.
\end{enumerate}
Furthermore, all the conversions are effective.
\end{theorem}

\begin{remark}\label{rem:allstatesfinal} The original proof of \cite{carton2022continuous}
which transforms a \tDT{} into a \DSST{} only considers machines where all
states are final. Nevertheless, the proof can easily be adapted to
transducers with non-final states. Furthermore, given a \DSST{}
(possibly with non-final states) one can build an equivalent \DSST{}
where all states are final by {\cite[Lemma D.1]{carton2022continuous}}
(the B\"uchi conditions are hidden in the fact that the output must be infinite).
All in all, all the models (with all states final or not) exactly capture
the class of deterministic regular functions.
\end{remark}


Finally, we recall the domains
of deterministic regular functions.
We say that a language is \emph{B\"uchi deterministic}
if  it is accepted by a deterministic B\"uchi automaton
(see e.g.\ \cite{perrin2004}).

\begin{proposition}[{\cite{carton2022continuous}}] \label{prop:dom-reg} If $f$ is deterministic
regular, then $\Dom(f)$ is B\"uchi deterministic.
\end{proposition}

\section{Composition and decomposition theorems}
  \label{sec:decomp}

In this section, we show that deterministic regular functions
are closed under composition, and that conversely they
can be written as the composition of some ``basic'' functions.

It is known since \cite{chytil1977serial} (resp.\ \cite{alur2012regular})
that the class of regular functions of finite (resp.\ infinite) words
is closed under composition. We transport this result
to deterministic regular functions
of infinite words in \cref{thm:closure-comp}. However,
its proof is not an immediate extension of the regular case,
and it illustrates the main difficulty of this paper:
since look-aheads are not allowed,
it is complex for a \tDT{} to check if some property
happens \emph{after} its current position.

\begin{theorem}
        \label{thm:closure-comp}
    Deterministic regular functions are
    (effectively) closed under composition.
\end{theorem}

\begin{proof}[Proof idea.]
The approach is to compose the two transducers directly (using a product construction); the difficulty in the composition of two computations arises when one transducer is moving forward and the other backward. In that case, we need to rewind the computation of the transducer that moves backward by one computation step.

To recover the previous configuration look-ahead comes in handy.
As mentioned above, (infinite) look-aheads are not permitted, but we use a weaker form of \emph{finite} look-aheads (to be introduced in \cref{sec:twoway}) which does not increase the expressiveness of deterministic two-way transducers over infinite words (and can be effectively removed), see \cref{thm:laremoval}. 
Finite look-aheads account for non-local but finite properties.
The look-ahead we define basically re-traces the computation that the two-way transducer has taken so far.
Note that this is indeed a finite property as only a prefix of the input has been visited by the computation of the two-way transducer.
%
\end{proof}


As an easy consequence of \cref{thm:closure-comp},
let us observe that deterministic regular functions
(effectively) preserve B\"uchi deterministic languages by inverse image.
Analogue results hold for regular functions of finite (resp.\ infinite) words
with regular languages.
 
\begin{proposition} If $f : \Sigma^\omega \parfonc \Gamma^\omega$ is
deterministic regular and $L \subseteq \Gamma^\omega$
is B\"uchi deterministic, then $f^{-1}(L) \subseteq \Sigma^\omega$ is
(effectively) B\"uchi deterministic.
\end{proposition}

\begin{proof} The function $f \circ \ide_L$ (where
$\ide_L : \Gamma^\omega \parfonc \Gamma^\omega$ is
the identity function restricted to $L$) is deterministic regular.
It domain $f^{-1}(L)$ is 
B\"uchi deterministic by \cref{prop:dom-reg}.
\end{proof}

Let us now focus on the converse of  \cref{thm:closure-comp},
i.e.\ showing that any deterministic regular function
can be written as a composition of ``basic'' functions.
As mentioned in introduction, regular functions of finite
words can be written as compositions of
$\mapcr$ (see \cref{ex:mapcr-fini}) and sequential functions
(computed by one-way transducers).

\begin{theorem}[{\cite[Theorem 13]{bojanczyk2020single}}]
\label{theo:KR-finite} Over finite words, a function is regular
if and only if it can (effectively) be written as
a composition of $\mapcr$
and sequential functions.
\end{theorem}

To state our similar result for deterministic regular functions
of infinite words, we first recall formally the definition 
of sequential functions of infinite words.

\begin{definition}[Sequential functions]
\label{def:sequential}
A \emph{deterministic one-way transducer}
is a \tDT{} $(\Sigma,\Gamma, Q, q_0, F, \delta, \lambda)$ such
that for all $q \in Q$ and $\sigma \in (\Sigma\uplus \{\lmark\})$,
$\delta(q,\sigma)$ has shape $(\_, \rmove$) (when defined).
The class of (partial) functions over infinite words
computed by one-way deterministic
transducers is called \emph{sequential functions
of infinite words.}
\end{definition}

\begin{example} Any function that replaces some letter
of its input by another letter is sequential.
The functions $\replace$ and $\mapcr$
of \cref{ex:replace,ex:mapcr} are \emph{not} sequential
(this can be shown using a pumping argument). Observe that $\replace$
can be written as the composition of: a sequential function
that replaces each $\sigma_i \in \{1,2\}$ by
$\sigma_i \hash{}$, the function $\mapcr$,
and finally a sequential function that uses the first copy
of each block to determine the value of $\sigma_i$,
and transforms the (mirror) second copy accordingly.
\end{example}

Now, we state the decomposition result,
that also uses $\mapcr$ from \cref{ex:mapcr}.
Its proof is somehow technical and it illustrates
once more the main difficulty of this paper:
deterministic regular functions are not able
to check many properties about the ``future''.

\begin{theorem}
        \label{theo:KR-detreg}
        A function is deterministic regular
        if and only if it can (effectively) be written as a composition of
        $\mapcr$  and sequential functions of infinite words.
\end{theorem}

\begin{proof}[Proof idea]
In the case of finite words, the proofs of
\cite{bojanczyk2018polyregular, bojanczyk2020single}
rely on Simon's factorization forests theorem \cite{simon1990factorization}.
They first build a factorization forest,
and then use its structure to simulate
the runs of a transducer. Furthermore, over finite words,
such forests can be computed by a \emph{rational function},
which is a composition of sequential functions and $\mapcr$.
We follow a similar proof sketch for infinite words,
but the main issue is that factorization forests can no
longer be computed by a composition
of sequential functions and $\mapcr$
(their structure may depend on regular
properties of the input). Thus we use instead a 
weakened version of forests, introduced
by Colcombet  under the name of
\emph{forward Ramseyan splits} \cite{colcombet2007combinatorial}.
Such splits can be computed with a sequential function.
Our new techniques show how to simulate the runs of a transducer
by using a forward Ramseyan split.
\end{proof}

\section{Guarded MSO-transductions}
\label{sec:MSO}

In this section, we define the logic \MSO over finite and infinite
words, as well as \MSO-transductions, and its guarded fragment. We also
state the logical characterization of deterministic regular functions
(Theorem~\ref{thm:main}).

\subparagraph{MSO on infinite words}

Infinite words over
$\Sigma$ are seen as structures of domain $\mathbb{N}$, over 
the signature $\mathcal{W}_\Sigma = \{ S(x,y), (\sigma(x))_{\sigma\in \Sigma}\}$
which consists of the successor predicate $S(x,y)$,
naturally interpreted as the successor over $\mathbb{N}$, and unary
predicates $\sigma(x)$ for all $\sigma\in \Sigma$, interpreted as the
set of positions labelled $\sigma$. Given an infinite word
$u\in\Sigma^\omega$, we denote by $G_u$ the structure it induces, and 
just $u$ when it is clear  that
$u$ denotes the structure $G_u$.

Monadic second-order formulas are defined as first-order logic
formulas, which can additionally use quantifiers $\exists X,\forall X$ over sets of positions, 
and membership atomic formulas of the form $x\in X$, where
$x$ is a first-order variable while $X$ is a set variable. We denote
by $\MSO[\Sigma,S,\leq]$ (or just \MSO when the predicates are clear
from the context), the set of monadic second-order formulas over the
word signature $\mathcal{W}_\Sigma$ extended with the order predicate
$\leq$ (interpreted by the natural order on $\mathbb{N}$).  It is
well-known that the predicate $\leq$ is syntactic sugar. The semantics is defined as expected (details can be found in \cite{thomas1997languages,courcelle2012graph}
for instance). For a formula $\phi$ with sets of free first-order and set
variables $\overline{x},\overline{X}$ (we use the tuple notation which
implicitly assumes an order between variables), we may write it
$\phi(\overline{x},\overline{X})$ to explicit the free variables of $\phi$.
We also denote by $Free(\phi)$ the free (first-order and set) variables of
$\phi$.%
Given a word $w$, an $n$-tuple of positions
$\overline{p}$ of $w$ and an $m$-tuple $\overline{P}$ of sets of
positions of $w$, we write $w\models \phi(\overline{p},\overline{P})$
to mean that the structure induced by $w$ is a model of $\phi$ under
assignments $\overline{p}$ and $\overline{P}$.

\begin{example}\label{ex:mso}
    The formula $\first(x) = \forall y\cdot \neg S(y,x)$ is
    satisfied by any word and position $x$ such that $x$ is the first
    position to the left. 
\end{example}

Over an alphabet $\Sigma$, any \emph{closed} formula $\phi\in\MSO$
\emph{defines} a regular language $L_\phi = \{ u
\in\Sigma^\omega\mid u \models \phi\}$. By B\"uchi-Elgot-Trakhtenbrot's
theorem~\cite{tra61short,Bu60,Elg61}, it is known $\MSO$ defines precisely
the class of regular languages over alphabet~$\Sigma$: for any
language $L$ over $\Sigma$, $L$ is regular if and only if $L = L_\phi$ for
some $\phi\in\MSO$. \MSO formulas can also be interpreted over finite
word structures, whose domains are the (finite) set of word positions.
It is also well-known that a language of finite words is regular iff
it is MSO-definable.

\subparagraph{\MSO-transductions of infinite words}\label{subsec:MSOT}

\MSO-transductions define transformations of graph structures, and have been studied in the
context of finite words by Engelfriet and Hoogeboom in~\cite{engelfriet2001mso}
(see also \cite{courcelle2012graph} for a more recent introduction
to \MSO-transductions). The main result of~\cite{engelfriet2001mso} is a
B\"uchi-like theorem: a function of finite words is \MSO-definable if and only if it is
regular (i.e.\ recognizable by a deterministic two-way transducer). This
result was then lifted to functions of infinite words
in~\cite{alur2012regular}, but deterministic two-way
transducers may need infinite look-aheads to capture the full expressive power of
\MSO-transductions.

In an \MSO-transduction, the output word structure is defined via an \MSO
interpretation over a fixed number $k$ of copies of the input word (seen as
a structure). Therefore, the nodes of the output word are copies $1$ to $k$
of the nodes of the input word.  Output nodes are pairs $(i,c)$ (often
denoted $i^c$), for every copy~$c$ and input node~$i$.

The output label and successor predicates are defined by \MSO formulas with
one and two free first-order variables respectively, interpreted over the
input structure. For instance, over the output alphabet $\Gamma = \{a,b\}$,
to set all the output labels to $a$, one just specifies the formulas
$\phi_a^c(x) = \top$ and $\phi_b^c(x)=\bot$ for all copies~$c$.  The output
successor predicate relates input nodes of possibly different copies, and
is therefore defined by formulas of the form $\phi_S^{c,d}(x,y)$, indexed
by copies $c,d \in \{1,\dots,k\}$.

Finally, there is one distinguished copy~$c_0$ together with a formula
$\phi_{\fst}^{c_0}(x)$, which must be satisfied by at most one node $x$.
Intuitively, if the output structure is a word, this formula defines the
first node of the output word. The domain of the output structure is
composed of all nodes that can be reached from the initial node $x^{c_0}$
by following multiple successor edges. In general, the output structure of
an input word $u$ by an MSO-transduction $\mathcal{T}$ might not be an
infinite word structure, in which case $u$ is not in the domain of the
function defined by $\mathcal{T}$.

Formally, an \emph{\MSO-transduction} over an input alphabet $\Sigma$
and output alphabet $\Gamma$ is a tuple
$\mathcal{T} = (k, (\phi_{\gamma}^c(x))_{1\leq c\leq k,\gamma\in\Gamma}, (\phi_{S}^{c,d}(x,y))_{1\leq
  c,d\leq k},c_0,\phi_{\fst}^{c_0}(x))$
where $k\in\mathbb{N}\setminus\{0\}$, $1\leq c_0\leq k$ and for all
input $u\in\Sigma^\omega$, there is at most one position~$i$ such that
$u\models \phi_{\fst}^{c_0}(i)$. We may omit $c_0$ in the tuple
above.

We now formally define the semantics of \MSO-transductions. Let
$u\in\Sigma^\omega$ and $N\subseteq \mathbb{N}\times \{1,\dots,k\}$. We
first define the set of output nodes that can be reached from $N$ in zero
or more steps. We let $\textsf{Post}^0_{u}(N) = N$ and for all $\ell>0$,
\[
\textsf{Post}^\ell_u(N) = \{ j^d\mid \exists i^c\in
\textsf{Post}^{\ell-1}_u(N)\cdot u\models \phi_S^{c,d}(i,j)\}
\text{ and } \textsf{Post}^*_u(N) = \bigcup_{\ell\geq 0} \textsf{Post}^{\ell}_u(N)
\]

Given an \MSO-transduction $\mathcal{T}$ as above, and input word
$u\in\Sigma^\omega$, the \emph{output structure}, denoted $\trans(u)$, is the
structure over signature $\mathcal{W}_\Gamma$ defined by the following
interpretation:
\begin{itemize}
  \item the domain is $D = \textsf{Post}^*_u(\{i^{c_0}\mid u\models
    \phi_{\fst}^{c_0}(i)\})$ (note that the argument of
    $\textsf{Post}^*_u$ is either empty or a singleton)
  \item a node $i^c\in D$ is labelled $\gamma\in\Gamma$ if $u\models
    \phi_\gamma^c(i)$
  \item a node $j^d$ is a successor of a node $i^c$ if $u\models
    \phi_S^{c,d}(i,j)$. 
\end{itemize}

The output structure $\trans(u)$ may not be a word structure. For instance,
a node might have multiple labels, $\trans(u)$ may contain cycles, or
branching. So we restrict semantically the function defined by $\trans$ to word
structures. Formally, the \emph{function defined by $\trans$} is the
function $\sem{\trans}:\Sigma^\omega\parfonc \Gamma^\omega$ whose graph is:
\[
\{(u,v)\in \Sigma^\omega\times\Gamma^\omega\mid G_v\text{ (the
  structure associated with $v$) is
  isomorphic to } \trans(u)\}
\]

We denote by $\MSOT$ the set of $\MSO$-transductions and say that a
function $f:\Sigma^\omega \parfonc \Gamma^\omega$ is $\MSOT$-definable
if $f = \sem{\trans}$ for some $\trans\in\MSOT$.

\begin{figure*}[t]
\centering

\begin{subfigure}{1\textwidth}

\begin{tikzpicture}[->,>=stealth',shorten >=1pt,auto,scale=0.75,every node/.style={scale=0.75}]

\tikzstyle{alivenode}=[circle,fill=black!80,thick,inner sep=1pt,minimum size=4mm]

\tikzstyle{deadnode}=[circle,fill=black!20,thick,inner sep=0pt,minimum size=4mm]

\node at (-1,1.5)  {{\footnotesize input}};
\node at (-1,1) {{\footnotesize word}};
\node at (-1,0) {copy $1$} ;
\node at (-1,-1.5) {copy $2$} ;

\node [alivenode] (a1) at (0,1) {{\color{white} $a$}} ;
\node [alivenode](a2) at (2,1) {{\color{white} $c$}} ;
\node [alivenode](a3) at (4,1) {{\color{white} $a$}} ;
\node [alivenode](a4) at (6,1) {{\color{white} $b$}} ;
\node [alivenode](a5) at (8,1) {{\color{white} $a$}} ;
\node [alivenode](a6) at (10,1) {{\color{white} $c$}} ;
\node [alivenode](a7) at (12,1) {{\color{white} $a$}} ;
\node [alivenode](a8) at (14,1) {{\color{white} $c$}} ;
\node (a9) at (16,1) {{$\ldots$}} ;

  \draw[->]   (a1) -- node[above] {$S$} (a2) ;
  \draw[->]   (a2) -- node[above] {$S$} (a3) ;
  \draw[->]   (a3) -- node[above] {$S$} (a4) ;
  \draw[->]   (a4) -- node[above] {$S$} (a5) ;
  \draw[->]   (a5) -- node[above] {$S$} (a6) ;
  \draw[->]   (a6) -- node[above] {$S$} (a7) ;
  \draw[->]   (a7) -- node[above] {$S$} (a8) ;

\node (a1) at (0,1.5) {\begin{scriptsize}$1$\end{scriptsize}} ;
\node (a2) at (2,1.5) {\begin{scriptsize}$2$\end{scriptsize}} ;
\node (a3) at (4,1.5) {\begin{scriptsize}$3$\end{scriptsize}} ;
\node (a4) at (6,1.5) {\begin{scriptsize}$4$\end{scriptsize}} ;
\node (a5) at (8,1.5) {\begin{scriptsize}$5$\end{scriptsize}} ;
\node (a6) at (10,1.5) {\begin{scriptsize}$6$\end{scriptsize}} ;
\node (a7) at (12,1.5 ) {\begin{scriptsize}$7$\end{scriptsize}} ;
\node (a8) at (14,1.5 ) {\begin{scriptsize}$8$\end{scriptsize}} ;

\node [alivenode] (x1) at (0,0) {{\color{white} $a$}} ;
\node [alivenode] (x2) at (2,0) {{\color{white} $c$}} ;
\node [alivenode] (x3) at (4,0) {{\color{white} $a$}} ;
\node [alivenode] (x4) at (6,0) {{\color{white} $b$}} ;
\node [alivenode] (x5) at (8,0) {{\color{white} $a$}} ;
\node [alivenode] (x6) at (10,0) {{\color{white} $c$}} ;
\node [alivenode] (x7) at (12,0) {{\color{white} $a$}} ;
\node [alivenode] (x8) at (14,0) {{\color{white} $c$}} ;
\node  (x9) at (16,0) {$\ldots$} ;

\node [alivenode] (y1) at (0,-1.5) {{\color{white} $a$}} ;
\node [deadnode] (y2) at (2,-1.5) {{\color{white} $c$}} ;
\node [alivenode] (y3) at (4,-1.5) {{\color{white} $a$}} ;
\node [deadnode] (y4) at (6,-1.5) {{\color{white} $b$}} ;
\node [deadnode] (y5) at (8,-1.5) {{\color{white} $a$}} ;
\node [deadnode] (y6) at (10,-1.5) {{\color{white} $c$}} ;
\node [deadnode] (y7) at (12,-1.5) {{\color{white} $a$}} ;
\node [deadnode] (y8) at (14,-1.5) {{\color{white} $c$}} ;
\node  (y9) at (16,-1.5) {$\ldots$} ;

  \draw[->]   (x1) -- node[pos=0.2,right] {$\phi_S^{1,2}$} (y1) ;
  \draw[->]   (y1) -- node[below] {$\phi_S^{2,1}$} (x2) ;

  \draw[->]   (x2) -- node[below] {$\phi_S^{1,1}$} (x3) ;

  \draw[->]   (x3) -- node[pos=0.2,right] {$\phi_S^{1,2}$} (y3) ;
  \draw[->]   (y3) -- node[below] {$\phi_S^{2,1}$} (x4) ;

  \draw[->]   (x4) -- node[below] {$\phi_S^{1,1}$} (x5) ;
  \draw[->]   (x5) -- node[below] {$\phi_S^{1,1}$} (x6) ;
  \draw[->]   (x6) -- node[below] {$\phi_S^{1,1}$} (x7) ;
  \draw[->]   (x7) -- node[below] {$\phi_S^{1,1}$} (x8) ;

  \draw[color=black!20,->]   (y2) -- node[below] {$\phi_S^{2,1}$} (x3) ;
  \draw[color=black!20,->]   (y4) -- node[below] {$\phi_S^{2,1}$} (x5) ;
  \draw[color=black!20,->]   (y5) -- node[below] {$\phi_S^{2,1}$} (x6) ;
  \draw[color=black!20,->]   (y6) -- node[below] {$\phi_S^{2,1}$} (x7) ;

  \draw[color=black!20,->]   (y7) -- node[below] {$\phi_S^{2,1}$} (x8) ;



\fill[blue!20,rounded corners, fill opacity=0.2] (-0.3,0.4) rectangle (16.3,-1.8);

\end{tikzpicture}
\caption{Input and output structures of the \MSO-transduction for the
function $\double$ of \cref{ex:intro1} on input word
$acab(ac)^\omega$\label{subfig:msot1}}
\end{subfigure}

\begin{subfigure}{1\textwidth}
\begin{tikzpicture}[->,>=stealth',shorten >=1pt,auto,scale=0.75,every node/.style={scale=0.75}]

\tikzstyle{alivenode}=[circle,fill=black!80,thick,inner sep=1pt,minimum size=4mm]

\tikzstyle{deadnode}=[circle,fill=black!20,thick,inner sep=0pt,minimum size=4mm]

\node at (-1,1.5)  {{\footnotesize input}};
\node at (-1,1) {{\footnotesize word}};
\node at (-1,0) {copy $1$} ;
\node at (-1,-1.5) {copy $2$} ;

\node [alivenode] (a1) at (0,1) {{\color{white} $a$}} ;
\node [alivenode](a2) at (2,1) {{\color{white} $b$}} ;
\node [alivenode](a3) at (4,1) {{\color{white} $1$}} ;
\node [alivenode](a4) at (6,1) {{\color{white} $b$}} ;
\node [alivenode](a5) at (8,1) {{\color{white} $a$}} ;
\node [alivenode](a6) at (10,1) {{\color{white} $2$}} ;
\node [alivenode](a7) at (12,1) {{\color{white} $b$}} ;
\node [alivenode](a8) at (14,1) {{\color{white} $2$}} ;
\node (a9) at (16,1) {{$\ldots$}} ;

  \draw[->]   (a1) -- node[above] {$S$} (a2) ;
  \draw[->]   (a2) -- node[above] {$S$} (a3) ;
  \draw[->]   (a3) -- node[above] {$S$} (a4) ;
  \draw[->]   (a4) -- node[above] {$S$} (a5) ;
  \draw[->]   (a5) -- node[above] {$S$} (a6) ;
  \draw[->]   (a6) -- node[above] {$S$} (a7) ;
  \draw[->]   (a7) -- node[above] {$S$} (a8) ;

\node (a1) at (0,1.5) {\begin{scriptsize}$1$\end{scriptsize}} ;
\node (a2) at (2,1.5) {\begin{scriptsize}$2$\end{scriptsize}} ;
\node (a3) at (4,1.5) {\begin{scriptsize}$3$\end{scriptsize}} ;
\node (a4) at (6,1.5) {\begin{scriptsize}$4$\end{scriptsize}} ;
\node (a5) at (8,1.5) {\begin{scriptsize}$5$\end{scriptsize}} ;
\node (a6) at (10,1.5) {\begin{scriptsize}$6$\end{scriptsize}} ;
\node (a7) at (12,1.5 ) {\begin{scriptsize}$7$\end{scriptsize}} ;
\node (a8) at (14,1.5 ) {\begin{scriptsize}$8$\end{scriptsize}} ;

\node [alivenode] (x1) at (0,0) {{\color{white} $a$}} ;
\node [alivenode] (x2) at (2,0) {{\color{white} $b$}} ;
\node [alivenode] (x3) at (4,0) {{\color{white} $1$}} ;
\node [alivenode] (x4) at (6,0) {{\color{white} $b$}} ;
\node [alivenode] (x5) at (8,0) {{\color{white} $a$}} ;
\node [alivenode] (x6) at (10,0) {{\color{white} $2$}} ;
\node [alivenode] (x7) at (12,0) {{\color{white} $b$}} ;
\node [alivenode] (x8) at (14,0) {{\color{white} $2$}} ;
\node  (x9) at (16,0) {$\ldots$} ;

\node [deadnode] (y1) at (0,-1.5) {{\color{white} $a$}} ;
\node [deadnode] (y2) at (2,-1.5) {{\color{white} $b$}} ;
\node [deadnode] (y3) at (4,-1.5) {{\color{white} $1$}} ;
\node [alivenode] (y4) at (6,-1.5) {{\color{white} $b$}} ;
\node [alivenode] (y5) at (8,-1.5) {{\color{white} $a$}} ;
\node [deadnode] (y6) at (10,-1.5) {{\color{white} $2$}} ;
\node [alivenode] (y7) at (12,-1.5) {{\color{white} $b$}} ;
\node [deadnode] (y8) at (14,-1.5) {{\color{white} $2$}} ;
\node  (y9) at (16,-1.5) {$\ldots$} ;

  \draw[->]   (x1) -- node[below] {$\phi_S^{1,1}$} (x2) ;
  \draw[->]   (x2) -- node[below] {$\phi_S^{1,1}$} (x3) ;
  \draw[->]   (x3) -- node[below] {$\phi_S^{1,1}$} (x4) ;
  \draw[->]   (x4) -- node[below] {$\phi_S^{1,1}$} (x5) ;
  \draw[->]   (x5) -- node[below] {$\phi_S^{1,2}$} (y4) ;
  \draw[->]   (y4) -- node[below] {$\phi_S^{2,2}$} (y5) ;
  \draw[->]   (y5) -- node[below] {$\phi_S^{2,1}$} (x6) ;
  \draw[->]   (x6) -- node[below] {$\phi_S^{1,1}$} (x7) ;
  \draw[->]   (x7) -- node[left] {$\phi_S^{1,2}$} (y7) ;
  \draw[->]   (y7) -- node[below] {$\phi_S^{2,1}$} (x8) ;

  \draw[color=black!20,->]   (y1) -- node[below] {$\phi_S^{2,2}$} (y2) ;
  \draw[color=black!20,->]   (y2) -- node[below] {$\phi_S^{2,2}$} (y3) ;
  \draw[color=black!20,->]   (y3) -- node[below] {$\phi_S^{2,2}$} (y4) ;
  \draw[color=black!20,->]   (y6) -- node[below] {$\phi_S^{2,2}$} (y7) ;



\fill[blue!20,rounded corners, fill opacity=0.2] (-0.3,0.4) rectangle (16.3,-2.1);

\end{tikzpicture}
\caption{Input and output structures of the \MSO-transduction for the
function $\conddouble$ of \cref{ex:intro1bis} on input word
$ab1ba2b2\dots$\label{subfig:guardedMSO}}
\end{subfigure}

\end{figure*}

\begin{example}\label{ex:msotrans}
    We consider again the function $\double$ of \cref{ex:intro1},
    illustrated on \cref{subfig:msot1} and show how to define it
    with an \MSO-transduction. Since some $a$ must be duplicated, two
    copies are needed, so $k=2$.  Labels are preserved: $\phi_{\sigma}^c(x) = \sigma(x)$ for all
    $c\in\{1,2\}$ and $\sigma\in\Sigma$. The first copy $c_0$ is $1$,
    and $\phi_{\fst}^{c_0}(x) = \first(x)$.  The successor formulas distinguish if
    there is a $b$ in the future or not. First, from the 2nd
    to the 1st copy, there is always a successor relation from a node to
  its successor in copy $1$:
  $\phi_S^{2,1}(x,y) = S(x,y)$. There is a successor from $x^1$ to
  $y^2$ if $x = y$, $x$ is labelled $a$ and there is a $b$ in the
  remaining infinite suffix starting at $x$:
  $\phi_S^{1,2}(x,y) =
  a(x)\wedge (x=y)\wedge \exists z\cdot x\leq z\wedge b(z)$.
  On the first copy, it depends on the label of the
  input: $\phi_S^{1,1}(x,y) = S(x,y)\wedge (a(x)\rightarrow (\forall
  z\geq x\cdot \neg b(z))$. On the  second copy, there is never a
  predicate edge: $\phi_S^{2,2}= {\perp}$. On \cref{subfig:msot1},
  the interpretation of those formulas is depicted, in bold if they
  are part of the output word, in light grey otherwise. One
  can see that the output structure induced by all the descendants of
  the first node (by the transitive closure of the successor relation)
  is isomorphic to the structure
  $G_{aacaab(ac)^\omega}$.

  The function $\conddouble$ of
  \cref{ex:intro1bis}, illustrated in \cref{subfig:guardedMSO}, is
  definable by an \MSOT with two copies ($k=2$). Formulas
  $\phi_{\fst}^{c_0}$ and $\phi_\sigma^c$ are the same as for $\double$. Then:
     \begin{align*}
       \phi_S^{1,1}(x,y) = &\ \phi_S^{2,2}(x,y) =  S(x,y)\wedge \neg
                             2(y) \qquad\qquad \phi_S^{2,1}(x,y) =  S(x,y)\wedge 2(y) \\
       \phi_S^{1,2}(x,y) = & \exists \gv\cdot y<x\leq \gv\wedge 2(\gv)
                          \wedge  \forall z\leq y\cdot (S(z,y)\rightarrow (1(z)\vee
         2(z))) \wedge \\ & \forall t\cdot (y\leq t\leq x)\rightarrow (a(t)\vee b(t))
  \end{align*}
\end{example}

The class of regular functions of infinite words has been defined
in~\cite{alur2012regular} as the class of functions
recognizable by deterministic two-way transducers extended with
regular (infinite) look-ahead: to take a transition, such a transducer
can query a regular oracle on the infinite current suffix (given as a
deterministic parity automaton for example). Equivalently, this class
corresponds to functions recognizable by (deterministic) \SST:
they work as \DSST{} but are not forced to output the content of a
special register infinitely often. Instead, the output of a run
depends on the set of states that are seen infinitely often along that
run, and can be ``computed'' only once the infinite input has been
processed (see~\cite{alur2012regular}) for more
details.  The following provides a logical characterization of the class
of regular functions:

\begin{theorem}[\cite{alur2012regular}]
   A function $f: \Sigma^\omega \parfonc \Gamma^\omega$ is regular if and only if it is \MSOT-definable.
\end{theorem}

The definition of \MSOT in~\cite{alur2012regular} is slightly
different, but equivalent, to the definition we take in
this paper.

\subparagraph{Guarded \MSO-transductions of infinite words}

Guarded \MSO formulas are a syntactical restriction of \MSO
formulas. This restriction requires all the free variables and
quantifiers to be guarded by a first-order
variable $\gv$, in the sense that quantifiers should only talk about
positions which are \emph{before} $\gv$ (i.e.\ smaller than
$\gv$). Intuitively, the satisfiability of a guarded formula on an
infinite word only depends on the finite prefix up to position $\gv$.  
Formally, given two first-order variables $x$ and $\gv$, we let
$\guard(x,\gv)$ be the
formula $x\leq \gv$ ($x$ is guarded by $\gv$), and for a set variable $X$,
we let $\guard(X,\gv)$ be the formula $\forall x\in X, \guard(x,\gv)$.
Then, an \MSO formula $\varphi$ is \emph{guarded by some variable
$\gv$} if it is equal to $\psi(\gv)\wedge \bigwedge_{\alpha\in Free(\psi)}\guard(\alpha,\gv)$
for some $\psi(\gv)$ such that all its quantified subformulas, i.e.\ subformulas
of the form $QX\cdot \psi'$ or $Qx\cdot \psi'$ for some
$Q\in\{\exists,\forall\}$, are in one of the following forms:
\[
\begin{array}{llll}
(1)\ \forall x\cdot
  \guard(x,\gv)\rightarrow \zeta &  (2)\ \exists
x\cdot \guard(x,\gv)\wedge \zeta & (3)\ \forall X\cdot \guard(X,\gv)\rightarrow
  \zeta &  (4)\ \exists
          X\cdot \guard(X,\gv)\wedge \zeta
          \end{array}
\]

An \MSO formula is \emph{guarded} if it is of the form $\exists
\gv\cdot \varphi$ where $\varphi$ is guarded by $\gv$. 
We denote by $\GMSO$ the set of guarded
$\MSO$-formulas. For conciseness, we may write $\forall x:\gv\cdot \zeta$ instead of
$\forall x\cdot \guard(x,\gv)\rightarrow \zeta$, and $\exists x:\gv\cdot \zeta$ instead of
$\exists x\cdot \guard(x,\gv)\wedge \zeta$ (and similarly for set
variables).

\begin{example}\label{ex:guardedMSO}
    All the formulas of the \MSO-transduction of
    \cref{ex:msotrans} defining the function $\double$ are guarded, or
    trivially equivalent to a guarded formula. For example, the
    formula $\first(x)$ is equivalent to the guarded formula
    $\exists \gv\cdot x\leq \gv\wedge \forall y\leq \gv\cdot \neg
    S(y,x)$.

    The order predicate $x\leq y$ is definable by the guarded formula
    $\exists \gv\cdot x\leq \gv\wedge y\leq \gv\wedge y=\gv$. Since
    $\neg (x\leq y)$ is equivalent to $y\leq x\wedge y\neq x$, we
    easily get
    that any \GMSO-formula $\phi$ is equivalent to an \GMSO-formula
    $\psi$ in which the
    order predicate is only used to guard quantifiers, by existentially quantifying a global
    guard, guarding all the local guards used to define the atomic
    formulas of the form $z\leq t$ occurring in $\phi$ (assumed to
    occur positively). 
 \end{example}



\begin{remark}
    \GMSO formulas only talk about prefixes, in the following sense: If $\varphi = \exists \gv\cdot \psi(\gv)$ is a closed guarded formula
    and $w\in\Sigma^\omega$,  then $w\models \varphi$ if and only if there exists
    a finite prefix $u$ of $w$ such that $u\models \psi(\ell)$, where
    $\ell$ is the last position of $u$. This allows us to get the
    following immediate characterization: A language $L\subseteq \Sigma^\omega$ is $\GMSO$-definable if and only if there exists a regular language
    $F\subseteq \Sigma^*$ such that $L = F\Sigma^\omega$.
\end{remark}

\begin{definition}[Guarded \MSO-transductions]
A \emph{guarded \MSO-transduction} (\GMSOT) is an \MSO-transduction all
formulas of which are guarded.
\end{definition}

\begin{example}\label{ex:gmsot}
As explained in \cref{ex:guardedMSO}, all formulas of the \MSO-transduction of
    \cref{ex:msotrans} defining $\double$ are guarded, or
    trivially equivalent to a guarded formula. 

\end{example}

We can now state the logical characterization of deterministic regular
functions: 

\begin{theorem}[Logical characterization]\label{thm:main}
   A function $f: \Sigma^\omega \parfonc \Gamma^\omega$ is
   deterministic regular if and only if it is \GMSOT-definable.
\end{theorem}

The proof is given  in
\cref{sec:proofmain}. As an application of this
result,  since deterministic regular functions are (effectively) closed under
composition by \cref{thm:closure-comp}, we obtain that \GMSOT
are (effectively) closed under composition as well. This is a well-known
result for \MSOT over finite strings~\cite{engelfriet2001mso}, infinite strings~\cite{alur2012regular} and
more generally any structure~\cite{Cou94}, yet with purely logic-based
and direct proofs, while we use here involved automata-based
arguments (look-ahead removal). Indeed, composition closure of \MSOT
is obtained by formula substitutions. To compose two
\MSOT $\trans_2\circ \trans_1$, the predicates occurring in  $\trans_2$
are substituted by their definition in $\trans_1$. %
Such a direct proof idea does not work in the guarded
fragment \GMSOT, as guarded formulas are not closed under
negation. 


We conclude this section by discussing an
alternative definition of \GMSO-transductions, denoted
\GMSOTO, where instead of defining the output successor relation,
it requires to define the total order $\leq$ of the output structure with \GMSO formulas.
This however allows to define non-computable functions, as stated by
the following proposition proved in Appendix:

\begin{proposition}\label{prop:notcomput}
    There exists an $\GMSOTO$ which defines an uncomputable function.
\end{proposition}

Without the guarded restriction, the two definitions both define the
class of regular functions of infinite words.

\section{Two-way transducers with finite look-ahead}\label{sec:twoway}

We extend deterministic two-way transducers with finite look-ahead. Transitions are
additionally labelled by a regular language of \emph{finite
  words}, called (finite) look-ahead. A transition with look-ahead $L$
can only be taken if the remainder of the input
sequence has a prefix that belongs to $L$. Such a finite prefix is called a
\emph{look-ahead witness} for $L$. To ensure determinism, if several
look-aheads succeed, it is required that there is a unique shortest look-ahead witness.
The transducer follows the transition which minimizes the length of the
witness. 
If no look-aheads succeed the computation fails.

\begin{definition}[Finite look-ahead]
A \emph{deterministic two-way transducer with finite look-ahead
(\tDTLA)} is a tuple $\trans = (\Sigma,\Gamma,Q,q_0,F,\delta,\lambda)$
where $\Sigma,\Gamma,Q,q_0,F,\lambda$ are defined as for deterministic two-way
transducers w/o look-ahead, $\delta$ is a transition function $Q \times (\Sigma \uplus \{{\lmark}\}) \times \mathcal R^*(\Sigma))
        \parfonc Q \times \{\rmove, \lmove\}$ where $\mathcal R^*(\Sigma)$
        is the set of all regular languages of \emph{finite} words
        over $\Sigma$. The function $\delta$ is required to have finite
        domain. The \emph{look-ahead} for a transition
        $(q,\sigma,L) \mapsto (q,d)$ is $L$. 
        Furthermore, we require that if $\delta(q,\sigma,L)$ and
        $\delta(q,\sigma,L')$ are defined, then $L \cap L' =
        \emptyset$ for all $L,L' \in \mathcal R^*(\Sigma)$, $q \in Q$
        and $\sigma \in \Sigma$.
        Finally, it is assumed that the look-ahead languages are represented by deterministic finite automata. 
\end{definition}

The semantics of a deterministic two-way transducer with finite look-ahead remains unchanged compared to the model without look-ahead.
The only difference in the presence of look-ahead is when a transition is enabled:
A transition with look-ahead $L$ can only be taken if the remainder of the input sequence has a prefix that belongs to $L$.
Formally, in a configuration $(q,i)$ over input $u$, a transition of
the form $\delta(q,\sigma,L)$ where $L \subseteq \Sigma^*$ is enabled if $u[i] = \sigma$ and there exists some $i < j$ such that $u[i{+}1{:}j] \in L$. The word $u[i{+}1{:}j]$ is called a
\emph{witness} for $L$. 
To ensure determinism, whenever the
transducer is in a configuration $(q,i)$, if several look-aheads
$L_1,\dots,L_k$ are enabled, the triggered transition is the unique (ensured by the disjointness requirement)
transition with shortest witness.

\subparagraph{Removing finite look-ahead}

We know that infinite look-ahead is strictly more expressive than finite look-ahead.
The natural question is how much expressiveness is gained by adding finite look-ahead to deterministic two-way transducers w/o look-ahead. 
As already explained in the introduction, any function defined by such a transducer is (Turing machine) computable:
A Turing machine can memorize where it is in the input, verify which look-ahead succeeds, and continue the computation from the memorized position.
A two-way transducer does not have the ability to memorize a position arbitrarily far away in the input.
Hence, verifying (in the absence of some look-ahead ``oracle'') that some finite prefix of the remainder of the input is a witness for some look-ahead and returning to a specific position becomes a problem to be solved.
This problem is not unique to two-way transducers over infinite words, it also appears when some regular property of the remainder of a finite input word must be checked and subsequently the two-way transducer must return to the position it has been in before checking the property.
On finite words, this task can be handled using the Hopcroft-Ullman \cite{aho1969general} or the improved tree-outline construction \cite{dartois2017reversible}.
However, these constructions rely on the fact that the input word is finite.
We prove that this task can be also accomplished for infinite words using different techniques.

In the following, we show that no expressiveness is gained by allowing finite look-ahead.

\begin{theorem}[Finite look-ahead removal]\label{thm:laremoval}
  Given a $\tDTLA$, one can effectively construct an equivalent $\tDT$.
\end{theorem}

\begin{proof}[Proof sketch.]
The proof is divided into two parts.
The main part is to translate a given \tDTLA{} into an equivalent \BSST{} with bounded copy.
We then use \cref{theo:2dt-dsst} to obtain an equivalent $\tDT$.
Given a deterministic two-way transducer \emph{without} look-ahead, the
standard approach to obtain an equivalent \SST{} is to simulate the right-to-right
runs of the deterministic two-way transducer on the so-far
read prefix of the infinite input, store their outputs in
registers and compose these registers in the right way (with the output of the ``main'' left-to-right run) to re-create the output of the two-way transducer.
Since the two-way transducer is deterministic there is a global bound on the
number of different right-to-right runs on any prefix of the input.
The constructions presented in \cite{alur2012regular,dartois2016aperiodic,carton2022continuous} are all built on this idea.
In \cite{alur2010expressiveness}, equivalence between \SST{} and two-way transducers on finite words is shown but the work exhibits no direct translation.


Our goal is to design a similar construction for deterministic two-way
transducers \emph{with} finite look-ahead.  The main difficulty is that
there is no global bound on the number of different runs that can occur on
a prefix.  The reason is that such a transducer is non-deterministic in the
sense that it must make a decision which look-ahead will succeed.  This
choice is verified a posteriori when a look-ahead witness is seen.

Hence, we extend the standard construction to go from a deterministic
two-way transducer to an SST by additionally taking the look-ahead choices
into account.  This approach results in a tree structure representation of
the possible runs (similar to a standard run-tree of a non-deterministic
automaton, here the non-determinism is the look-ahead choice).  A branch in
such a tree corresponds to a possible run and the nodes additionally contain
information to detect when look-ahead choices succeed or are doomed to
fail.  The size of the tree representations is kept bounded by sharing
information and a relevant pruning strategy.  The strategy takes care of
removing branches whose look-ahead choices \emph{cannot succeed} and
(prefixes of) branches where the look-ahead choices \emph{already have
  succeeded}.  Applying this construction to a deterministic two-way
transducer without look-ahead yields the standard translation construction. 
\end{proof}

\section{Logic-transducer correspondence: proof of
  \texorpdfstring{\cref{thm:main}}{Theorem~\ref{thm:main}}}\label{sec:proofmain}

In this section, we give an overview of the proof of the logical characterization of \cref{thm:main}. We first prove that
any deterministic regular function is $\GMSOT$-definable. The proof is
standard and uses same ideas as for regular functions of finite
words~\cite{engelfriet2001mso} and infinite words~\cite{alur2012regular}.

\begin{lemma}\label{lem:trans2log}
    If a function $f : \Sigma^\omega\parfonc \Gamma^\omega$ is
    deterministic regular, then it is $\GMSOT$-definable.
\end{lemma}

\begin{proof}
    The main idea is to define in \GMSOT{} the runs of a \tDT{}. Each
    copy of the \GMSOT{} represents a state of the \tDT{}, and
    there is a successor edges between node $x^{p}$ to node $y^{q}$,
    where $x,y$ are input positions and $p,q$ are states, if and only
    if there exists a \emph{finite} run from configuration $(p,x)$ to
    configuration $(q,y)$ which produces output symbols only in
    configuration $(p,x)$ and $(q,y)$. This property can be expressed
    by an \GMSO formula.
\end{proof}

Proving the converse of \cref{lem:trans2log} is more involved. We
first go to an intermediate model with \MSO instructions, in the spirit
of~\cite{engelfriet2001mso}, called \emph{jumping}
\GMSO-transducers, proved to be equivalent to
\tDT. It is a finite-state
model which can $(i)$ test \GMSO
properties of the current position (called look-around), $(ii)$ test safety
constraints defined by \MSO formulas, and $(iii)$ jump from one
position to another one with binary \GMSO formulas. Formally, it has a
finite set of states (all final), and transitions are of the form
$p\xrightarrow{\phi_{\la}(x)\mid w,\phi_{\mv}(x,y),\phi_{\saf}(x)} q$
where $p,q$ are states, $\phi_{\la},\phi_{\mv}$ are 
\GMSO formulas, $\phi_{\saf}$ is an \MSO formula, and $w$ is a
finite word. Look-around occurring on transitions with same source
state are assumed to be pairwise disjoint (their conjunction is not
satisfiable). The initial configuration is $(q_0,0)$ where $q_0$ is
the initial state. Whenever it is in a configuration $(q,i)$, over an
infinite word $u\in\Sigma^\omega$, it enables the transitions whose
look-around $\phi_{\la}(i)$ holds on $u$, and select the transition with
shortest witness. Call $t$ this transition. It
triggers $t$ only if there exists $j$ such that
$\phi_{\mv}(i,j)$ holds and for all $k\geq i$, $u[{:}k]\models
\phi_{\saf}(i)$ (otherwise the computation fails). It then
outputs $\gamma$ and moves to some position $j$ such that
$\phi_{\mv}(i,j)$ holds. Note that there could be several $j$, and
therefore several runs on the same input in general. We thus make the
following \emph{assumption}, which can be described informally as
follows: for any reachable configuration of the transducer from the
initial configuration, there is always a unique $j$. Formally, for all infinite sequence of
configurations $(q_0,i_0=0)(q_1,i_1)(q_2,i_2)\dots$, for all $k\geq
0$, for any transition $t$ triggered from configuration $(q_k,i_k)$ to
$(q_{k+1},i_{k+1})$, if $\phi_{\mv}(x,y)$ is the jumping formula of
$t$, then $i_{k+1}$ is the unique position such that
$\phi_{\mv}(i_k,i_{k+1})$ holds. As for two-way transducers, a sequence of configurations
$(q_0,i_0=0)(q_1,i_1)\dots$ is
\emph{accepting} if $\lim_{k\rightarrow\infty} i_k = \infty$ and it
produces an infinite word.

We show that this model defines deterministic regular functions:
\begin{lemma}\label{lem:jumping}\label{coro:jumpTodetregful}
    Any jumping \GMSO-transducer defines a deterministic regular function. 
\end{lemma}

\begin{proof}[Sketch of proof]
    The proof goes in two steps. First, it is shown that
    jumping \GMSO-transducers are equivalent to \emph{walking}
    \GMSO-transducers, i.e.\ \GMSO-transducers which moves (backward or
    forward) between successive positions. This step is standard
    (it appears e.g.\ in~\cite{engelfriet2001mso} in the non-guarded
    setting). Then, walking \GMSO-transducers are shown to be
    equivalent to an extension of \tDT{} with finite look-around and
    safety constraints, then proved to be equivalent to \tDT{} by
    transforming look-arounds into look-aheads, and then removing
    look-aheads (based on the techniques of \cref{sec:twoway}) and safety constraints.
\end{proof}


\begin{lemma}\label{lem:gmsot2detreg}
    Any \GMSO-transduction is equivalent to a jumping \GMSO-transducer. 
\end{lemma}
\begin{proof}

    Let $\trans =
    (k,(\phi_\gamma^c)_{c\in[k],\gamma\in\Gamma},(\phi_S^{c,d})_{c,d\in[k]},\phi_{\fst}^{c_0}(x))$
    be an $\GMSOT$ defining $f$. We construct a jumping
    \GMSO-transducer $\trans'$ equivalent to $\trans$. The set of states of $\trans'$
    is $\{0,1\dots,k\}$. In state $0$, $\trans'$ first jumps to the initial
    position, i.e.\ the position $y$ which satisfies
    $\phi_{\fst}^{c_0}(y)$ and moves to state $c_0$. This is done by a
    transition going from state $0$ to state $c_0$, with the trivial
    look-around and safety constraint $\top$, and the move $\phi_{\mv}(x,y){:}=\first(x)\wedge
    \phi_{\fst}^{c_0}(y)$. Then, it follows the successor relation of $\trans$, and uses the label
    formulas to determine which label to output. Using safety
    constraints, $\trans'$ also makes sure that the output graph structure is
   a word structure. In particular, they express that for any
   reachable node, there is exactly one label and at most one
   successor. There is no need to check that there is \emph{at least} one
   successor, because if there is none, then the run of $\trans'$ stops and
   the input is not accepted, which is consistent with the semantics
   of $\trans$ (the input is also rejected by $\trans$ in that case). There is also no need to check that there is no cycle,
   because if there is some, then $\trans'$ will never visit all input
   positions, and hence the input will be rejected, which is again
   consistent with the semantics of $\trans$. Formally, for all copies $c,d\in\{1,\dots,k\}$ and output label
   $\gamma$, since $\phi_S^{c,d}(x,y)$ and $\phi_\gamma(x)$ are
   guarded, there are of the form $\phi_S^{c,d}(x,y) = \exists \gv\cdot\psi_S(x,y,\gv)$ and
   $\phi_\gamma(x) = \exists\gv\cdot\psi_\gamma(x,\gv)$.  Then we add
   the following transition to $\trans'$, from $c$ to $d$:
   \[
   c\xrightarrow{\phi_{\la}(x){:}=\exists \gv\exists
     z\leq \gv\cdot \psi_S^{c,d}(x,z,\gv)\wedge
     \psi_\gamma^c(x,\gv)\wedge \textsf{disj}_{c,d,\gamma}(x,\gv) \mid \gamma,\phi_{\mv}(x,y){:}=\phi_S^{c,d}(x,y),\phi_{\saf}(x)} d
   \]
   in which $\textsf{disj}_{c,d,\gamma}(x,\gv) = \forall \gv'\leq
   \gv\cdot \bigwedge_{\gamma'\neq \gamma} \neg
   \psi_{\gamma'}^c(x,\gv')\wedge \bigwedge_{d'\neq d}\forall z'\leq \gv'\cdot \neg\psi_S^{c,d'}(x,z',\gv')$
   ensures disjointness of the look-around, and $\phi_{\saf}(x)$ equals
   \[
   \begin{array}{lr}
     (\bigwedge_{d'\neq d} \forall y\cdot
     \neg \phi_S^{c,d'}(x,y)) \wedge & \text{no successor of
                                $x$ in any  copy $d'\neq d$}\\

     (\forall y\forall y'\cdot 
   (\phi_S^{c,c}(x,y)\wedge \phi_S^{c,c}(x,y'))\rightarrow y=y')\wedge
                                     & \text{at most one successor of $x$
                                                     in copy $c$}\\
     (\bigwedge_{\gamma'\neq \gamma} \neg \phi^c_{\gamma'}(x) & \text{no
                                                      other label
                                                       for $x$}
     \end{array}
     \]
     At this point, we remind the reader that safety constraints
     are not required to be defined by guarded formulas, as they are
     regular properties of finite words. However, the look-around and jumping formulas must be
     guarded, and it is indeed the case in the transition above.

     Finally, note that $\trans'$ satisfies
     the requirement that on infinite sequences of configurations $(q_0,i_0)\dots$,
     for all $k\geq 0$, $i_{k+1}$ is the unique successor of $i_k$ by
     the jumping formula. Indeed, if a sequence of
     configurations of $\trans'$ is infinite, it implies that
     all safety constraints are satisfied, and they precisely make
     sure that there is no branching.
   \end{proof}

   As a corollary of Lemmas~\ref{lem:gmsot2detreg} and~\ref{coro:jumpTodetregful}, we obtain the converse direction of \cref{thm:main}:

   \begin{corollary}
       Any \GMSOT-definable function $f$ is deterministic regular. 
\end{corollary}



\section{Conclusion}

In this paper, we have shown that the class of deterministic regular functions
is characterized  by computational models such as deterministic two-way transducers,
deterministic two-way transducers with \emph{finite} (regular) look-aheads,
B\"uchi \SST, by the logical formalism of \emph{guarded}
\MSO-transductions, and by finite compositions of sequential functions
and $\mapcr$. The transformations between those models are effective. 
We have also shown that it is closed under composition,
by extending to infinite words the known composition closure of
deterministic two-way transducers, yet with new proof
techniques. It is also conjectured that the class of deterministic
regular functions is equal to the class of \emph{continuous} regular
functions (for the Cantor topology). It is already known that it
includes the continuous letter-to-letter rational functions~\cite{filiot2021synthesizing}
and the strictly larger class of continuous rational functions~\cite{carton2022continuous}. We believe that the new
characterizations proved in this paper, will be helpful towards
solving this open and seemingly difficult problem. All this, together with the fact that
deterministic regular functions are computable, unlike regular
functions, shows the robustness of this class.


\bibliography{biblio}

\newpage

\appendix

\input{app-decomp.tex}

\input{app-mso}

\input{appendix-sarah.tex}

\input{app-proof-mso}

\end{document}

%% file: app-decomp.tex
\section{Decomposition: proof of \texorpdfstring{\cref{theo:KR-detreg}}{Theorem~\ref{theo:KR-detreg}}}

\label{proof:KR}

In this section, we show that a function is deterministic regular
if and only if it can be (effectively) written as a composition of
sequential and $\mapcr$ functions. The right-to-left implication
is clear, since deterministic regular functions
are closed under composition (\cref{thm:closure-comp})
and the sequential and $\mapcr$ functions are deterministic regular.
It remains to show how a deterministic regular function
can be decomposed using these basic functions,
and this is the purpose of the current \cref{proof:KR}.

\subsection{A class of functions closed under composition}

We first introduce a class of functions $\Class$ which is
closed under composition and
goes both from finite words to finite words
and infinite words to infinite words
(i.e.\ of type $(\Sigma^* \parfonc \Gamma^*) \cup (\Sigma^\omega \parfonc \Gamma^\omega)$).
Our goal is to show that any deterministic regular function
can be computed as the restriction of a function of $\Class$
to infinite words. However, in the proof we sometimes need
to apply these functions to either finite or infinite words. 
Indeed, we shall build a ``map'' operation working on finite or infinite factors
of a given word, and for which there is no way to ``decide'' at runtime if the current
factor is finite or infinite.

Recall that one-way transducers and sequential functions over infinite words
have been presented in \cref{def:sequential}. We generalize this
definition in order to be able compute \emph{with the same machine}
both over finite and infinite words.

\begin{definition}[Sequential functions of finite and infinite words]
\label{def:sequential-finin}
A \emph{deterministic one-way transducer over finite and
infinite words} is a tuple $(\Sigma,\Gamma, Q, q_0, F_{*}, F_{\omega},\delta, \lambda)$ such
that:
\begin{itemize}
\item $(\Sigma,\Gamma, Q, q_0, F_{\omega}, \delta, \lambda)$ is a 
one-way deterministic transducer over infinite words;
\item $(\Sigma,\Gamma, Q, q_0, F_{*}, \delta, \lambda)$
is a  one-way deterministic transducer over
finite words, that is $F_{*} : Q \parfonc \Gamma^*$
is a final output function. The output over a finite input defined
over finite words by concatenating the output produced
along a run and adding $F_{*}(q)$ (if defined,
otherwise the output is undefined) where $q$ is the last
state of the run.
\end{itemize}
Such a machine describes a function of type
$(\Sigma^* \parfonc \Gamma^*) \cup (\Sigma^\omega \parfonc \Gamma^\omega)$.
The class of (partial) functions
computed by one-way deterministic
transducers over finite and infinite words
is called \emph{sequential functions
of finite and infinite words.}
\end{definition}

\begin{remark}
It is essential to have the \emph{same} transducer
that defines \emph{both} the function over finite words
and the function over infinite words.
\end{remark}

\begin{definition}
\label{def:class}
The class $\Class$ of functions of type
$(\Sigma^* \parfonc \Gamma^*) \cup (\Sigma^\omega \parfonc \Gamma^\omega)$
is the closure under composition of the following functions:
\begin{itemize}
\item sequential functions over finite and infinite words;
\item for any fresh symbol $\hash{} \not \in \Sigma$
the function $\mapcr : ((\Sigma \uplus \{\hash{}\})^* \fonc (\Sigma \uplus \{\hash{}\})^*)
\cup ((\Sigma \uplus \{\hash{}\})^\omega \fonc (\Sigma \uplus \{\hash{}\})^\omega)$
defined for $u_i \in \Sigma^*$ by:
\begin{itemize}
\item $u_1 \hash{} \cdots \hash{} u_n
\mapsto u_1\hash{} \mirror{u_1} \hash{} \cdots \hash{} u_n\hash{} \mirror{u_n}$
when the input is finite;
\item $u_1 \hash{} \cdots \hash{} u_i\hash{}  \cdots
\mapsto u_1 \hash{}\mirror{u_1} \hash{} \cdots\hash{} u_i \hash{} \mirror{u_i} \hash{} \cdots $
if the infinite input contains infinitely many $\hash{}$;
\item $u_1 \hash{} \cdots \hash{} u_n \hash{} u
\mapsto u_1 \hash{} \mirror{u_1} \hash{}  \cdots
\hash{}  u_n \hash{} \mirror{u_n} \hash{}  u $
if $u \in \Sigma^\omega$
\end{itemize}
\end{itemize}
\end{definition}

\begin{remark} If the context is not clear,
we shall say that $\mapcr{}$ ``has separator $\hash{}$''
to mean that the symbol $\hash{}$ is used to separate
the factors.
\end{remark}

Let us now describe some properties of this class $\Class$.

\begin{example} \label{ex:mapc}
The (total) function $\mapc: (\Sigma \uplus \{\hash{}\})^{\infty} \fonc  (\Sigma \uplus \{\hash{}\})^{\infty}$
is defined in a similar way
as $\mapcr$ from \cref{def:class}, except that each
copy mirror factor $\mirror{u_i}$ is replaced by a simple copy of $u_i$.
This function $\mapc$ belongs to $\Class$.
Indeed, we apply $\mapcr$ twice, which outputs 
a word of shape $u_1 \hash{} \mirror{u_1} \hash{} \mirror{u_1} \hash{} u_1
\hash{} u_2 \cdots$. It is easy to remove the useless pieces using
a sequential function.
\end{example}

\begin{remark} One can also define the function $\mapr$
where each factor is only reversed but not copied before.
However, this function cannot be defined
on infinite words with finitely many $\hash{}$ symbols
(and not even extended to a continuous function with total domain).
\end{remark}


Finally, we show that the class $\Class$ is closed
under a ``map'' operation, that applies a given function
to the factors of a word delimited by a specific symbol.
We give in \cref{lem:then} a more precise statement, 
that enables us to apply different functions on the first factors.

\begin{lemma}
\label{lem:then}
Let $f_1, f_2, \dots, f_n : \Sigma^{\infty} \parfonc \Gamma^{\infty} \in \Class$
and $\hash{}$ be a fresh symbol. One can effectively build in $\Class$ a function 
$\then{f_1\hash{} f_2 \hash{} \cdots \hash{} f_n\hash{}} : \finin{(\Sigma \uplus \{\hash{}\})}
\parfonc \finin{(\Gamma \uplus \{\hash{}\})}$
such that:
\[\then{f_1\hash{} f_2\hash{} \cdots \hash{} f_n\hash{}} (u_1 \hash{} u_2 \cdots) =  f_1(u_1)  \hash{} f_2(u_2)
  \hash{} \cdots \hash{} f_n(u_n) \hash{} f_{n} (u_{n+1})  \hash{} f_{n} (u_{n+2}) \cdots\]
whenever $u_1 \in \Dom(f_1), u_2 \in \Dom(f_2), \dots$.
\end{lemma}

\begin{remark} Several elements are left implicit in the formulation of
\cref{lem:then}:
\begin{itemize}
\item if there are $k \le n$ factors in the input, then $\then{f_1 \hash{} \cdots \hash{} f_n\hash{}}$
only applies $f_1,\dots, f_k$;
\item the last factor (if it exists) can be infinite;
\item if the input is infinite, we must have
$f_1(u_1) \hash{} f_2(u_2) \hash{} f_3(u_3) \cdots \in (\Gamma\uplus \{\hash{}\})^{\omega}$
for the output to be defined.
\end{itemize}
\end{remark}

\begin{proof} We only treat the case $n=1$, which corresponds
to the classical ``map'' operation. The other cases can be treated
in a similar way, using sequential functions to drop specific marks
on the $n$ first pieces. Let $f \defined f_1$. We show the result by induction on the construction 
of $f$. If $f$ is a sequential function then we build a sequential
function $\then{f\hash{}}$ described by a one-way transducer similar to that
of $f$ does, except if a $\hash{}$ is seen,
in which case it produces the (finite) final output of the transducer
in the current state, and goes back to the initial state
to pursue its computation.
If $f = g \circ h$ the result is clear by induction hypothesis.
If $f$ is $\mapcr$ with separator $\#$ (necessarily $\# \neq \hash{}$),
we apply the sequential function that
turns each $\hash{}$ into $ \hash{} \#$.
Then we apply $\mapcr$ with separator $\#$
on the  whole input. We conclude
by applying a sequential function that removes
the $\hash{} \hash{}$ factor each times it sees one,
and in this case replaces the next $\#$ by a $\hash{}$.
\end{proof}

\subsection{Forward Ramseyan splits}

Now, we show how to decompose a deterministic
regular function as a function of $\Class$.
Following the decomposition results for regular \cite{bojanczyk2018polyregular}
or polyregular \cite{bojanczyk2020single} functions over finite words,
our first goal is to split the input using a structure that is 
similar to Simon's factorization forests \cite{simon1990factorization}.
Indeed, the runs of a two-way transducer behave in a simple
way on such splits. We present in \cref{def:split} a weakened version of  factorization
forests, named \emph{forward Ramseyan splits}
and introduced by Colcombet in \cite{colcombet2007combinatorial}.
We shall recall Colcombet's main result in  \cref{theo:colcombet}: such
splits can be computed by a sequential function.

If $\Sigma$ is an alphabet, we consider an infinite number of fresh
symbols $\hash{1}, \hash{2}, \dots \not \in \Sigma$.
For all $k \ge 0$, we define the alphabet $\Hash{k} \defined \{\hash{1}, \dots, \hash{k} \}$.

\begin{definition}[Forward Ramseyan split \cite{colcombet2007combinatorial}]
\label{def:split}
Let $\mu : \Sigma^* \fonc M$ be a morphism into a finite monoid
and $k \ge 0$.  Let $u \in \finin{\Sigma}$.
We say that a word $\frs \in \finin{(\Sigma \uplus \Hash{k})}$
is a \emph{forward Ramseyan $\mu$-split} over $u$ of height
$k \ge 0$ if it can be defined by induction as follows:
\begin{itemize}
\item either $k = 0$ and $\frs = u = a \in \Sigma$ or $\frs = u = \movi$;
\item or $\frs = \frs_1 \hash{k} \cdots \hash{k} \frs_n$
where $u = u_1 \cdots u_n$,
for all $1 \le i \le n$, $\frs_i$ is a $\mu$-split
of height $k{-}1$ of  $u_i \in \finin{\Sigma} \smallsetminus \{\varepsilon\}$,
and for all $1 \le i,j < n$
$\mu(u_i) \mu(u_j) = \mu(u_i)$
(note that $u_i,u_j \in \Sigma^+$);
\item or $\frs = \frs_1 \hash{k}  \cdots \hash{k} \frs_i \hash{k} \cdots$
where $u = u_1 \cdots u_i \cdots \in \Sigma^{\omega}$,
for all $1 \le i$, $\frs_i$ is a $\mu$-split
of height $k{-}1$ of $u_i \in \Sigma^+$,
and for all $1 \le i,j$, $\mu(u_i) \mu(u_j) = \mu(u_i)$.
\end{itemize}
\end{definition}

\begin{remark}
\label{rem:ff-weaker}
In particular the condition $\mu(u_i) \mu(u_j) = \mu(u_i)$ implies
that $\mu(u_i)$ is an idempotent. Furthermore, it means that
the $\mu(u_i)$ are $\mathcal{L}$-equivalent in the sense
of Greene's relations (see e.g.\ \cite{colcombet2007combinatorial}).
In Simon's usual factorization forests (also known as Ramseyan $\mu$-splits), the condition
is stronger: $\mu(u_i) =\mu(u_i)^2 = \mu(u_j)$ (i.e.\ the idempotents are equal).
\end{remark}

\begin{remark} In the second case, there is no assumption
on the value $\mu(u_n)$, even if $u_n \in A^+$. In particular,
if $n=2$ there is no assumption on $\mu(u_2)$.
\end{remark}

From now on, the morphism $\mu$ will always be implicity
fixed, and we denote by $\FRS{k}(u)$ the set
of $\mu$-splits of height at most $k \ge 0$ over the word $u \in \finin{\Sigma}$.

\begin{theorem}[{\cite[Theorem 1]{colcombet2007combinatorial}}]
\label{theo:colcombet}
Let $\mu : \Sigma^* \fonc M$ be a  morphism into a finite monoid.
There exists a bound $N$ and a sequential function
$\spli : \Sigma^\omega \fonc (\Sigma \cup \Hash{k})^\omega$ such that for all
$u \in \Sigma^\omega$, $\spli (u) \in \FRS{N}(u)$.
The construction is effective.
\end{theorem}

\subsection{Transition monoid of a two-way transducer}

The goal of this section is to recall the well-known notion
of \emph{transition morphism} $\mu$ of a two-way transducer (see e.g.
 \cite{dave2018regular}) and to relate it to forward Ramseyan $\mu$-splits.
 In the rest of \cref{proof:KR}, we 
 consider a fixed deterministic two-way transducer
 $\trans = (\Sigma,\Gamma,Q,q_0, F, \delta, \lambda)$.
We let $\oleft{Q} \defined \{\oleft{q}: q \in Q\}$
and $\oright{Q} \defined \{\oright{q}: q \in Q\}$ be two disjoint copies of $Q$.
To simplify the notations, we assume without losing generalities
that $\lmark \in \Sigma$.

Recall that the transition function $\delta$ has type
$Q \times \Sigma \parfonc Q \times \{\lmove, \rmove\}$.
We extend the transition function to
$\delta : \oright{Q} \cup \oleft{Q} \times \Sigma^* \parfonc \oright{Q} \cup \oleft{Q} $
and similarly $ \oright{Q} \cup \oleft{Q} \times \Sigma^* \parfonc \Gamma^*$.
Intuitively, $\delta(\oleft{q},u) = \oright{p}$ and  $\lambda(\oleft{q},u) = v$ 
means that the longest run over $u$
that starts in the rightmost position of $u$ 
will eventually leave $u$ ``on the right'' in state $p$,
and the concatenation of its output is $v$.
We assume that $\delta(\_, \varepsilon)$ is the identity
function and that $\lambda(\_, \varepsilon) = \varepsilon$.
Finally, we also extend $\delta$
to infinite words $u \in \Sigma^\omega$ by defining
$\delta(\oright{q},u) \defined \oleft{p}$ if the longest
run leaves $u$ ``on the left''
and $\delta(\oright{q},u) \defined \omega$
if this run is infinite and visits arbitrary large positions.
Note that $\delta(\oleft{q},u)$ would not make sense.
The function $\lambda$ is extended accordingly.

The following result is folklore about two-way transducers
(over finite words, but it does not change anything
since we deal with transitions on finite words here).

\begin{proposition}[Transition morphism, see e.g.\ {\cite{dave2018regular}}]
\label{prop:transition-morphism}
One can effectively build a finite monoid $M$ and a
morphism $\mu : \Sigma^* \fonc M$, called the
 \emph{transition morphism} of $\trans$, such that
for all $u \in \Sigma^*$ and $\oleftright{q} \in \oright{Q} \cup \oleft{Q}$,
$\delta(\oleftright{q},u)$ only depends
on $\oleftright{q}$ and $\mu(u)$.
\end{proposition}

\begin{figure}[h!]

\centering
\begin{tikzpicture}{scale=1}

        \newcommand{\coulun}{red}
        \newcommand{\coulde}{blue}
        \newcommand{\coultr}{green}


        \draw (0,1.7) -- (10.5,1.7);
        \draw (0,2.2) -- (10.5,2.2);
        \draw (0,2.2) -- (0,1.7);

        \node[above] at (1,1.7) {$u_1$};
        \draw[dotted,  thick] (2,2.3) -- (2,-0.2);
        \node[above] at (3.5,1.7) {$u_2$};
        \draw[dotted,  thick] (5,2.3) -- (5,-0.2);
        \node[above] at (6,1.7) {$u_3$};
        \draw[dotted,  thick] (7,2.3) -- (7,-0.2);
        \node[above] at (8.75,1.7) {$u_4$};

        \draw[-,thick,\coulun](0,1.3) -- (2,1.3);
                \draw[-,thick](0,1.4) -- (0,1.2);
                \node[above] at (0.15,1.2) {\small $\oright{q}$};
        \draw[-,thick,\coulde](2,1.3) -- (2.5,1.3);
                \draw[-,thick](2,1.4) -- (2,1.2);
                \node[above] at (2.15,1.2) {\small $\oright{p}$};
        \draw[-,thick,\coulde] (2.5,1.3) arc (90:-90:0.1);
        \draw[-,thick,\coulde](2.5,1.1) -- (1.5,1.1);
        \draw[-,thick,\coulde] (1.5,1.1) arc (90:270:0.1);
        \draw[-,thick,\coulde](1.5,0.9) -- (5,0.9);
        \draw[-,thick,\coulun](5,0.9) -- (5.5,0.9);
                \draw[-,thick](5,1) -- (5,0.8);
                \node[above] at (5.15,0.8) {\small $\oright{p}$};
        \draw[-,thick,\coulun] (5.5,0.9) arc (90:-90:0.1);
        \draw[-,thick,\coulun](5.5,0.7) -- (4,0.7);
        \draw[-,thick,\coulun] (4,0.7) arc (90:270:0.1);
        \draw[-,thick,\coulun](4,0.5) -- (6,0.5);
        \draw[-,thick,\coulun] (6,0.5) arc (90:-90:0.1);
        \draw[-,thick,\coulun] (6,0.3) -- (4,0.3);
        \draw[-,thick,\coulun] (4,0.3) arc (90:270:0.1);
        \draw[-,thick,\coulun] (4,0.1) -- (7,0.1);
        \draw[-,thick,\coulde] (7,0.1) -- (10.5,0.1);
                \draw[-,thick](7,0.2) -- (7,0);
                \node[above] at (7.15,0.0) {\small $\oright{p}$};

        \node[above] at (1,0.9) {\textcolor{\coulun}{$\rho_1$}};
        \node[above] at (3.5,0.85) {\textcolor{\coulde}{$\rho_{2}$}};
        \node[above] at (6,0.45) {\textcolor{\coulun}{$\rho_{3}$}};
        \node[above] at (8.75,0.05) {\textcolor{\coulde}{$\rho_{4}$}};

\end{tikzpicture}

\caption{\label{fig:shape-run-forest} Shape of a run $\rho$
in \cref{lem:run-idempotents} (i.e.\ when $\mu(u_i) \mu(u_j) = \mu(u_i)$ for all $i,j$).}

\end{figure}
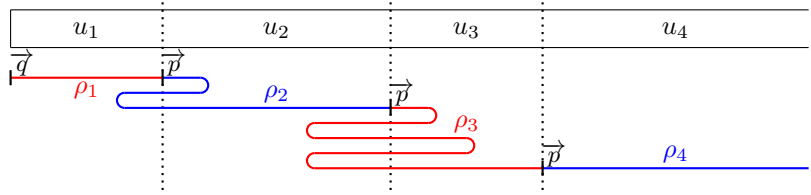

From now on, $\mu : \Sigma^* \fonc M$ is assumed
to be the transition morphism of $\trans$. Now, we show
that along a forward Ramseyan $\mu$-split built, the 
runs of $\trans$ can be described in a rather simple way. This
result is inspired by analog results for factorization forests
over finite words (see e.g.\ \cite[Lemma 6.8]{bojanczyk2018polyregular}),
but the current result is weaker due to the fact that our $\mu$-splits
are weaker than factorization forests (see \cref{rem:ff-weaker}).
The runs $\rho_1, \rho_2, \dots$
mentioned in \cref{lem:run-idempotents}
are depicted in \cref{fig:shape-run-forest}.

\begin{lemma}[Runs and forward Ramseyan splits]
\label{lem:run-idempotents}
Let \mbox{$k \ge 1$}, $u = u_1 u_2 \cdots \in \finin{\Sigma}$ and
$\frs = \frs_1 \hash{k} \frs_2 \cdots
\in \FRS{k}(u)$ be a $\mu$-split following the factorization
$u_1, u_2, \dots$.
Assume that $\delta(\oright{q},u_1) = \oright{p}$.
Let $\rho$ be the longest (finite or infinite) run of $\trans$ along $u$
that starts in configuration $(q,1)$. Then
$\rho$ can be factored under the form
$\rho = \rho_1 \fonc{} \rho_2 \fonc{} \cdots$
where:
\begin{enumerate}
\item \label{it:run:i} for all $i \ge 1$
such that $u_i$ is defined, $\rho_i$ starts in the first configuration
of $\rho$ that visits $u_i$;
\item \label{it:run:p} for all $i \ge 2$, $\rho_i$ begins with a configuration
of the form $(p, \_)$ (i.e.\ it starts in $\oright{p}$);
\item \label{it:run:2} for all $i \ge 2$ such that $u_{i+1}$ is defined,
$\rho_i$ only visits the positions of $u_i$ and $u_{i-1}$
(it cannot go back to $u_{i-2}$).
\end{enumerate}
\end{lemma}

\begin{proof}
We have $\delta(\oright{q},u_1) = \oright{p}$.
For all $i \ge 2$ such that $u_{i}$ is defined,
we must have finite words $u_1, \dots, u_{i-1} \in \Sigma^+$ and
$\mu(u_1) \mu(u_2) \cdots \mu(u_{i-1}) = \mu(u_1)$,
therefore $\delta(\oright{q},u_1 u_2 \cdots u_{i-1}) =
\delta(\oright{q}, u_1)= \oright{p}$.
This means that the factor $u_{i}$ is visited by $\rho$,
and furthermore that this visit starts in state $p$,
giving \cref{it:run:i,it:run:p} by defining $\rho_i$
accordingly. For \cref{it:run:2}, let $i \ge 3$ (for $i=2$
the result is obvious) be such that  $u_{i+1}$ is defined,
we show that $\rho_i$ only visits
$u_i$ and $u_{i-1}$. First, observe that
this run does not visit $u_{i+1}$ by construction
of $\rho_{i+1}$. Let us consider the state $r$
seen in the last visit of the first position of $u_{i-1}$ in $\rho_{i-1}$.
Since $\mu(u_{i-1} u_i) = \mu(u_{i-1})$, we have
$\delta(\oright{r}, u_{i-1} u_i) = \delta(\oright{r}, u_{i-1})
= \oright{p}$ (the last equality follows from \cref{it:run:p},
because it describes the beginning of $\rho_{i}$).
This means that when starting from $r$ in the first position
of $u_{i-1}$, $\trans$ will execute the end of $\rho_{i-1}$,
then $\rho_i$, and it will eventually leave $u_{i-1} u_i$ ``by the right''.
Hence the run $\rho_{i}$ stays in $u_{i-1}u_i$,
until it goes to $u_{i+1}$ in state $p$ (and this is
by construction the beginning of $\rho_{i+1}$).
\end{proof}

\begin{remark}
If $u_{n+1}$ is not defined (i.e.\ the split was of
shape $\frs_1 \hash{k} \cdots \hash{k} \frs_n$)
with $u_n \in \finin{\Sigma}$, there is no
reason why $\rho_n$ should visit only
$u_{n-1}$ and $u_n$. Indeed, we have no
information on $\mu(u_n)$ (even if it is defined), hence we cannot describe
the behavior of $\trans$ within $u_n$. As a consequence,
the last $\rho_n$ (if it exists) cannot be controlled, and
\cref{lem:run-idempotents} does not imply that $\delta(\oright{q}, u) = \omega$ if $u \in A^\omega$,
nor $\delta(\oright{q}, u) = \oright{p}$ if $u \in A^+$.
\end{remark}

\begin{remark} Observe that in \cref{fig:shape-run-forest},
the run $\rho_2$ visits $3$ times the first position of $u_2$,
whereas $\rho_3$ visits $5$ times the first position of $u_3$.
In the case of Simon's usual factorization forests (i.e.\ $\mu(u_1) = \mu(u_2) = \cdots$)
such a difference ($3 \neq 5$) cannot occur due to the fact that 
all factors have the same idempotent image.
\end{remark}

Recall that \cref{theo:KR-finite} states that any regular function
over finite words can be written as the restriction of a function of $\Class$
to finite words. Using this result, we claim in \cref{lem:left-finite} that
the runs of $\trans$ that start on the right of a finite word
can be simulated by a function of $\Class$.
We even replace the input word $u$ by a $\mu$-split $\frs \in \FRS{k}(u)$
(this is not necessary here, but it will be simpler to
handle the more precise $\frs$ in the rest of
the proof).

\begin{lemma}
\label{lem:left-finite}
For all $k \ge 0$, one can effectively build a function
$\runleft{k} : (\Sigma \cup \Hash{k} \cup \oleft{Q})^*
\parfonc (\Sigma \cup \Hash{k} \cup \Gamma \cup  \oleft{Q})^*  \in \Class$ such that for all
$u \in \Sigma^+$, $\frs \in \FRS{k}(u)$ and $q \in Q$:
\begin{itemize}
\item if $\delta(\oleft{q},u) = \oright{p}$ and
$\lambda(u \oleft{q}) = v$ then 
$\runleft{k}(\frs \oleft{q}) = v \frs \oright{p} $;
\item if $\delta(\oleft{q},u) = \oright{p}$ and
$\lambda(\oleft{q},u) = v$ then 
$\runleft{k}(\frs \oleft{q}) = v \oleft{p} \frs $.
\end{itemize}
\end{lemma}

\begin{proof} Such a function (from finite words to finite words)
can be computed by a two-way transducer
that ignores the symbols of $\Hash{k}$.
The result follows from \cref{theo:KR-finite}.
\end{proof}

\subsection{Main inductive proof}
\label{ssec:main-indu}

The heart of the proof is to show \cref{lem:right-runs}  by induction on 
$k \ge 0$. It is roughly an analogue of \cref{lem:left-finite},
but for the runs that start on the left of a finite
or infinite word. Due to the fact that words can
be infinite ``on the right'' but not ``on the left'',
there is a fundamental dissymmetry between the
two aforementioned lemmas.

\begin{lemma}
\label{lem:right-runs}
For all $k \ge 0$, one can effectively build
a function $\runright{k} \in \Class$ such that
for all $q \in Q$, $u \in \Sigma^\infty$
and $\frs \in \FRS{k}(u)$:
\begin{itemize}
\item if $\delta(\oright{q},u) = \oleft{q}$
and $\lambda(\oright{q},u) = v \in \Gamma^*$ 
then $\runright{k}(\oright{q} \frs) = v \oleft{q} \frs$;
\item if $\delta(\oright{q},u) = \omega$
and $\lambda(\oright{q},u) = v \in \Gamma^\omega$
then   $\runright{k}(\oright{q} \frs) = v$;
\item if $u \in \Sigma^+$, $\delta(\oright{q},u) = \oright{q}$
and $\lambda(\oright{q}u) = v \in \Gamma^*$  then
$\runright{k}(\oright{q} \frs) = v \frs \oright{q}$.
\end{itemize}
\end{lemma}

\begin{remark} If we do not fall in one of the three cases
(e.g.\ $\delta(\oright{q},u) = \omega$ but  $\lambda(\oright{q}u) \in \Gamma^*$),
the value of $\runright{k}$ does not matter (it may even be undefined).
\end{remark}

Now, let us explain
why \cref{theo:KR-detreg} immediately follows from this result.

\begin{proof}[Proof of \cref{theo:KR-detreg}]
By \cref{theo:colcombet}, one can effectively build
a function $\spli : \Sigma^\omega \fonc (\Sigma \cup \Hash{k})^\omega$
such that for all $u \in \Sigma^\omega$,
$\spli (u) \in \FRS{N}(u)$. We postcompose $\spli$
with the function that adds a the letter $\oright{q_0}$
in the beginning of its input, and then with $\runright{N}$.
This composition produces the same output as $\trans$
whenever the input has an accepting run whose output is infinite.
However, it may be defined on inputs that are no in the
domain of $\trans$. But recall that this domain
$L$ is a Büchi deterministic language by \cref{prop:dom-reg}.
Therefore, it is sufficient to precompose our construction
by the restriction of the identity function to $L$
(which is a sequential function over infinite words).
\end{proof}

The rest of \cref{ssec:main-indu} is devoted to the
proof of \cref{lem:right-runs}.
We show this result by induction on $k \ge 0$.
For $k = 0$, necessarily $u = \frs = a$
or $u = \frs = \varepsilon$ and the result
follows by building a sequential function
that hardcodes the functions $\delta$ and
$\lambda$.

Now, assume that the function $\runright{k}$ in $\Class$
is built for some $k \ge 0$. We describe how
to build the function $\runright{k+1}$ in $\Class$.
Let  $u \in A^{\infty}$ and $\frs \in \FRS{k{+}1}(u)$.
Without loss of generalities (up to first applying
a sequential function turning $ \hash{k+1}$ into  $\hash{}$),
we assume that $\oright{q}\frs$ is of shape
$\oright{q}\frs_1 \hash{} \frs_{2}  \hash{} \cdots$
where $\frs_i \in \FRS{k}(u_i)$, $u_i \in A^{\infty}$
and $u = u_1 u_2 \cdots$.

\begin{enumerate}

\item We first apply the function $\then{\runright{k}\hash{}\ide\hash{}}$ (where $\ide$
stands for the identity function) built from \cref{lem:then}, which outputs
the following (with $v_1 \defined \lambda(\oright{q}u_1)$):
\begin{itemize}
\item if $\delta(\oright{q},u_1) = \omega$
(necessarily $u = u_1 \in \Sigma^\omega$) and $v \in \Gamma^\omega$,
then $v_1$ (nothing left to do);
\item if $\delta(\oright{q}, u_1) = \oleft{p_1}$, then
$v_1 \oleft{p_1} \frs_1 \hash{}  \frs_2 \cdots $
(nothing left to do);
\item if $\delta(\oright{q}, u_1) = \oright{p}$
(necessarily $u_1 \in \Sigma^+$), then
$v_1  \frs_1 \oright{p} \hash{}  \frs_2 \cdots $.
\end{itemize}

\item We then apply a sequential function that
checks if a $ \oright{p} \hash{}$ occurs, removes
the $\oright{p}$ and replaces each
letter $\hash{}$ by $\hash{} \oright{p}$, thus it produces:
\begin{itemize}
\item if $\delta(\oright{q}, u_1) = \omega$ and $v_1 \in \Gamma^\omega$,
then $v$  (nothing left to do);
\item if $\delta(\oright{q}, u_1) = \oleft{p_1}$, then
$v_1 \oleft{p_1} \frs_1 \hash{}  \frs_2 \cdots $
 (nothing left to do);
\item if $\delta(\oright{q}, u_1) = \oright{p}$,
then
$v_1  \frs_1 \hash{} \oright{p}  \frs_2 \hash{} \oright{p} \frs_3 \hash{} \oright{p} \cdots $.
In this case, following the notations of \cref{lem:run-idempotents}, then $v_1$ is the output
produced along $\rho_1$, and for all $i \ge 2$ $\rho_i$ starts in state
$p$. Now, the idea is to simulate \emph{in parallel, using a ``map''}
(and this is the key idea of this proof)
all the runs $\rho_i$ for $i \ge 2$ since all of them
begin in state $p$.
\end{itemize}

\item \label{it:right}
We thus want to apply $\runright{k}(\oright{p}\frs_{i})$ on all the appropriate factors.
For this, we first write a $\#$ symbol before each
$\oright{p}$ (this operation will enable to leave the factors
without $\oright{p}$ unchanged). We build $\tryright \defined \then{\ide \# \runright{k}\#}$
(recall that if it does not meet a $\#$, this function will
just behave as the identity function $\ide$).
Then we apply apply the function $\then{\tryright\hash{}}$
on the whole input:
%
%
%
\begin{itemize}
\item the word is not modified if there was no
letter $\oright{p}$ (first cases above);
\item if the input was  $v_1  \frs_1 \hash{} \oright{p}
 \frs_2 \hash{} \oright{p} \frs_3 \hash{} \oright{p} \cdots $,
 we obtain a word of shape $v_1  \frs_1 \hash{} 
 w_2 \hash{} w_3 \hash{} \cdots$. Let us
 describe formally the $w_i$ for $i \ge 2$,
 depending on the runs $\rho_i$ of \cref{lem:run-idempotents}:
\begin{itemize}
\item if $u_{i+1}$ is defined, two cases occur:
  \begin{itemize}
  \item either $\rho_i$ never visits the last position of $u_{i-1}$,
  which means that it leaves $u_i$ ``directly by the right'',
  and in state $p$ by \cref{lem:run-idempotents}. In this
  case we have  $w_i = v_i \frs_i \oright{p}$
  where $v_i$ is the output produced
  along the  whole $\rho_i$;
  \item or $\rho_i$ visits the last position of $u_{i-1}$,
  and the first visit of this position is done in state $p_i \in Q$,
  then  $w_i = v_i \oleft{p_{i}}\frs_i$ where $v_i \in \Gamma^*$
  is output of $\trans$ along $\rho_i$ from its beginning to this first visit;
  \end{itemize}
\item else if $u_n \in \finin{\Sigma}$ is the last factor,
three cases can occur which produce
either $w_n = v_n \frs_n \oright{r}$ (similar to the first case above)
or $w_n = v_n\oleft{p_n} \frs_i $ (similar to the second case)
or $w_n = v_n \in \Gamma^\omega$ ($\rho_n$ visits arbitrarily large positions
of $u_n \in \Sigma^\omega$)
\end{itemize}
\end{itemize}

\item \label{it:clean} Then, we apply a sequential function that replaces
each subword $\oright{s} \hash{}$ by $\hash{}$
(in order to remove the $\oright{p} \hash{}$
in the factors, since the corresponding $\rho_i$
have been fully simulated).

\item \label{it:dupli} Now, our goal is to simulate the next part
of the runs $\rho_i$ that were not completed,
i.e.\ deal with the factors of shape
$w_i = v_i \oleft{p_{i}}\frs_i$. In this case,
we need to compute $\runleft{k}(\frs_{i-1} \oleft{p_{i}})$.
We first build a function that copies the $\frs_{i-1}$
in the next factor, if it contains some $\oleft{s}$.

\begin{claim}
\label{cla:behind}
Let $\#$ be a fresh symbol. We can effectively build a function of $\Class$
that takes
as input the word built so far and turns
each factor $w_i$ of shape $v_i \oleft{p_i}\frs_i$ for $i \ge 2$
into $ v_i  p_i  \#\frs_{i-1} \oleft{p_i}\# \frs_i$,
and leaves the other factors unchanged.
\end{claim}

\begin{proof} Since all factors have shape
either $v_i \oleft{p_i} \frs_i$, or $v_i \frs_i$
or $v_n \frs_n \oright{r}$ or $v_n \in \Gamma^\omega$,
we can first apply a sequential function
that adds a $\#$ symbol right before reach $\frs_i$.
We then apply $\mapc$ with separator $\#$,
giving factors of shape $\# \frs_{i-1}\hash{} v_i (\oleft{p_i}?) \#
\frs_{i-1}\hash{} v_i (\oleft{p_i}?) \#$. We then apply a sequential
function that uses the first $\frs_{i-1}$ for the previous factor,
then outputs $v_i$, then $p_i \# \frs_{i_1}$ if there is a $\oleft{p_i}$,
ignores the next $v_i$ and ends with $\oleft{p_i}$.
It is easy to see that this function behaves as expected
if $v_i$ or $\frs_{i-1}$ is infinite.
\end{proof}

Finally, we apply the function described in \cref{cla:behind} to the  whole input.

\item \label{it:left} It remains to apply $\runleft{k}(\frs_{i-1} \oleft{p_{i}})$
on the appropriate factors.
For this, we build $\tryleft \defined \then{\ide \# \runleft{k} \# \ide\#}$
(recall that if it does not meet a $\#$, this function will
just behave as the identity function $\ide$)
and then we apply the function $\then{\tryleft\hash{}}$
on the whole word:
\begin{itemize}
\item the factors that do not contain $\#$ are not modified
(i.e.\ the factors where there was previously no $\oleft{s} \in \oleft{Q}$,
plus the first one that may contain a $\oleft{s}$);
\item otherwise, each factor of shape
$ v_i  p_i  \#\frs_{i-1} \oleft{p_i}\# \frs_i$
 for $i \ge 2$ is now turned into:
\begin{itemize}
\item either $v_i  p_i  \# v'_i  \frs_{i-1} \oright{p'_i}\# \frs_i$.
It means that after its stay in $u_{i-1}$, $\rho_i$ goes back
to $u_i$ in state $p'_i$, and that the output of $\trans$ along
this portion of $\rho_i$ in $u_{i-1}$ is  $v'_i$. In this case
we want to concatenate $v_i$ and $v'_i$;
\item or $v_n  p_n  \# v'_n \oleft{p'_n} \frs_{n-1} \# \frs_n$.
In this case, it means that $\rho_{n}$ will visit $u_{n-2}$.
According to \cref{lem:run-idempotents}, this is only possible if
$u_n \in \Sigma^\omega$ was the last factor. In this case the $\mu$-split
cannot help us controlling $\rho_n$, but this very
particular case can occur only once in the  whole process
and will be treated in \cref{it:left-right}.
\end{itemize}
\end{itemize}

\item \label{it:last} Now, let us remove the $\#$ and the
useless copies of factors.

\begin{claim} \label{cla:norm}
One can effectively build in $\Class$ a function $\norm$
over the alphabet $\Sigma \cup \Gamma \cup Q \cup \oleft{Q} \cup \oright{Q}
\cup \{\#\}$ that behaves as follows:
\begin{itemize}
\item if its input does not contain $\#$, it behaves as the identity function;
\item if its input has shape
$v_i  p_i  \# v'_i  \frs_{i-1} \oright{p'_i}\# \frs_i$,
it is mapped to
$v_i  v'_i \oright{p'_i} \frs_i$;
\item if its input has shape $v_n  p_n  \# v'_n \oleft{p'_n} \frs_{n-1} \# \frs_n$,
it is mapped to $v_n  p_n \frs_n$
\end{itemize}
\end{claim}

\begin{proof} We first replace the second $\#$ (if it exists) by a $\hash{}$
and then apply a $\mapc$ with separator $\hash{}$. Thus if the input
had the second (resp.\ the third) shape it is now
$v_i  p_i  \# v'_i  \frs_{i-1} \oright{p'_i}\hash{}
v_i  p_i  \# v'_i  \frs_{i-1} \oright{p'_i}\hash{}
\frs_i\cdots$ (resp.
$v_n  p_n \# v'_n \oleft{p'_n} \frs_{n-1}\hash{}
v_n  p_n  \# v'_n \oleft{p'_n} \frs_{n-1}\hash{}
\frs_n \cdots$). We finally apply a sequential function that
outputs what it sees until a factor $s \#$ with $s \in Q$.
In this case it reads the next factor between $\#$
and $\hash{}$ (without writing) to determine whether
the input has the second or the third shape,
and then it behaves accordingly on the last piece.
This process leaves unchanged
the (finite or infinite) words without $\#$.
\end{proof}

We then apply the function $\then{\norm\hash{}}$ to our  whole input. Observe
that the last case of \cref{cla:norm}, we have undone
the computation of $\runleft{k}(\frs_{i-1} \oleft{p_i})$
and the state $p_i$ no longer has an over-arrow: it is somehow
``frozen'' (but recall that this situation can occur only for the last
factor) and the rest of $\rho_i$ will be simulated in \cref{it:left-right}.

\item \label{it:iterate} By iterating $|Q|$ times the steps described in \cref{it:right,it:clean,it:dupli,it:left,it:last},
one can simulate the runs $\rho_i$ for $i \ge 2$ (except possibly the last one,
cf. above). Indeed, these runs visit at most $|Q|$ times
the first position of their respective $u_i$.
As a result, we obtain an output that has one of the following shapes
(where $v_1 \defined \lambda(\oright{q}u_1)$):
\begin{itemize}
\item if $\delta(\oright{q}, u_1) = \omega$ and $v_1 \in \Gamma^\omega$,
then $v_1$ (nothing left to do);
\item if $\delta(\oright{q}, u_1) = \oleft{p_1}$, then
$v_1 \oleft{p_1} \frs_1 \hash{}  \frs_2 \cdots $
(nothing left to do);
\item if $\delta(\oright{q}, u_1) = \oright{p}$
(necessarily $u_1 \in A^+$), then two main cases possible cases occur:
\begin{itemize}
\item no state was ``frozen'' (i.e.\ each $\rho_i$
has been entirely simulated), and we get a word
of shape $v_1 \frs_1 \hash{} v_2 \frs_2 \hash{} \cdots$
(possibly ending with a $v_i \in \Gamma^\omega$ or with
a $\oright{r}$ if the input was finite), where each
$v_i$ is exactly the output of $\trans$ along $\rho_i$
of \cref{lem:run-idempotents}. In this case it
remains to regroup the $v_i$ (and the $\frs_i$
if the input is finite);
\item or a state $s \in Q$ was ``frozen'' at some point
(in the last factor) and we obtain a word of shape
$v_1 \frs_1 \hash{} \cdots \hash{} v_n s \frs_n$,
where the $v_i$ for $i < n$ are the output of $\trans$
along $\rho_i$, and $v_n$ is the output along $\rho_n$
until some visit of the last position of $u_{n-1} \in \Sigma^+$,
which is done in state.
In this case, we still have some output to produce.
\end{itemize}
\end{itemize}

\item \label{it:normal} Now, we claim that we can apply a function
of $\Class$ so that the result is:
\begin{itemize}
\item if $\delta(\oright{q}, u_1) = \omega$ and
$v_1 \defined \lambda(\oright{q}u_1) \in \Gamma^\omega$,
then $v_1$ (nothing to do);
\item if $\delta(\oright{q}, u_1) = \oleft{p_1}$
and $v_1 \defined \lambda(\oright{q}u_1) \in \Gamma^*$, then
$v_1 \oleft{p_1} \frs_1 \hash{}  \frs_2 \cdots $
(nothing to do);
\item else if $\delta(\oright{q}, u_1) = \oright{p}$:
\begin{itemize}
\item either no state was frozen during the process and we
have either $v \in \Gamma^\omega$ (infinite input)
or $v \frs_1 \hash{} \cdots \hash{}\frs_n \oright{s}$
(finite input) where $v \defined \lambda(\oright{q}u)$
(nothing to do);
\item or a state was frozen and we obtain
$v \# \frs_1 \hash{} \cdots \hash{}   \frs_{n-1} \hash{} \oleft{s}\frs_n$
where $u = u_1 \cdots u_n$ with $u_n \in \finin{\Sigma}$
and  $v$ is the output produced along the run $\rho$
of \cref{lem:run-idempotents} until some visit in the last position of 
$u_{n-1}$, which is done in state $s \in Q$.
\end{itemize}
\end{itemize}

To create such a function of $\Class$, we first add a $\#$
before any $s, \oleft{s}$ or $\oright{s}$ (if it exists).
We then apply $\then{\mapc \# \ide \#}$
and use the first copy to produce the $v_1 v_2 \cdots v_n$
(with a final $\#$ if required) and the second one for
the $\frs_1 \hash{} \cdots \hash{} \frs_{n-1}$.
Finally we add the $\oleft{s} \frs_n$ or $\frs_n \oright{s}$.
Observe that the construction can be extended
in a correct way if there are no
 $s, \oleft{s}$ nor $\oright{s}$: the input
was infinite and we only produce  the $v_i$.

\item \label{it:left-right} Finally, let us explain briefly how to deal with the last
case of \cref{it:normal}, that is
when we have  $v \# \frs_1 \hash{} \cdots \hash{}  \frs_{n-1} \hash{} \oleft{s}~\frs_n$
with $u_n \in \finin{\omega}$. We first transform the $\hash{} \oleft{s}$ into $\oleft{s}\#$.
Then we apply the function
$\then{\ide \# \runleft{k+1} \# \ide\#}$ in order to simulate a
run on the prefix (beware that we need $\runleft{k+1}$ and not $\runleft{k}$ here).
The result either gives  an output and a state $\oleft{t}$ on the left
or $\oright{t}$ on the right. If there is a $\oright{t}\#$,
we replace it by $\# \oright{t}$ and then
apply $\then{\ide \# \ide \# \runright{k}\#}$ to simulate a run on the suffix.
Then we regroup the outputs, in a manner similar  to \cref{it:normal}.
Since the first position of $u_n$ is visited at most $|Q|$ times,
we can simulate the  whole run $\rho$ by iterating these operations
$|Q|$ times.

\end{enumerate}

%% file: app-mso.tex
\section{Guarded MSO-transductions}\label{app:mso}

\subsection{\texorpdfstring{\GMSOT}{GMSOT} with total order, and proof
of Proposition~\ref{prop:notcomput}}\label{subsec:ordermsot}

In the definition of \MSOT and \GMSOT of Sections~\ref{subsec:MSOT}, it is not required to
define the total order $\leq$ of the output structure. We consider
here \MSO-transductions where the total order is required to be defined, instead of the
successor.  While this does not change the expressiveness
of \MSOT, as the successor is expressible from the order and
vice-versa, the picture is different with the guarded fragment.  We
prove that this modification allows one to define functions which are
not deterministic regular.

In this section, we write $\GMSOT[S]$  instead
of $\GMSOT$ to emphasize the fact that only successor is
defined. We denote \GMSOTO the class of
guarded \MSOT with order, defined exactly as $\GMSOT[S]$ except that
instead of having formulas for successors, there are formulas
$\phi_\leq^{c,d}(x,y)$ for the order, and any copies $c,d$. The
semantics is as follows. Let $T\in\GMSOTO$, over input
alphabet $\Sigma$ and output alphabet $\Gamma$.  On input
$u\in\Sigma^\omega$, $T$ defines an output structure that we
denote $T(u)$, over the signature
$\{\leq(x,y),(\gamma(x))_{\gamma\in\Gamma}\}$. We then denote by
$T(u)|_S$ the structure over signature
$\{S(x,y),(\gamma(x))_{\gamma\in\Gamma}\}$ obtained from $T(u)$ by
interpreting $S$ has the one-step successor of the interpretation
of $\leq$ in $T(u)$. Note that at this point, the structure $T(u)|_S$
might not correspond to a word structure, it is just \emph{some}
structure over $\{S(x,y),(\gamma(x))_{\gamma\in\Gamma}\}$. As for
$\GMSOT[S]$, we define the semantics of $T$ over word structures only:
\[
\sem{T} = \{ (u,v)\in\Sigma^\omega\times\Gamma^\omega\mid
T(u)|_S\text{ is isomorphic to } G_v\}
\]

\begin{example}\label{ex:gmsoto}
    Let $\Sigma=\Gamma=\{a,b\}$ and $\textsf{er}_b : \Sigma^*\rightarrow \Sigma^*$ the (erasing) morphism defined by
    $\textsf{er}_b(a) = a$ and $\textsf{er}_b(b) = \epsilon$. We
    define a function $h:\Sigma^\omega\parfonc\Gamma^\omega$ on
    inputs
    of the form $bub^\omega$, for $u\in\{a,b\}^*$, by $h(bub^\omega) =
    b\textsf{er}_b(u)b^\omega$. The function $h$ is not continuous
    (and so not deterministic regular): $\lim_{n\infty}
    h(b^nab^\omega) = bab^\omega\neq h(\lim_{n\infty} b^nab^\omega) =
    h(b^\omega) = b^\omega$. We show that it is definable by a
    $1$-copy \GMSOTO. An example of output structure  on input
    $bbabaab^\omega$ is given below (we depict only the successor predicate and not
    the order):

    \vspace{2mm}

\begin{center}
    \begin{tikzpicture}[->,>=stealth',shorten >=1pt,auto,scale=0.5,every node/.style={scale=0.75}]

\tikzstyle{alivenode}=[circle,fill=black!80,thick,inner sep=1pt,minimum size=4mm]

\tikzstyle{deadnode}=[circle,fill=black!10,thick,inner sep=0pt,minimum size=4mm]

\node at (-1,1) {{\footnotesize input}};
\node at (-1,-0.7) {{\footnotesize output}} ;

\node [alivenode] (a1) at (0,1) {{\color{white} $b$}} ;
\node [alivenode](a2) at (2,1) {{\color{white} $b$}} ;
\node [alivenode](a3) at (4,1) {{\color{white} $a$}} ;
\node [alivenode](a4) at (6,1) {{\color{white} $b$}} ;
\node [alivenode](a5) at (8,1) {{\color{white} $a$}} ;
\node [alivenode](a6) at (10,1) {{\color{white} $a$}} ;
\node [alivenode](a7) at (12,1) {{\color{white} $b$}} ;
\node [alivenode](a8) at (14,1) {{\color{white} $b$}} ;
\node (a9) at (15,1) {{$\ldots$}} ;

  \draw[->]   (a1) --  (a2) ;
  \draw[->]   (a2) --  (a3) ;
  \draw[->]   (a3) --  (a4) ;
  \draw[->]   (a4) --  (a5) ;
  \draw[->]   (a5) --  (a6) ;
  \draw[->]   (a6) --  (a7) ;
  \draw[->]   (a7) --  (a8) ;

\node (a1) at (0,1.5) {\begin{scriptsize}$1$\end{scriptsize}} ;
\node (a2) at (2,1.5) {\begin{scriptsize}$2$\end{scriptsize}} ;
\node (a3) at (4,1.5) {\begin{scriptsize}$3$\end{scriptsize}} ;
\node (a4) at (6,1.5) {\begin{scriptsize}$4$\end{scriptsize}} ;
\node (a5) at (8,1.5) {\begin{scriptsize}$5$\end{scriptsize}} ;
\node (a6) at (10,1.5) {\begin{scriptsize}$6$\end{scriptsize}} ;
\node (a7) at (12,1.5 ) {\begin{scriptsize}$7$\end{scriptsize}} ;
\node (a8) at (14,1.5 ) {\begin{scriptsize}$8$\end{scriptsize}} ;

\node [alivenode] (x1) at (0,-0.7) {{\color{white} $b$}} ;
\node [alivenode] (x2) at (2,-0.7) {{\color{white} $b$}} ;
\node [alivenode] (x3) at (4,-0.7) {{\color{white} $a$}} ;
\node [alivenode] (x4) at (6,-0.7) {{\color{white} $b$}} ;
\node [alivenode] (x5) at (8,-0.7) {{\color{white} $a$}} ;
\node [alivenode] (x6) at (10,-0.7) {{\color{white} $a$}} ;
\node [alivenode] (x7) at (12,-0.7) {{\color{white} $b$}} ;
\node [alivenode] (x8) at (14,-0.7) {{\color{white} $b$}} ;
\node  (x9) at (15,-0.7) {$\ldots$} ;

\draw[->] (x1) to [bend right=25] (x6);
  \draw[->]   (x6) --  (x5) ;
  \draw[->]   (x5) to [bend left=30]  (x3) ;
  \draw[->]   (x3) to [bend left=50]  (x2) ;
  \draw[->]   (x2) to [bend left]  (x4) ;
  \draw[->]   (x4) to [bend left=20]   (x7) ;
  \draw[->]   (x7) to  (x8) ;


\fill[blue!20,rounded corners, fill opacity=0.2] (-0.3,0.4) rectangle (15.3,-2.4);

\end{tikzpicture}
\end{center}

To define the formulas, we use the guarded formula $\first(x) =
\exists \gv\forall
y\leq x\leq \gv\cdot y=x$ which holds true if $x$ is the first input position. Note that its negation is equivalent to the
guarded formula $\exists y\leq x\cdot y\neq
x$. Then, the first output element is defined by $\phi_{\fst}(x) =
\first(x)\wedge b(x)$, which holds true if $x$ is the first
input element \emph{and} it is labeled $b$. If the first element is
not labelled $b$, then the latter formula is not satisfied, and by definition of
the semantics of \GMSOT, the output structure is empty. So, it
is not isomorphic to any infinite word structure and therefore the input word is
not in the domain. The label formulas preserve the labels:
$\phi_\sigma(x) = \sigma(x)$ for all $\sigma\in\{a,b\}$.

Let us define the order formula
$\phi_{\leq}(x,y)$. If $x$ is the first input node, then it is smaller
than any $y$. If $x$ is the second node, then it is smaller than any
$b$ which appears later. If $x$ is labelled $a$, then it is smaller
than any $b$ but the first one, and to any other $a$ that appears to
its left:
\[
\begin{array}{llllllll}
  \phi_\leq(x,y) & = & (b(x)\wedge b(y))\rightarrow x\leq y\wedge \\
                 & & (b(x)\wedge a(y))\rightarrow \first(x))
                     \wedge \\
  & & a(x)\rightarrow (b(y)\wedge\neg \first(x))\vee (a(y)\wedge
      y\leq x)
\end{array}
\]
    Note that all the formulas above are guarded. We do not explicitly
    write them in guarded form, but it suffices to use $max(x,y)$ as
    guard.

                         \end{example}

This proves \cref{prop:notcomput}.


%% file: appendix-sarah.tex
\newcommand{\register}[1]{\draw[<-,thick,draw=gray] () to[out=north,in=south] node[fill=lightgray, rectangle, rounded corners, yshift=1em] {#1} +(0,2em);}

\newcommand{\registerright}[2]{
\draw[<-,thick,draw=black,dotted] () to[out=south,in=south] node[yshift=-1em] {#1} +(0,-2em);
\draw[<-,thick,draw=gray] +(0,-3em) to[out=south,in=north] node[fill=lightgray, rectangle, rounded corners, yshift=-1em] {#2} +(0,-5em);
}

\forestset{look-ahead tree/.style={for tree={thick, rectangle, rounded corners, draw, l sep=15pt, grow'=0, edge+={thick, draw},}}}

\forestset{look-ahead tree with registers/.style={for tree={thick, rectangle, rounded corners, draw, l sep=20pt, s sep = 3em, grow'=0, edge+={thick, draw},}}}

\forestset{look-ahead tree with registers small/.style={for tree={thick, rectangle, rounded corners, draw, l sep=5pt, s sep = 3em, grow'=0, edge+={thick, draw},}}}

\newcommand{\lasum}[1]{}
\newcommand{\qok}{{q_{\Sigma^*}}}
\newcommand{\qa}{q_{a?}}
\newcommand{\qb}{q_{b?}}
\newcommand{\qc}{q_{c?}}
\newcommand{\pa}{p_{a}}
\newcommand{\pb}{p_{b}}
\newcommand{\pc}{p_{c}}

\newcommand{\set}[1]{\{#1\}}
\newcommand{\up}{\mathsf{up}}
\newcommand{\delf}{\mathsf{del}_{\mathsf{fail}}}
\newcommand{\dels}{\mathsf{del}_{\mathsf{success}}}
\newcommand{\summary}{\operatorname{\textsf{\small summary}}}
\newcommand{\mer}{\mathsf{merge}}
\newcommand{\outp}{\mathsf{out}}

\newcommand{\LAT}{\ensuremath{\mathrm{LAT}}\xspace}

\section{Removal of finite look-ahead}
\label{app:twoway}

In the remainder this section let $\trans = (\Sigma,\Gamma,Q,q_0,F,\delta,\lambda)$ be a \tDTLA.
Wlog., we assume that only transitions moving to the right use non-trivial look-ahead languages and transition going to the left use the trivial look-ahead language $\Sigma^*$ (aka.\ no look-ahead).

This can be achieved by replacing every transition that moves to the left using some non-trivial look-ahead by a transition going right (with this look-ahead) and two transitions moving left using no look-ahead.
A convenient side effect of this construction is that for a state either only right or only left transitions are enabled.

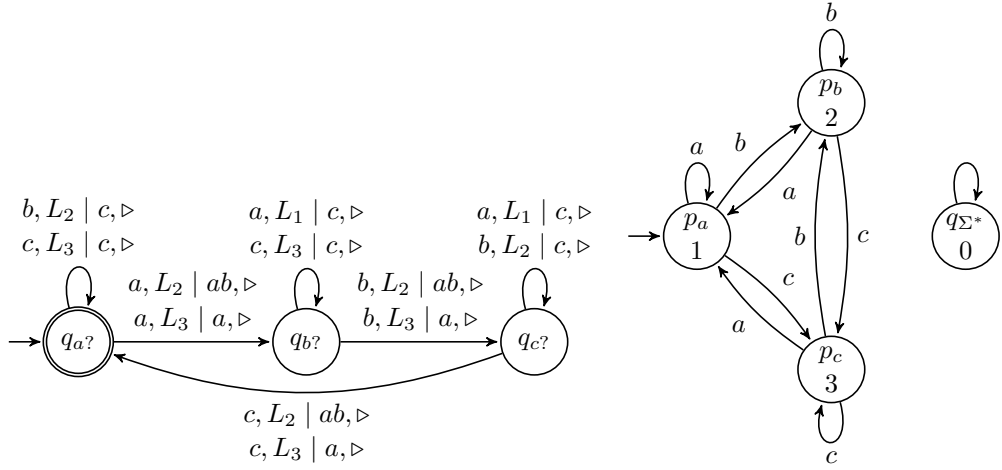
\begin{figure}[h!t]
\begin{tikzpicture}[->, >=stealth', shorten >=1pt, auto, node distance=3cm, semithick, initial text={}]
\tikzstyle{every state}=[align=center]
\node[initial,state,accepting] (A)              {$\qa$};
\node[state]                    (B) [right of=A] {$\qb$};
\node[state]                    (C) [right of=B] {$\qc$};

\path (A) edge []  node[align=center] {$a,L_2 \mid ab, \rmove$ \\ $a,L_3 \mid a, \rmove$} (B)
          edge [loop above] node[align=center] {$b,L_2 \mid c, \rmove$ \\ $c,L_3 \mid c, \rmove$} (A)
      (B) edge []  node[align=center] {$b,L_2 \mid ab, \rmove$ \\ $b,L_3 \mid a, \rmove$} (C)
          edge [loop above] node[align=center] {$a,L_1 \mid c, \rmove$ \\ $c,L_3 \mid c, \rmove$} (B)
      (C) edge [bend left=20]  node[align=center] {$c,L_2 \mid ab, \rmove$ \\ $c,L_3 \mid a, \rmove$} (A)
          edge [loop above] node[align=center] {$a,L_1 \mid c, \rmove$ \\ $b,L_2 \mid c, \rmove$} (C);
\end{tikzpicture}
\begin{tikzpicture}[->, >=stealth', shorten >=1pt, auto, node distance=2.5cm, semithick, initial text={}]
  \tikzstyle{every state}=[align=center,inner sep=-1pt,ellipse]
  \node[initial,state] (A)                    {$\pa$ \\ $1$};
  \node[state]         (B) [above right of=A] {$\pb$ \\ $2$};
  \node[state]         (C) [below right of=A] {$\pc$ \\ $3$};
  \node[state]         (E) [below right of=B] {$\qok$ \\ $0$};
  
  \path (A) edge [bend left=10]  node {$b$} (B)
            edge [bend left=10]  node {$c$} (C)
            edge [loop above] node {$a$} (A)
        (B) edge [bend left=10]  node {$a$} (A)
            edge [bend left=10]  node {$c$} (C)
            edge [loop above] node {$b$} (B)
        (C) edge [bend left=10]  node {$b$} (B)
            edge [bend left=10]  node {$a$} (A)
            edge [loop below] node {$c$} (C)
        (E) edge [loop above] node {}    (E);
  \end{tikzpicture}
\caption{On the left-hand side is a deterministic two-way transducer with finite look-aheads, on the right side is the corresponding look-ahead automaton.}
\label{fig:automata-look-ahead-info}
\end{figure}

Let $L_1,\dots,L_n$ be the regular look-ahead languages that occur in $\delta$.
Wlog., we assume that these look-ahead languages can be accepted using the same DFA with different final states.
Formally, let $\mathcal A = (\Sigma,Q_\mathcal A,q_0^\mathcal A,F_\mathcal A,Q_\mathcal A,\delta_\mathcal A)$ be a DFA and $c: Q_\mathcal A \to \mathcal P(\{1,\cdots,n\})$ be a (multi-)coloring of its states.
Let $F_i = \{ q \in Q_\mathcal A \mid c(q) = i \}$ for all $1 \leq i \leq n$, and let $\mathcal A_i = (A,Q_\mathcal A,q_0^\mathcal A,F_\mathcal A,F_i,\delta_\mathcal A)$.
We assume that $L(\mathcal A_i) = L_i$.
We refer to $\mathcal A$ as \emph{look-ahead automaton}.
We extend its states by a designated sink state $\qok$ with color 0 to model the trivial look-ahead language $\Sigma^*$ (aka.\ no look-ahead).
Let $C$ be the set of colors $\{0,1,\dots,n\}$.
See \cref{fig:automata-look-ahead-info} for an example.

\subsection{Look-ahead trees}

\begin{figure}[p]

\vspace*{1em}
\begin{subfigure}{0.45\textwidth}
\begin{forest}
  look-ahead tree with registers
  [ {$\qa,\set{(\qok,0)}$}, fill=lime 
  ] {\register{$\varepsilon$}} {\lasum{$\set{(\qok,0)}$}} 
\end{forest}
\caption{starting look-ahead tree}
\label{subfig:lat-a}
\end{subfigure}
\begin{subfigure}{0.5\textwidth}
\begin{forest}
  look-ahead tree with registers
  [ {$\qa,\set{(\qok,0)}$}, fill=lime 
      [ {$\qb,\set{(\pa,2)}$} 
      ] {\register{$ab$}} 
      [ {$\qb,\set{(\pa,3)}$} 
      ] {\register{$a$}} 
  ] {\register{$\varepsilon$}} 
\end{forest}
\caption{after application of $\up(\cdot,a)$; no more changes after application of $\dels(\cdot)$, $\delf(\cdot)$, and $\mer(\cdot)$}
\label{subfig:lat-b}
\end{subfigure}

\vspace*{1em}
\begin{subfigure}{\textwidth}
\begin{forest}
  look-ahead tree with registers
  [ {$\qa,\set{(\qok,0)}$}, fill=lime 
      [ {$\qb,\set{(\pa,2)}$} 
          [ {$\qb,\set{(\pa,1)}$}, fill=lime 
          ] {\register{$c$}} 
      ] {\register{$ab$}} 
      [ {$\qb,\set{(\pa,3)}$} 
          [ {$\qb,\set{(\pa,1)}$}, fill=lime 
          ] {\register{$c$}} 
      ] {\register{$a$}} 
  ] {\register{$\varepsilon$}} 
\end{forest} 
\caption{after application of $\up(\cdot,a)$; no more changes after application of $\dels(\cdot)$, $\delf(\cdot)$, and $\mer(\cdot)$}
\label{subfig:lat-c}
\end{subfigure}

\vspace*{1em}
\begin{subfigure}{\textwidth}
\begin{forest}
  look-ahead tree with registers
  [ {$\qa,\set{(\qok,0)}$}, fill=lime 
      [ {$\qb,\set{(\pc,2)}$} 
          [ {$\qb,\set{(\pa,1)}$}, fill=lime 
              [ {$\qc,\set{(\pa,2)}$} 
              ] {\register{$ab$}} 
              [ {$\qc,\set{(\pa,3)}$} 
              ] {\register{$a$}} 
          ] {\register{$c$}} 
      ] {\register{$ab$}} 
      [ {$\qb,\set{(\pc,3)}$}, fill=lime 
          [ {$\qb,\set{(\pa,1)}$}, fill=lime, tikz={\node [draw,ellipse,inner sep=-1pt,fit to=tree,label=below:\emph{remaining part}] {};} 
              [ {$\qc,\set{(\pa,2)}$} 
              ] {\register{$ab$}} 
              [ {$\qc,\set{(\pa,3)}$} 
              ] {\register{$a$}}  
          ] {\register{$c$}} 
      ] {\register{$a$}} 
  ] {\register{$\varepsilon$}} 
\end{forest}
\caption{after application of $\up(\cdot,b)$; result after application of $\dels(\cdot)$ is marked; the output $\varepsilon \cdot a$ from the to-be-removed part is appended to $\out$}
\label{subfig:lat-d}
\end{subfigure}

\caption{
  Overview of how to construct look-ahead trees based on the transducer given in \cref{fig:automata-look-ahead-info}.
  The \cref{subfig:lat-a,subfig:lat-b,subfig:lat-c,subfig:lat-e,subfig:lat-g,subfig:lat-i,subfig:lat-j} depict look-ahead trees, the \cref{subfig:lat-d,subfig:lat-f,subfig:lat-h} depict intermediate construction steps.
  Nodes that are drawn with lime background have succeeded.
  Nodes drawn with grey background do not belong to the tree; they visualize registers that are associated to the nodes of the tree.
  The second part of this figure can be found on \cpageref{fig:ex-look-ahead-tree-pt2}.
}
\label{fig:ex-look-ahead-tree-pt1}
\end{figure}

\begin{figure}[p]
\ContinuedFloat

\vspace*{1em}
\begin{subfigure}{\textwidth}
\begin{forest}
  look-ahead tree with registers
  [ {$\qb,\set{(\pa,1)}$}, fill=lime 
      [ {$\qc,\set{(\pa,2)}$} 
      ] {\register{$ab$}} 
      [ {$\qc,\set{(\pa,3)}$} 
      ] {\register{$a$}}  
  ] {\register{$c$}} 
\end{forest}
\caption{after application of $\dels(\cdot)$; no more changes after application of $\delf(\cdot)$ and $\mer(\cdot)$}
\label{subfig:lat-e}
\end{subfigure}

\vspace*{1em}
\begin{subfigure}{\textwidth}
\begin{forest}
  look-ahead tree with registers
  [ {$\qb,\set{(\pa,1)}$}, fill=lime 
      [ {$\qc,\set{(\pb,2)}$}, fill=lime 
        [ {$\qc,\set{(\pa,1)}$}, fill=lime, tikz={\node [draw,ellipse,inner sep=-1pt,fit to=tree,label=below right:\emph{\phantom{bla bla}remaining part},xshift=1em] {};} 
          [ {$\qc,\set{(\pa,2)}$} 
          ] {\register{$c$}}  
        ] {\register{$c$}}  
      ] {\register{$ab$}} 
      [ {$\qc,\set{(\pb,3)}$} 
        [ {$\qc,\set{(\pa,1)}$}, fill=lime 
          [ {$\qc,\set{(\pa,2)}$} 
          ] {\register{$c$}}  
      ] {\register{$c$}}  
      ] {\register{$a$}} 
  ] {\register{$c$}} 
\end{forest} 
\caption{after application of $\up(\cdot,a)$, $\dels(\cdot)$, $\delf(\cdot)$, $\mer(\cdot)$ and $\up(\cdot,b)$; result after application of $\dels(\cdot)$ is marked; the output $c \cdot ab$ from the to-be-removed part is appended to $\out$}
\label{subfig:lat-f}
\end{subfigure}

\vspace*{1em}
\begin{subfigure}{\textwidth}
\begin{forest}
  look-ahead tree with registers
        [ {$\qc,\set{(\pa,1)}$}, fill=lime 
          [ {$\qc,\set{(\pa,2)}$} 
          ] {\register{$c$}}  
        ] {\register{$c$}}  
\end{forest} 
\caption{after application of $\dels(\cdot)$; no more changes after application of $\delf(\cdot)$ and $\mer(\cdot)$}
\label{subfig:lat-g}
\end{subfigure}

\vspace*{1em}
\begin{subfigure}{\textwidth}
\begin{forest}
  look-ahead tree with registers
        [ {$\qc,\set{(\pa,1)}$}, fill=lime 
          [ {$\qc,\set{(\pa,2)}$} 
            [  {$\qc,\set{(\pa,1)}$}, fill=lime, tikz={\node [draw,ellipse,inner sep=-1pt,fit to=tree,label=below:\emph{merge}] {};} 
              [  {$\qc,\set{(\pa,1)}$}, fill=lime 
              ] {\register{$c$}}  
            ] {\register{$c$}}  
          ] {\register{$c$}}  
        ] {\register{$c$}}  
\end{forest} 
\caption{after application of $\up(\cdot,a)$, $\dels(\cdot)$, $\delf(\cdot)$, $\mer(\cdot)$, $\up(\cdot,a)$, $\dels(\cdot)$, and $\delf(\cdot)$; nodes that will be affected by $\mer(\cdot)$ are marked}
\label{subfig:lat-h}
\end{subfigure}

\vspace*{1em}
\begin{subfigure}{\textwidth}
  \begin{forest}
    look-ahead tree with registers
          [ {$\qc,\set{(\pa,1)}$}, fill=lime 
            [ {$\qc,\set{(\pa,2)}$} 
              [  {$\qc,\set{(\pa,1)}$}, fill=lime 
              ] {\register{$cc$}}  
            ] {\register{$c$}}  
          ] {\register{$c$}}  
  \end{forest} 
\caption{after application of $\mer(\cdot)$}
\label{subfig:lat-i}
\end{subfigure}

\vspace*{1em}
\begin{subfigure}{\textwidth}
  \begin{forest}
    look-ahead tree with registers
          [ {$\qc,\set{(\pa,1)}$}, fill=lime 
            [ {$\qc,\set{(\pa,2)}$} 
              [  {$\qc,\set{(\pa,1)}$}, fill=lime 
              ] {\register{$cccccccccccc$}}  
            ] {\register{$c$}}  
          ] {\register{$c$}}  
  \end{forest} 
\caption{after $10 \times$ application of $\up(\cdot,a)$, $\dels(\cdot)$, $\delf(\cdot)$ and $\mer(\cdot)$}
\label{subfig:lat-j}
\end{subfigure}

\caption{
  Overview of how to construct look-ahead trees based on the transducer given in \cref{fig:automata-look-ahead-info}.
  The \cref{subfig:lat-a,subfig:lat-b,subfig:lat-c,subfig:lat-e,subfig:lat-g,subfig:lat-i,subfig:lat-j} depict look-ahead trees, the \cref{subfig:lat-d,subfig:lat-f,subfig:lat-h} depict intermediate construction steps.
  Nodes that are drawn with lime background have succeeded.
  Nodes drawn with grey background do not belong to the tree; they visualize registers that are associated to the nodes of the tree.
  The first part of this figure can be found on \cpageref{fig:ex-look-ahead-tree-pt1}.
}
\label{fig:ex-look-ahead-tree-pt2}
\end{figure}

We introduce some notation.

\begin{definition}[look-ahead information]
 A \emph{look-ahead information} is a tuple of the form $(p,c)$, where $p \in Q_\mathcal A$, and $c \in C$.
 The intended meaning of such a tuple is that the look-ahead automaton is currently in state $p$ and we want to reach a state with color $c$.

 We say that a look-ahead information $(p,c)$ has \emph{succeeded} if $c \in c(p)$.

 Given a look-ahead information $(p,c)$ and a letter $\sigma \in \Sigma$, we define 
 $\up((p,c),\sigma) = (\delta_\mathcal A(p,\sigma),c)$ if $(p,c)$ has not succeeded, otherwise 
 $\up((p,c),\sigma)$ = $(p,c)$ (because once we know if a look-ahead information has succeeded, we want to freeze its status).

 Let $I$ denote the set of all look-ahead information.
 Given a subset $M \subseteq I$ and a letter $\sigma \in \Sigma$, we define $\up(M,\sigma) = \set{ \up(i,\sigma) \mid i \in M}$.
\end{definition}

We introduce the concept of a look-ahead trees which we inductively define.
Look-head trees are trees whose nodes have labels from $Q \times I$, i.e.\ they store a state of the two-way transducer and a set of look-ahead information.
A visualization is given in \cref{fig:ex-look-ahead-tree-pt1,fig:ex2-look-ahead-tree-pt1}.

\begin{definition}[look-ahead tree]
  This set of single-node trees are look-ahead trees: $\{(q,\set{(\qok,0)}) \mid q \in Q\}$.

  Given a look-ahead tree $t$ and a letter $\sigma \in \Sigma$, applying the following operations ($t_1 = \up(t,\sigma)$ then $t_2 = \dels(t_1)$ then $t_3 = \delf(t_2)$ and finally $t' = \mer(t_3)$) defined below yields a new look-ahead tree $t'$.

  \paragraph*{Update}

  The first step of the update process is to update all look-ahead information wrt.\ the letter $\sigma$.
  Therefore, we re-label every node.
  A label of the form $(q,M)$ is replaced by $(q,\up(M,\sigma))$.

  We say that a nodes with label $(q,M)$ has \emph{succeeded} if all its look-ahead information stored in $M$ have succeeded.
  A path of in the tree is called \emph{successful} if all its nodes have succeeded.

  The second step of the update process is to add new nodes to the tree.
  For a state $q$, we collect the set of $\rmove$-transitions that are potentially enabled with the letter $\sigma$.
  Let $S_q = \set{ i \mid \delta(q,\sigma,L_i) \neq \bot}$.

  For a leaf with label $(q,M)$ we add the following nodes as its children if $S_q \neq \emptyset$.
  For all $i \in S_q$, we add a child with label $(q',\set{(q_0^\mathcal A,i)}$ if $\delta(q,\sigma,L_i) = (q',\rmove)$.

  Otherwise (if $S_q = \emptyset$), we add a child with label ``fail''.

  We refer to this process as $\up(\cdot,\sigma)$.

  \paragraph*{Delete}
  The purpose of the delete step is two-fold.

  On the one hand, we want to remove a path from the root to a leaf if it is successful as this indicates the path stores info about the accepting run (we have seen witnesses for the look-aheads on this path).
  On the other hand, we want to remove paths of which we know that they do not store info about an accepting run.

  After the delete step, the look-ahead tree only contains nodes that store info about runs where some look-head still has to be verified.

  As the fist step, if the tree contains a node $n$ which is not the root node $r$ and the path from $r$ to $n$ is successful, and $n$ has no children that have succeeded, then $n$ becomes the new root node.
  Note that the assumption of disjointness for the look-aheads ensures that there is at most one such path. 

  We refer to this process as $\dels(\cdot)$.

 \paragraph*{}
  As the second step, we remove nodes that do not store info about the accepting run.
  Therefore, the following steps are done:
  \begin{itemize}
    \item If a leaf is labeled ``fail'', we delete this leaf and all its ancestors until the ancestor has at least two children.
    This indicates that this run of the two-way transducer has failed.
    \item If two paths from the root to a leaf describe the same situation (to be explained) we can delete them both.
    More concretely, given a path $\pi = n_1n_2\cdots n_i$ from the from the root to leaf such that the label of the node $n_j$ is $(q_j,M_j)$ for all $1 \leq j \leq i$.
    We denote by $\summary(\pi)$ the set $\bigcup_{j=1}^n M_j$.
    It summarizes the status of the look-aheads that have to hold on this path.
    Let $\pi_1$ and $\pi_2$ be two paths from the root to two leafs such that the leaf labels are $(q,M)$ and $(q,M')$, respectively, and $\summary(\pi_1) = \summary(\pi_2)$.
    By construction, these paths describe \emph{different} runs of the two-way transducer, but they have the same look-ahead status.
    That means that both of these runs must either fail or succeed.
    It is not possible that both runs will succeed as this contradicts the fact the two-way transducer is deterministic.
    Hence, both of these runs will fail and we can remove their representation from the look-ahead tree.
    To do so, whenever there are two paths whose leafs store the same state of the two-way transducer and their look-ahead summary is equal, we delete this leaf and all its ancestors until the ancestor has at least two children.
  \end{itemize}
  
  We refer to this process as $\delf(\cdot)$.

  \paragraph*{Merge}
  Finally, if the tree contains a linear branch segment of length at least 3 meaning that there are nodes $n_1,n_2,\ldots,n_j$ such that $j>3$ and $n_{i+1}$ is the child of $n_{i}$ and $n_i$ has no other children for $1 \leq i < j$, then this branch is merged as follows:
  The nodes $n_3,\ldots, n_{j-1}$ are removed, and the look-ahead information from the deleted nodes is pushed to $n_j$.
  Let $M_3,\ldots,M_{j_1} \subseteq I$ be their sets of look-ahead information, then we replace the look-ahead information $M_j$ of node $n_j$ by $\bigcup_{i=3}^{j} M_i$.
  Note that after a applying a merge, another linear branch segment might have been created, so the merge operation must be repeated until no changes occur.

  We refer to this process as $\mer(\cdot)$.
\end{definition}

\begin{lemma}
  The set of look-ahead trees is finite.
\end{lemma}

\begin{proof}
This follows directly from the inductive definition.
The set of look-ahead summaries is finite, and linear branch segments are bounded in height.
Hence, the set of look-ahead trees is finite.
\end{proof}

\subsection{Construction of a Büchi deterministic SST}

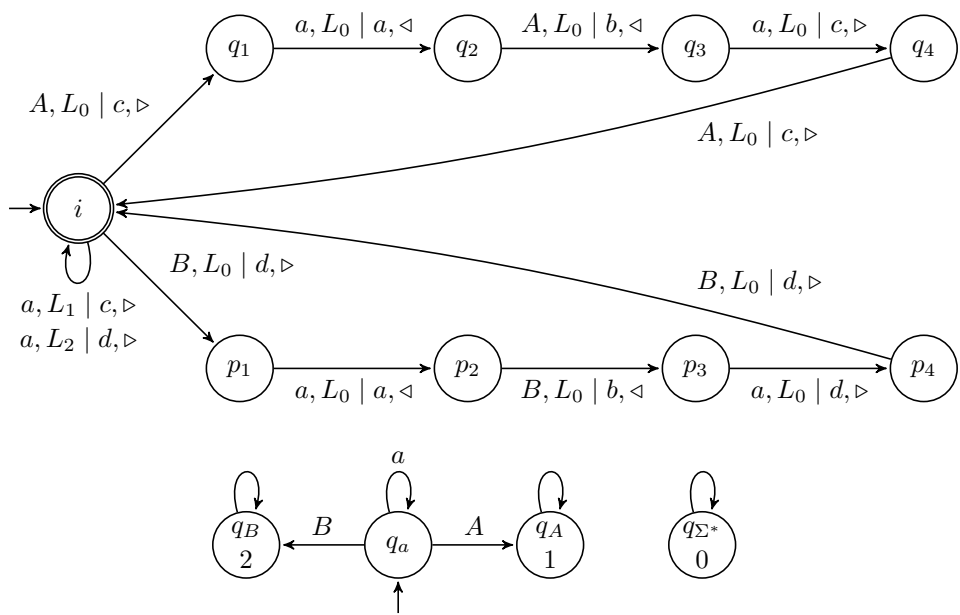
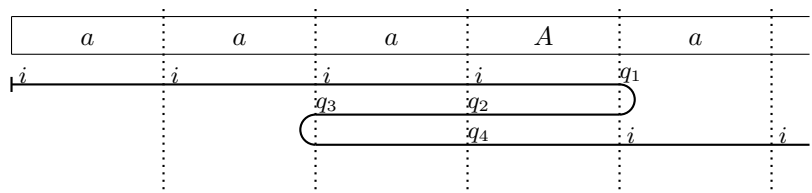
\begin{figure}[h!t]

  \begin{subfigure}{\textwidth}
  \centering
  \begin{tikzpicture}[->, >=stealth', shorten >=1pt, auto, node distance=3cm, semithick, initial text={}]
  \tikzstyle{every state}=[align=center]
  \node[initial,state,accepting] (A)              {$i$};
  \node[state]                    (B) [above right of=A] {$q_1$};
  \node[state]                    (C) [right of=B] {$q_2$};
  \node[state]                    (D) [right of=C] {$q_3$};
  \node[state]                    (E) [right of=D] {$q_4$};
  \node[state]                    (B2) [below right of=A] {$p_1$};
  \node[state]                    (C2) [right of=B2] {$p_2$};
  \node[state]                    (D2) [right of=C2] {$p_3$};
  \node[state]                    (E2) [right of=D2] {$p_4$};

  \path (A) edge []  node[align=center] {$A,L_0 \mid c, \rmove$} (B)
            edge []  node[align=center] {$B,L_0 \mid d, \rmove$} (B2)
            edge [loop below] node[align=center] {$a,L_1 \mid c, \rmove$ \\ $a,L_2 \mid d, \rmove$} (A)
        (B) edge []  node[align=center] {$a,L_0 \mid a, \lmove$} (C)
        (C) edge []  node[align=center] {$A,L_0 \mid b, \lmove$} (D)
        (D) edge []  node[align=center] {$a,L_0 \mid c, \rmove$} (E)
        (E) edge [bend left=5]  node[align=center,near start] {$A,L_0 \mid c, \rmove$} (A)
        (B2) edge []  node[align=center,swap] {$a,L_0 \mid a, \lmove$} (C2)
        (C2) edge []  node[align=center,swap] {$B,L_0 \mid b, \lmove$} (D2)
        (D2) edge []  node[align=center,swap] {$a,L_0 \mid d, \rmove$} (E2)
        (E2) edge [bend right=5]  node[align=center,near start,swap] {$B,L_0 \mid d, \rmove$} (A);
  \end{tikzpicture}

  \vspace*{1em}
  \centering
  \begin{tikzpicture}[->, >=stealth', shorten >=1pt, auto, node distance=2cm, semithick, initial text={}]
    \tikzstyle{every state}=[align=center,inner sep=-1pt,ellipse]
    \node[initial,state,initial where=below] (A)                    {$q_a$};
    \node[state]         (B) [right of=A] {$q_A$ \\ $1$};
    \node[state]         (C) [left of=A] {$q_B$ \\ $2$};
    \node[state]         (E) [right of=B] {$\qok$ \\ $0$};
    
    \path (A) edge []  node {$A$} (B)
              edge []  node[swap] {$B$} (C)
              edge [loop above] node {$a$} (A)
          (B) edge [loop above] node {} (B)
          (C) edge [loop above] node {} (C)
          (E) edge [loop above] node {}    (E);
    \end{tikzpicture}
    \caption{Top: a deterministic two-way transducer with finite look-ahead. Bottom: the corresponding look-ahead automaton.}
  \end{subfigure}
  
    \vspace*{1em}
    \begin{subfigure}{\textwidth}
      \centering
      \begin{tikzpicture}{scale=1}


              \draw (0,1.7) -- (10.5,1.7);
              \draw (0,2.2) -- (10.5,2.2);
              \draw (0,2.2) -- (0,1.7);

              \node[above] at (1,1.7) {$a$};
              \draw[dotted,  thick] (2,2.3) -- (2,-0.2);
              \node[above] at (3,1.7) {$a$};
              \draw[dotted,  thick] (4,2.3) -- (4,-0.2);
              \node[above] at (5,1.7) {$a$};
              \draw[dotted,  thick] (6,2.3) -- (6,-0.2);
              \node[above] at (7,1.7) {$A$};
              \draw[dotted,  thick] (8,2.3) -- (8,-0.2);
              \node[above] at (9,1.7) {$a$};
              \draw[dotted,  thick] (10,2.3) -- (10,-0.2);

              \draw[-,thick](0,1.4) -- (0,1.2);
              \draw[-,thick,](0,1.3) -- (2,1.3);
              \draw[-,thick,](2,1.3) -- (4,1.3);
              \draw[-,thick,](4,1.3) -- (6,1.3);
              \draw[-,thick,](6,1.3) -- (8,1.3);
  
              \draw[-,thick,] (8,1.3) arc (90:-90:0.2);
              \draw[-,thick,](8,0.9) -- (4,0.9);
              \draw[-,thick,] (4,0.9) arc (90:270:0.2);
              \draw[-,thick,](4,0.5) -- (10.5,0.5);
  
              \node[above] at (0.15,1.2) {\small $i$};
              \node[above] at (2.15,1.2) {\small $i$};
              \node[above] at (4.15,1.2) {\small $i$};
              \node[above] at (6.15,1.2) {\small $i$};
              \node[above] at (8.15,1.2) {\small $q_1$};
              \node[above] at (6.15,0.8) {\small $q_2$};
              \node[above] at (4.15,0.8) {\small $q_3$};
              \node[above] at (6.15,0.4) {\small $q_4$};
              \node[above] at (8.15,0.4) {\small $i$};
              \node[above] at (10.15,0.4) {\small $i$};

      \end{tikzpicture}
      \caption{\label{fig:run} Run of the transducer on the input $aaaAa\cdots\in \Sigma^\omega$. Recall that the transducer always reads the symbol to the right, even when it moves to the left.}  
    \end{subfigure}

  \caption{A deterministic two-way transducer with finite look-ahead and a corresponding run.}
  \label{fig:look-ahead-example}
  \end{figure}

  \begin{figure}[p]
    
    \vspace*{1em}
    \begin{subfigure}{0.3\textwidth}
    \begin{forest}
      look-ahead tree with registers
      [ {$i,\set{(\qok,0)}$}, fill=lime 
      ] {\register{$\varepsilon$}} 
    \end{forest} 
    \caption{starting look-ahead tree}
    \label{subfig:lat2-a}
    \end{subfigure}
    \begin{subfigure}{0.65\textwidth}
    \begin{forest}
      look-ahead tree with registers
      [ {$i,\set{(\qok,0)}$}, fill=lime 
          [ {$i,\set{(q_a,1)}$} 
            [ {$i,\set{(q_a,1)}$} 
            ] {\register{$c$}}  
            [ {$i,\set{(q_a,2)}$}, fill=pink 
            ] {\register{$d$}}  
          ] {\register{$c$}} 
          [ {$i,\set{(q_a,2)}$} 
            [ {$i,\set{(q_a,1)}$}, fill=pink 
            ] {\register{$c$}}  
            [ {$i,\set{(q_a,2)}$} 
            ] {\register{$d$}}  
          ] {\register{$d$}} 
      ] {\register{$\varepsilon$}} 
    \end{forest} 
    \caption{after application of $\up(\cdot,a)$, $\dels(\cdot)$, $\delf(\cdot)$, $\mer(\cdot)$, $\up(\cdot,a)$ and $\dels(\cdot)$; the red nodes indicate branches where the look-ahead info summary is the same, since the state in these nodes is also the same, these branches must be deleted}
    \label{subfig:lat2-b}
    \end{subfigure}

    \vspace*{1em}
    \begin{subfigure}{\textwidth}
    \begin{forest}
      look-ahead tree with registers
      [ {$i,\set{(\qok,0)}$}, fill=lime 
          [ {$i,\set{(q_a,1)}$} 
            [ {$i,\set{(q_a,1)}$} 
            ] {\register{$c$}}  
          ] {\register{$c$}} 
          [ {$i,\set{(q_a,2)}$} 
            [ {$i,\set{(q_a,2)}$} 
            ] {\register{$d$}}  
          ] {\register{$d$}} 
      ] {\register{$\varepsilon$}} 
    \end{forest} 
    \caption{after application of $\delf(\cdot)$ and $\mer(\cdot)$}
    \label{subfig:lat2-c}
    \end{subfigure}

    \vspace*{1em}
    \begin{subfigure}{\textwidth}
    \begin{forest}
      look-ahead tree with registers
      [ {$i,\set{(\qok,0)}$}, fill=lime 
          [ {$i,\set{(q_A,1)}$}, fill=lime 
            [ {$i,\set{(q_A,1)}$}, fill=lime 
              [ {$q_1,\set{(\qok,0)}$}, fill=lime 
              ] {\register{$c$}}  
            ] {\register{$cc$}}  
          ] {\register{$c$}} 
          [ {$i,\set{(q_A,2)}$} 
            [ {$i,\set{(q_A,2)}$} 
            [ {$q_1,\set{(\qok,0)}$}, fill=lime 
            ] {\register{$c$}}  
            ] {\register{$dd$}}  
          ] {\register{$d$}} 
      ] {\register{$\varepsilon$}} 
    \end{forest} 
    \caption{after application of $\up(\cdot,a)$, $\mer(\cdot)$, $\dels(\cdot)$, $\delf(\cdot)$, $\mer(\cdot)$, $\up(\cdot,A)$; this tree represents (some of the) left-to-right runs of the transducer from \cref{fig:look-ahead-example} on the prefix $aaaA$ of $aaaAa\cdots\in \Sigma^\omega$; left-to-right runs that are not represented have been identfied as doomed to fail in previous steps (see e.g.\ \cref{subfig:lat2-b})}
    \label{subfig:lat2-d}
    \end{subfigure}

    \caption{
      Overview of how to represent left-to-right runs and right-to-right runs using look-ahead trees based on the transducer given in \cref{fig:look-ahead-example} on the input $aaaAa\cdots \in \Sigma^\omega$.
      The \cref{subfig:lat2-a,subfig:lat2-c,subfig:lat2-e,subfig:lat2-f,subfig:lat2-h} represent objects that are stored in the state space of the SST that is constructed for the the transducer given in \cref{fig:look-ahead-example}.
      The \cref{subfig:lat2-b,subfig:lat2-d,subfig:lat2-g} represent intermediate construction steps to detemine a new state of the SST.
      All gray nodes in these figures represent registers and their contents.
      The second part of the figure can be found on \cpageref{fig:ex2-look-ahead-tree-pt2}.
    }
    \label{fig:ex2-look-ahead-tree-pt1}
    \end{figure}

\begin{figure}[p]
\ContinuedFloat

    \vspace*{1em}
    \begin{subfigure}{\textwidth}
      \begin{forest}
        look-ahead tree with registers
        [ {$q_3,\set{(\qok,0)}$}, fill=lime 
            [ {$q_4,\set{(\qok,0)}$}, fill=lime 
            ] {\register{$c$}} 
        ] {\register{$\varepsilon$}} {\registerright{$\set{q_2}$}{$b$}} 
      \end{forest}  
    \caption{This look-ahead tree and associated set represent a right-to-right from $q_2$ to $q_3$ of the transducer from \cref{fig:look-ahead-example} on the prefix $aaa$ of $aaaAa\cdots\in \Sigma^\omega$.}
    \label{subfig:lat2-e}
    \end{subfigure}

    \vspace*{1em}
    \begin{subfigure}{\textwidth}
    \begin{forest}
      look-ahead tree with registers
      [ {$q_3,\set{(\qok,0)}$}, fill=lime 
          [ {$q_4,\set{(\qok,0)}$}, fill=lime 
            [ {$i,\set{(\qok,0)}$}, fill=lime 
            ] {\register{$c$}} 
          ] {\register{$c$}} 
      ] {\register{$\varepsilon$}} {\registerright{$\set{q_1}$}{$ab$}} 
    \end{forest} 
    \caption{This look-ahead tree and associated set represent a right-to-right from $q_1$ to $q_4$ of the transducer from \cref{fig:look-ahead-example} on the prefix $aaaA$ of $aaaAa\cdots\in \Sigma^\omega$.}
    \label{subfig:lat2-f}
    \end{subfigure}

    \vspace*{1em}
    \begin{subfigure}{\textwidth}
    \scalebox{0.89}{
    \begin{forest}
      look-ahead tree with registers small
      [ {$i,\set{(\qok,0)}$}, fill=lime 
          [ {$i,\set{(q_A,1)}$}, fill=lime 
            [ {$i,\set{(q_A,1)}$}, fill=lime 
              [ {$q_1,\set{(\qok,0)}$}, fill=lime 
    [ {$q_3,\set{(\qok,0)}$}, fill=lime 
    [ {$q_4,\set{(\qok,0)}$}, fill=lime 
      [ {$i,\set{(\qok,0)}$}, fill=lime 
      ] {\register{$c$}} 
    ] {\register{$c$}} 
    ] {\register{$ab$}} 
              ] {\register{$c$}}  
            ] {\register{$cc$}}  
          ] {\register{$c$}} 
          [ {$i,\set{(q_A,2)}$} 
            [ {$i,\set{(q_A,2)}$} 
            [ {$q_1,\set{(\qok,0)}$}, fill=lime 
    [ {$q_3,\set{(\qok,0)}$}, fill=lime 
    [ {$q_4,\set{(\qok,0)}$}, fill=lime 
      [ {$i,\set{(\qok,0)}$}, fill=lime 
      ] {\register{$c$}} 
    ] {\register{$c$}} 
    ] {\register{$ab$}} 
            ] {\register{$c$}}  
            ] {\register{$dd$}}  
          ] {\register{$d$}} 
      ] {\register{$\varepsilon$}} 
    \end{forest}}
    \caption{The left-to-right runs represented in \cref{subfig:lat2-d} and the right-to-right run represented in \cref{subfig:lat2-f} can be connected. This shows the resulting representation.}
    \label{subfig:lat2-g}
    \end{subfigure}

    \vspace*{1em}
    \begin{subfigure}{\textwidth}
    \begin{forest}
      look-ahead tree with registers
      [ {$i,\set{(\qok,0)}$}, fill=lime 
      ] {\register{$c$}} 
    \end{forest} 
    \caption{The result of applying the operations $\dels(\cdot)$, $\delf(\cdot)$ and $\mer(\cdot)$ to the tree in \cref{subfig:lat2-g}. The output $\varepsilon \cdot c \cdot cc \cdot c \cdot ab \cdot c$ from the removed part has been added to the register $\out$.}
    \label{subfig:lat2-h}
    \end{subfigure}
    
    \caption{
      Overview of how to represent left-to-right runs and right-to-right runs using look-ahead trees based on the transducer given in \cref{fig:look-ahead-example} on the input $aaaAa\cdots \in \Sigma^\omega$.
      The \cref{subfig:lat2-a,subfig:lat2-c,subfig:lat2-e,subfig:lat2-f,subfig:lat2-h} represent objects that are stored in the state space of the SST that is constructed for the the transducer given in \cref{fig:look-ahead-example}.
      The \cref{subfig:lat2-b,subfig:lat2-d,subfig:lat2-g} represent intermediate construction steps to detemine a new state of the SST.
      All gray nodes in these figures represent registers and their contents.
      The first part of the figure can be found on \cpageref{fig:ex2-look-ahead-tree-pt1}.
    }
    \label{fig:ex2-look-ahead-tree-pt2}
    \end{figure}

Now that we have defined look-ahead trees, we design an SST (that has no look-ahead) that simulates the run(s) of the deterministic two-way transducer with finite look-ahead.

The main difficulty is that in order to simulate the run of the two-way transducer the streaming string transducer does not have a priori access to the information which look-ahead succeeds.
Hence, the SST must follow all possible runs and verify whether the assumed look-aheads are indeed enabled.
For each simulated transition of the transducer, this might take an unbounded amount of time.
Thus, it is not possible to store the followed runs in the state space of the SST until it clear which run is the right one.

In the section above, we introduced look-ahead trees.
In the following we give a construction that allows us to store an abstract representation of the followed runs using look-ahead trees.

Our goal is to design a \BSST{} $\mathcal S$, given an input word $u \in \Sigma^\omega$, that on a prefix $u'$ of $u$ maintains the (abstract) information which left-to-right runs on $u'$ are possible and which right-to-right runs are possible.
Furthermore, the output of these runs must be stored in appropriate registers, so that it can be used to build the output of the SST as soon as it is clear which of these (partial) runs are part of the (overall) run that the transducer takes.

Now, we describe the states of the SST, how a run is represented by a state, and which registers of the SST contain the output of the run and how to update the designated output register $\out$.

A visualization of these steps is given in \cref{fig:ex2-look-ahead-tree-pt2}.

Basically, for a prefix $u' \in \Sigma^*$ of $u \in \Sigma^\omega$, we want to represent all of its left-to-right and left-to-left runs and their store outputs on $u'$. 

We state more precisely what a left-to-right and a right-to-right run is.
Recall that by assumption in a configuration $(q,i)$ of $\trans$ on $u$ either only transitions moving right are enabled or one transition moving left (because there is no look-ahead to be verified when moving left).

\begin{definition}[left-to-right run, right-to-right run]
  On a prefix $u' \in \Sigma^*$ of $u \in \Sigma^\omega$, a \emph{left-to-right} (resp.\ \emph{right-to-right}) \emph{run} from $p$ to $q$ is a run starting in $(p,0)$ (resp.\ starting in $(p,|u'|)$) and ending in $(q,|u'|)$ such that in $(q,|u'|)$ only $\rmove$-transitions are enabled and no position $>|u'|$ is visited in the run.
Note that this definition enforces that right-to-right runs must take at least one transition.
\end{definition}

Let $s_0$ be the starting state of $\mathcal S$.
We first describe the state of $\mathcal S$ that is reached after reading a letter $\sigma \in \Sigma$ from $s_0$ and its associated runs on the prefix $\sigma$ and register contents.

Then, we describe how to determine from the state reached at $u' \in A^*$ the state reached at $u'\sigma$ for some $\sigma \in \Sigma$ as well as its associated runs and register contents.

After reading a, let the register $\out$ hold $\varepsilon$.
On the prefix $\sigma \in \Sigma$, we might have several left-to-right runs of the form $(q_0,0)(q',1)$ where $\delta(q_0^\mathcal A,\sigma,L) = (q',\rmove)$.

It is easy to see that these run are represented in the look-ahead tree $\up((q_0,\set{(\qok,0}),\sigma)$ by following the path from the root to the corresponding leaf.
To the root node of $t$ we associate a register (say $\mathfrak{r}$) with content $\varepsilon$.
To a leaf we associate a register with content $\lambda(\tau)$ if $\tau$ was the transition that enabled this move.

Then the output of a run $(q_0,0)(q',1)$ is given by $\cro{\out}\cro{\mathfrak{r}}\cro{\mathfrak{l}}$ where $\mathfrak{l}$ is the register that was added for the transition that lead to $(q',1)$.

Formally, to determine the right-to-right runs on some prefix, we need to know one more input letter as a right-move from position $i$ to $i-1$ depends on the input letter $i$.
So for ease of presentation, wlog., we assume that this piece of information is always given.

On the prefix $\sigma \in \Sigma$ (say the next letter is $\sigma'$), we might have right-to-right runs of the form $(q_1,1)(q_2,0)\cdots(q_n,1)$ where $n \geq 3$ is an odd number, a right-move must be made in $(q_n,1)$, and $\delta(q_i,\sigma') = (q_{i+1},\lmove)$ for all odd $i \leq n$ and $\delta(q_{i},\sigma,L_{j_i}) = (q_{i+1},\rmove)$ for all even $i \leq n$.
Such a run is represented by a pair of a look-ahead tree and a set of states as follows:
The look-ahead tree is $(q_n,\set{\bigcup_{i=2, i \text{ even}}^{n-2}(q_0^\mathcal A,j_i})$ and its associated state set is $\set{q_1}$.
We associate a register (say $\mathfrak{r}$) to the root node of the tree with content $\varepsilon$ and a register (say $\mathfrak{s}$) to $q_1$ that stores $\lambda(\tau_1)\lambda(\tau_2)\cdots\lambda(\tau_{n-1})$ where $\tau_1\cdots\tau_{n-1}$ is the sequence of transitions taken in the run $(q_1,1)(q_2,0)\cdots(q_n,1)$.
Then its output is given by $\cro{\mathfrak{s}}\cro{\mathfrak{r}}$.

Now we describe the general step.
Let $u' \in \Sigma^*$ be a non-empty prefix of $u \in \Sigma^\omega$, let $s$ be the state of $\mathcal S$ reached after reading $u'$, let the next input letter be $a$ and the next but one input letter $\sigma'$.

The state $s$ stores one look-ahead tree $t$ and possibly several pairs $(t_1,M_1),\ldots,(t_n,M_n) \in \mathrm{LAT} \otimes \mathcal P(Q)$, where $\mathrm{LAT}$ is the set of look-ahead trees and $\mathcal P(Q)$ is the powerset of $Q$ (the state set of the two-way transducer $\trans$).

We assume following invariants about $t, t_1,\ldots,t_n$.
For ease of presentation, if $\pi$ is a path in a tree let $\pi_{\mathfrak{R}}$ denote its associated sequence of registers $\mathfrak{r_1}\cdots\mathfrak{r_j}$.
We then use $\cro{\pi_{\mathfrak{R}}}$ to denote $\cro{\mathfrak{r_1}}\cdots\cro{\mathfrak{r_j}}$.

\begin{claim}\label[claim]{claim:represent}
In $t$, let $\pi$ be a path from the root to a leaf with label of the form $(q,X)$.
Then there exists a left-to-right of $\trans$ on $u'$ from $q_0$ to $q$ with output $\cro{\out}\cro{\pi_{\mathfrak{R}}}$.

In $(t_i,M_i)$, let $\pi$ be a path from the root to a leaf with label of the form $(q,X)$, and let $p \in M_i$ and $\mathfrak{r}$ be the register associated to $p \in M_i$.
Then there exists a right-to-right of $\trans$ on $u'$ from $p$ to $q$ with output $\cro{\mathfrak{r}}\cro{\pi_{\mathfrak{R}}}$.
\end{claim}

\begin{claim}\label[claim]{claim:out}
If $u \in \Dom(\trans)$ the output $\trans(v)$ can be written as $\cro{\out}v$ for some $v\in \Gamma^\omega$ and there exists a path $\pi$ from the root to a leaf in $t$ such that $\cro{\pi_\mathfrak{R}} \pref v$.
\end{claim}

Note that \cref{claim:represent,claim:out} are also valid for the state reached on a prefix $\sigma \in \Sigma$.

\subparagraph*{}
\noindent To determine the next state of $\mathcal S$, we construct 
\[
  \tilde t = \up(t,\sigma), \text{ and } \tilde t_i = \up(t_i,\sigma)
\]
for all $1 \leq i \leq n$.
To each of the newly inserted leaves we associate a register that holds the output of the transition that was used to construct this leaf.

Clearly, if $\pi$ is a path from the root to a leaf in $t$ and $\pi'$ is a path from the root to a leaf in $\tilde t$ such that $\pi \pref \pi'$:
If $\pi$ represents a left-to-right run of $\trans$ on $u'$ from $q_0$ to $q$ with output $\cro{\out}\cro{\pi_\mathfrak{R}}$, then $\pi'$ represents a left-to-right run of $\trans$ on $u'a$ from $q_0$ to $q'$ with output $\cro{\out}\cro{\pi'_{\mathfrak{R}}} = \cro{\out}\cro{\pi_{\mathfrak{R}}}\cro{\mathfrak{r}}$ where $\mathfrak{r}$ is the register associated to the leaf introduced for the transition $\delta(q,\sigma,L) = (q',\rmove)$.

For the trees $\tilde t_1,\ldots,\tilde t_n$ we update the associated sets $M_1,\ldots,M_n$.
Let $\tilde M_i$ be the set that contains $q$ if there is a state $p \in M_i$ and a transition $\delta(p,\sigma') = (q,\lmove)$.
Furthermore, let $\mathfrak{r}$ be the register associated to $p \in M_i$, we associate to $q \in \tilde M_i$ a register $\mathfrak{s}$ and let $\mathfrak{s} \mapsto v\mathfrak{r}$ where $v$ is output of the transition $\delta(p,\sigma') = (q,\lmove)$.

If the set $\tilde M_i$ is empty, then we remove the pair $(\tilde t_i,\tilde  M_i)$, because this indicates that it is not possible to extend the right-to-right runs of $\trans$ on $u'$ represented by $(t_i,M_i)$ to right-to-right runs of $\trans$ on $u'\sigma$.

It is easy to see that if $p \in M_i$ (with associated register $\mathfrak{r}$) and $q \in \tilde M_i$ (with associated register $\tilde{\mathfrak{r}}$) and $\delta(p,\sigma') = (q,\lmove)$ with output $o$, and $\pi$ is a path from the root to a leaf with label of the form $(p',X)$ in $t_i$ and $\pi'$ is a path from the root to a leaf with label of the form $(q',X')$ in $\tilde t_i$ and $\pi' \pref \pi$ and $\delta(p',\sigma,L) = (q',\rmove)$ with output $o'$:
Then if $\pi$ represents a right-to-right run of $\trans$ on $u'$ from $p$ to $q$ with output $\cro{r}\cro{\pi_\mathfrak{R}}$, then $\pi'$ represents a right-to-right run of $\trans$ on $u'\sigma$ from $p'$ to $q'$ with output $\cro{\tilde{\mathfrak{r}}}\cro{\pi'_\mathfrak{R}} = o\cro{r}\cro{\pi_\mathfrak{R}}o'$.

We now check if there is $(\tilde t_i,\tilde M_i)$ and $(\tilde t_j,\tilde M_j)$ such that these have to be combined in oder to (eventually) represent proper right-to-right runs.

This is the case if the following situation occurs.
There is a leaf with label $(q,X)$ in $\tilde t_i$ and in configuration $(q,|u'\sigma|)$ there is no right move enabled and there is $q \in \tilde M_j$. 
We make the following changes:
\begin{itemize}
  \item We change $(\tilde t_i,\tilde M_i)$ such that in $\tilde t_i$ the leaf with label $(q,X)$ gets (a copy of) $\tilde t_j$ as a subtree (i.e.\ we connect this leaf to the root of (a copy of) $\tilde t_j$).
  The content of the register associated to the (copy of) $\tilde t_j$, say $\mathfrak{r}$, is updated $\mathfrak{r} \mapsto \mathfrak{s}\mathfrak{r}$ where $\mathfrak{s}$ is the register associated to $q \in \tilde M_j$.
  \item From the set $M'_j$ the entry $q$ is deleted as we have connected the runs beginning in $q$ to the runs ending in $q$.
  If the set $\tilde M_j$ is empty after removing a state, the pair $(\tilde t_j,\tilde M_j)$ is no longer necessary and can be deleted.
\end{itemize}
These combination steps must be repeated until there are no more pairs that can be combined.
Let $(\hat t_1,\hat M_1),\ldots,(\hat t_m,\hat M_m)$ be the pairs that remain after these steps.
It is not difficult to see that these pairs are representations of right-to-right runs of $\trans$ on $u'\sigma$.
Furthermore note that there are possibly some right-to-right runs of $\trans$ on $u'\sigma$ that only visit the last letter $\sigma$.
These runs are not represented by $(\hat t_1,\hat M_1),\ldots,(\hat t_m,\hat M_m)$ as these represent runs that already existed and could be extended.
We have described for the prefix $\sigma$ how to get representations of right-to-right runs that visit only one letter.

Finally, it might be possible to combine runs represented by $\tilde t$ and runs represented by the pairs.

This is the case if there is a leaf with label $(q,X)$ in $\tilde t$ and in configuration $(q,|u'\sigma|)$ there is no right move enabled and there is $q \in M$ for some pair $(\bar t,M)$.
The combination is done in the same way as described above.  
Again, these combination steps must be repeated until $\tilde t$ can no longer be extended.
Let us denote by $\hat t$ the final result.
It is easy to see that $\hat t$ is a representation of left-to-right runs of $\trans$ on $u'a$.

There are three operations left to do.

From all trees, we remove those parts that represent successful run prefixes.

We distinguish whether we work on the tree $\hat t$ that is a representation of left-to-right runs, or on some tree $\bar t$ from a pair $(\bar t,M)$ that is a representation of right-to-right runs.

We consider the former case.
We construct $\dels(\hat t)$.
If a path segment $\pi$ of $\hat t$ has been removed in this operation (aka.\ a new root was chosen), then it is easy to see that $\cro{\out}\cro{\pi_{\mathfrak{R}}} \pref \trans(u)$ if $u \in \Dom(\trans)$.
We update $\out$ accordingly, let $\out \mapsto \out\  \mathfrak{r_1}\cdots\mathfrak{r_j}$ where $\mathfrak r_1\cdots\mathfrak r_j$ is the register sequence associated to the removed path $\pi$.

We consider the latter case.
We also construct $\dels(\bar t)$ for some $(\bar t,M)$.
The removed part of $\bar t$ models a partial run on which all look-aheads have been verified (aka.\ witnesses have been seen).
Hence, the look-ahead information is no longer required, we just need to remember the output of this partial run.
The output of this partial run is obtained by concatenating the output of the registers that are associated to the removed path.
We push the output of removed path to the new root node.
In the paragraph below we perform a similar operation for which we give a bit more formal details.

From all trees, we remove those parts that represent runs that fail.
This is done by applying the operations $\delf(\cdot)$.

Finally, to obtain proper look-ahead trees, we apply the operation $\mer(\cdot)$ to every tree.
Recall that a linear branch segment is merged by pushing all the look-ahead information to the last node.
We need to do the same for the associated output.
Hence, let $n_1\cdots n_j$ be the sequence of nodes that is merged such that $n_1$ and $n_j$ are kept, and let $\mathfrak{r}_1\cdots\mathfrak{r}_j$ be the associated register sequence, then we make the update $\mathfrak{r}_j \mapsto \mathfrak{r}_2\mathfrak{r}_3\cdots\mathfrak{r}_{j-1}\mathfrak{r}_j$.

Now, we have defined a look-ahead tree $t'$ and a set $\set{(t_1',M'_1),\ldots,(t'_k,M'_k)}$ of pairs of a look-ahead trees and a state set as well as register associations to every tree node resp.\ set entry and how to update them.
The next transition taken by $\mathcal S$ updates the registers as described and the reached state stores $t'$ and $\set{(t_1',M'_1),\ldots,(t'_k,M'_k)}$.
Note that the invariants \cref{claim:represent,claim:out} remain valid as outlined in the construction.

\subsection{Correctness proof}

Before we prove our main technical lemma about the construction, we argue that the number of pieces of information to be stored in a state of $\mathcal S$ is bounded as well as the number of used registers.

Recall that \LAT is the finite set of look-ahead information trees.
The information stored in a state is an element of $\LAT \times \mathcal P(\LAT \times \mathcal P(Q_\mathcal A))$.
Let $(t,\set{(t_1,M_1),\ldots,(t_n,M_n)})$ be a state.
The number of registers needed to store information relevant to this state is bounded by $b_1+b_2$, where $b_1$ is the number of nodes in the trees $t,t_1,\ldots,t_n$ and $b_2$ is $|M_1|+\cdots +|M_n|$.

\begin{lemma}\label{lem:look-ahead-removal}
Let $u \in \Sigma^\omega$, for $i \geq 1$, let $u_i$ denote the prefix of $u$ of length $i$.
Let $LR_i$ denote the set of left-to-right runs of $\trans$ on $u_i$, let $RR_i$ denote the set of right-to-right runs of $\trans$ on $u_i$.

If $u \in \Dom(\trans)$, let $\rho_u$ denote the accepting run of $\trans$ on $u$.
The sequence $\rho_u$ can be factorized such that we have factors that describe a left-to-right run on $u_i$ and possibly factors that describe right-to-right runs on $u_i$.
Let $LR_i^{\mathsf{acc}} \subseteq LR_i$ and $RR_i^{\mathsf{acc}} \subseteq RR_i$ denote these sets of runs.
The former set is a singleton.

Let $c_i$ be the configuration (current state and register contents) of $\mathcal S$ after reading $u_i$, let $s_i$ be the state of $\mathcal S$.
The following properties hold.
\begin{enumerate}
  \item\label[item]{i1} The state $s_i$ represents a subset of $LR_i$ and $RR_i$.
  \item\label[item]{i2} If $\rho \in LR_i \cup RR_i$ is represented, then for its associated sequence of registers $\pi$ holds that $\cro{\pi} = \outp(\rho)$ in $c_i$.
  \item\label[item]{i3} If $u \in \Dom(\trans)$, then $s_i$ represents a superset of $LR_i^{\mathsf{acc}}$ and $RR_i^{\mathsf{acc}}$.
  \item\label[item]{i4} If $u \in \Dom(\trans)$, then $\cro{\out} \pref \trans(u)$ in $c_i$.
\end{enumerate}
\end{lemma}

\begin{proof}
  It is easy to see, that \cref{i1,i2} are a direct consequence of \cref{claim:represent} and \cref{i3,i4} are a direct consequence of \cref{claim:out}.

  We proof the statement of \cref{claim:represent,claim:out} by induction over the length of the considered prefixes.

  For the induction base, the considered prefix is a letter $\sigma \in \Sigma$.
  We have explicitly given the reached state on $\sigma \in \Sigma$ and the register content $\cro{\out}$. Clearly \cref{claim:represent,claim:out} holds.

  For the induction step, let the claims be true for a prefix of length $i \geq 1$, we show that it is remains true for a prefix of length $i+1$.
  We have shown in the above construction that the claims remain valid when constructing the successor state and register updates.
\end{proof}

\begin{lemma}\label{lem:updateout}
  Let $u \in \Sigma^\omega$, if $u \in \Dom(\trans)$, then the register $\out$ is updated infinitely often (by appending a non-empty word) in the run of $\mathcal S$ on $u$.
\end{lemma}

\begin{proof}
First, we note that if $u \in \Dom(\trans)$, then $\trans(u)$ is infinite.

To show the desired statement, we revisit the construction of $\mathcal S$.
Recall that the register content of $\out$ is updated (by appending output) whenever the application of the operation $\dels(\cdot)$ removes a successful non-empty path $\pi$ from the tree that represents the left-to-right runs of $\trans$ on the current prefix of the input word.
Such a successful path can only exist again and again if the accepting run of $\trans$ on $u$ is represented.
This is guaranteed by \cref{i3} of \cref{lem:look-ahead-removal}.
The output that is appended is the content of the sequence of the registers associated to $\pi$.
By \cref{i4} of \cref{lem:look-ahead-removal} we know that we build $\trans(u)$ this way.
Since $\trans(u)$ is infinite, infinitely many of these removed paths must have a corresponding sequence of registers whose content forms a non-empty word.

It is left to show, that infinitely often $\dels(\cdot)$ removes a successful path.
Let $t_i$ denote the look-ahead tree representing the possible left-to-right runs after processing an infix of length $i \geq 1$.
Note that when computing the tree $t_{i+1}$ from $t_i$, the only operation that can change the \emph{number} of look-ahead information in a node is the operation $\mer(\cdot)$.
But by definition, this operation never changes the number of look-ahead information stored in the root and in its children.
Furthermore, note that all look-ahead information belonging to the accepting run eventually succeed (as this is the definition of the run being accepting).
Hence, as the accepting run is represented, updating the look-ahead information in the root and in the child corresponding to the accepting run will eventually result in these nodes having succeeded.
Then they are removed by the operation $\dels(\cdot)$.
\end{proof}

It remains to define the Büchi acceptance condition of $\mathcal S$.
Essentially, $\mathcal S$ keeps track of the run(s) of $\mathcal T$ on the input.
Whenever the register out is updated, another segment of the accepting run of $\mathcal S$ has been determined.
This run segment is then no longer stored in the state space of $\mathcal S$.
We need to change the SST such that whenever a run segment of the accepting run is removed, the SST visits a final state iff the run segment has seen a final state.
Note that this information is not directly recoverable from the stored abstraction but is easily added to the abstraction.

\begin{lemma}\label{lem:buechiacc}
  Let $u \in \Sigma^\omega$. The \BSST{} $\mathcal S$ has an accepting run on $u$ iff $u \in \Dom(\trans)$.
\end{lemma}

\begin{proof}
This is a direct consequence of the above definition of the Büchi acceptance condition which mimics the Büchi acceptance condition of the two-way transducer.
\end{proof}

\begin{lemma}\label{lem:boundedcopy}
  The \BSST{} $\mathcal S$ is bounded copy.
\end{lemma}

\begin{proof}
  Let $ \mathcal S = (\Sigma,\Gamma,Q_\mathcal S, q_0^\mathcal S, \delta_\mathcal S, \Regs, \out,
  \lambda')$.
  Our goal is to show that $\mathcal S$ is 1-bounded, i.e.\ for all $u \in
  \Sigma^\omega$ and $i \le j$ such that $\lambda'^u_i \circ \cdots \circ
  \lambda'^u_j$ is defined, this substitution is 1-bounded.
  A substitution $\tau$ is 1-bounded if for all $\reg, \regg \in \Regs$, $\reg$ occurs at most once
  in $\tau(\regg)$.
  Instead of a formal proof, we give an intuitive justification of this claim.

  First of all, we note that the only reason why the constructed \BSST{} is not copyless, is because of register updates that are made for registers associated to elements in $M$ in pairs of the form $(t,M)$ which are used to represent right-to-right runs on prefixes.

  Recall that the tree $t$ from a pair $(t,M)$ stores information about the transitions of right-to-right runs that move right, and the set $M$ simply stores states from which a right-to-right run is possible.
  Also recall that transitions that move to the left do not use look-ahead.
  This is the reason why we do not have a tree structure to represent a subsequent series of left-moves because no look-ahead has to be verified.
  Hence, no intermediate information must be kept. 
  
  So, recall that if $\mathcal S$ reaches a state $s_i$ on the prefix $u[1{:}i] \in \Sigma^*$ of the input $u \in \Sigma^\omega$ that contains a pair $(t,M)$ such that $p \in M$ then this means there is a right-to-right run of the transducer on $u[1{:}i]$ that begins in $p$.
  
  If in the next step $\mathcal S$ reaches a state $s_{i+1}$ on the the prefix $u[1{:}{i{+}1}] \in \Sigma^*$ that contains a pair $(t',M')$ which is the result of updating the pair $(t,M)$ and there is $p_1,p_2 \in M'$, because there are transitions $(p_1,\sigma) \mapsto (p,\lmove)$ and $(p_2,\sigma) \mapsto (p,\lmove)$ then this means there is a right-to-right run of the transducer on $u[1{:}{i{+}1}]$ that begins in $p_1$ resp.\ $p_2$.
  
  Recall that the register that stores output related to $p \in M$ for the pair $(t,M)$, say $\out_p^{(t,M)}$, is used to define the registers that store output related to $p_1,p_2 \in M'$ for the pair $(t',M')$, say $\out_{p_1}^{(t',M')}$ and $\out_{p_2}^{(t',M')}$.

  The key to prove that every substitution is 1-bounded is that in the end either only the right-to-right run of the transducer on $u[1{:}{i{+}1}]$ that begins in $p_1$ \emph{or} the run that begins in $p_2$ can be relevant for the accepting run of the transducer because the transducer is deterministic.
  So, going forward, the registers $\out_{p_1}^{(t',M')},\out_{p_2}^{(t',M')}$ will never contribute to the same register.
\end{proof}

We need one last auxiliary lemma before we can proof \cref{thm:laremoval}.
Namely, that we do not need the Büchi acceptance condition.

\begin{lemma}\label{lem:removebuechi}
  For every \BSST{} $\mathcal T_1$ there exists an equivalent \BSST{} $\mathcal T_2$ such that all its states are final.
\end{lemma}

\begin{proof}
The  \BSST{} $\mathcal T_1$ can easily be changed such that all of its states are final states.
Therefore, in $\mathcal T_2$, we need to introduce a new register, say $\mathfrak r$, whenever $\mathcal T_1$ would update the designated output register $\out$, the output is appended to $\mathfrak r$ instead.
Whenever $\mathcal T_1$ would visit a final state the register content in $\mathfrak r$ is appended to $\out$ and $\mathfrak r$ is cleared.

To see that $\sem{\mathcal T_1} = \sem{\mathcal T_2}$, it suffices to realize the following:
If an input word $u \in \Sigma^\omega$ is in the domain of $\sem{\mathcal T_1}$, then in the infinite run of $\mathcal T_1$ on $u$ 
\begin{itemize}
  \item $\mathcal T_1$ infinitely often visits a final state, and
  \item the length of the content in $\out$ tends towards infinity.
\end{itemize} 
Hence, in the infinite run of $\mathcal T_2$ on $u$ the register $\out$ is infinitely often updated and the length of the content tends to infinity.
If an input word $u \in \Sigma^\omega$ is not in the domain of $\sem{\mathcal T_1}$, then either the run visits final states only finitely often, or the length of the content in $\out$ does not tend towards infinity.
In the first case, the register $\out$ is only finitely many times updated in the run of $\mathcal T_2$ on $u$, so length of the content in $\out$ does not tend towards infinity (meaning the run is rejecting).
In the second case, the length of the content in $\out$ does not tend towards infinity anyway.
\end{proof}

Finally, we are ready to give the proof of look-ahead removal theorem.

\begin{proof}[Proof of \cref{thm:laremoval}]

  It is a direct consequence of \cref{i4} of \cref{lem:look-ahead-removal} and \cref{lem:updateout,lem:buechiacc} that $\sem{\trans} = \sem{\mathcal S}$, i.e.\ the deterministic two-way transducer with finite look-ahead and the deterministic Büchi SST recognize the same function.

  Since the deterministic SST is bounded copy as shown in \cref{lem:boundedcopy} an equivalent equivalent deterministic two-way transducer can be constructed as shown in \cite[Theorem 3.7]{carton2022continuous} (recall that we re-stated this in \cref{theo:2dt-dsst}).
  As already noted in \cref{rem:allstatesfinal}, the cited result makes the assumption that the bounded copy SST only has final states (which we can guarantee using \cref{lem:boundedcopy}.
  By construction, all states of the resulting deterministic two-way transducer are final states.
\end{proof}

\subsection{Extension with safety constraints}\label{sec:safety}

We add safety constraints to deterministic two-way
transducers with finite look-ahead. It is a \tDTLA{} $\trans$ with
an additional mapping $S$ from any transition $t$ to a regular
language of finite words $S(t)$. If $\trans$ triggers transition
$t$ at input position $i$ on a word $u\in\Sigma^\omega$, then it must
additionally makes sure that $u[i{:}j]\in S(t)$ holds for all $j\geq
i$, otherwise the run dies. We show that while safety constraints are
properties of infinite suffixes, they do not add extra computational
power to \tDTLA.

  A \emph{deterministic two-way transducer with finite look-ahead and
    safety constraints} (\tDTLAS) is a tuple
  $\trans' = (\Sigma,\Gamma,Q,q_0,F,\delta,\lambda,S)$ such that
  $\trans = (\Sigma,\Gamma,Q,q_0,F,\delta,\lambda)$ is a \tDTLA, called the underlying
  \tDTLA of $\trans'$, and $S : \delta
  \rightarrow \mathcal{R}^*(\Sigma)$ maps any transition to a regular
  language of \emph{finite} words, called a \emph{safety constraint}. 
  The semantics is defined as follows: on an input
  $u\in\Sigma^\omega$, $\trans'$ moves from a configuration
  $(q,i)$ to a configuration $(q,j)$, following a transition $t$,  if $(i)$ $\trans$ moves from
  $(q,i)$ to $(q,j)$ by transition $t$, and $(ii)$ $u[i{:}k]\in S(t)$
  for all $k\geq i$. It should be clear that due to this extra
  constraint, we have the following property about the functions
  $f_\trans$ and $f_{\trans'}$ recognized by $\trans$ and $\trans'$
  respectively: $\dom(f_{\trans'})\subseteq \dom(f_{\trans})$ and
  they both coincide on $\dom(f_{\trans'})$. Transitions are denoted
  $p\xrightarrow{\sigma,L\mid w,C} q$, whenever
  $t = (p,\sigma,L,q)\in\delta$ and $S(t) = C$. 
  Also, note that if for all transitions $t$, $S(t) = \Sigma^*$, then $\trans'$ is
  obviously equivalent to its underlying \tDTLA{} $\trans$.

\begin{lemma}\label{lem:twowaysafe}
A function is computed by a $\tDTLAS$ iff it is computed by a $\tDT$.
\end{lemma}

\begin{proof}
The proof is a slight extension of the proof of \cref{thm:laremoval}.
Hence, we only describe the changes.
We integrate the verification of the safety constraints into the translation from the transducer to an equivalent SST.
Therefore, we extend the nodes of look-ahead trees to also store information about the safety constraints.
Assume that the safety languages are given by deterministic finite automata, say the safety language $C_i$ is given by the DFA $\mathcal C_i$.
Whenever a new node in a look-ahead tree is created based on a transition $t$ with safety constraint $S(t) = C_i$ for some $i$, then the node stores the initial state of $\mathcal C_i$.
Whenever a look-ahead tree is updated based on a letter $\sigma \in \Sigma$ (denoted as $\up(\cdot,\sigma)$), then every stored state from a safety automaton is also updated according to $\sigma$.
If a state indicates that a safety constraint has been violated, the node is considered as failed and branches with the failed node can be removed from the tree.
When nodes of a tree are merged, their the safety information is also merged.
\end{proof}

\subsection{Extension with finite look-around}\label{sec:lookaround}

We consider here two-way transducers with the ability to test a finite
prefix from the current position (look-ahead), and to test the prefix
up to the current position (look-behind). Both are combined into a
\emph{look-around}, which is a regular language of pointed words,
where only one position is pointed, which corresponds to the current
position. Formally, for an alphabet $\Sigma$, we let $\dot{\Sigma} = \{
\dot{\sigma}\mid\sigma\in \Sigma\}$. A \emph{pointed word} is a finite word in
$\Sigma^*\dot{\Sigma}\Sigma^*$. A \emph{finite look-around} is regular language of pointed
words.

  A \emph{deterministic two-way transducer with finite look-around}
is a tuple of the form $\trans$ $=$ \mbox{$(\Sigma,\Gamma,Q,q_0,F,\delta,\lambda)$} defined as a \tDTLA{}
  but the type of $\delta$ is $Q \times (\Sigma \uplus \{{\lmark}\}) \times
  \mathcal{LAR}^*(\Sigma)
        \parfonc Q \times \{\rmove, \lmove\}$
        where $\mathcal{LAR}^*(\Sigma)$ is the set of finite look-arounds over
        $\Sigma$. As for \tDTLA, it is required that: $(i)$ $\text{dom}(\delta)$ is
        finite, $(ii)$ all the look-arounds occurring on transitions
        with same source state and same input symbol are disjoint. As for \tDTLA, to ensure determinism, when several look-arounds are
        successful, only transition with look-around having length-minimal
        witness is picked (it is unique by disjointness). If no
        look-around succeeds, the computation fails.
        More precisely, in a configuration $(q,i)$ on an input word $u$, if
        $u[i] = \sigma$, then the run follows the transition
        $\delta(q,\sigma,L)$ such that $L$ has minimal $j$ in the set
        $\{ j\mid \exists L'\cdot \delta(q,\sigma,L')\text{ is defined
          and } u[{:}i{-}1]\dot{\sigma}u[i{+1}{:}j]\in L'\}$.

        Likewise, we further extend this model to deterministic two-way transducers with finite
        look-around and safety constraints (\tDTLARS). Our goal is now
        to show that \tDT{} are as expressive as \tDTLARS. 
        
        \begin{lemma}\label{lem:twowaylarsafe}\label{coro:twowaylarsafefull}
            A function is computed by a $\tDTLARS$ iff it is computed
            by a $\tDT$.
        \end{lemma}

        \begin{proof}
            Intuitively, any look-around can be decomposed into a finite union of look-behind
            and look-ahead. We use this observation to prove that any $\trans\in\tDTLARS$ can be
            decomposed into:
            \begin{itemize}
              \item a \oDT which annotates the input with
                look-behind information (the set of states reached by the
                look-around automata on the prefix so far)
              \item a \tDTLAS which simulates $\trans$ on the
                annotated input 
            \end{itemize}

            In more formal details, let $\trans\in \tDTLARS$ over an input
            alphabet $\Sigma$. Let us assume that any look-around $L$ is
      given by a DFA $\aut_L$ with set of states $Q_L$ and set of final
      states $F_L$. Given a state $q\in Q_L$, we denote by ${\sf Right}_L^{q}$
      the set of words in $\Sigma^*$, accepted by $A_L$ where the initial state is
      replaced by $q$. Symmetrically, let ${\sf Left}_L^q$ be the set of
      words in $\Sigma^*$, accepted by $A_L$ where the set of final states is
      replaced by $\{q\}$. Clearly:
      \[
      L = \bigcup_{q\in Q_L,\sigma\in\Sigma}\bigcup_{p\in Q_L\cdot \delta_L(q,\dot{\sigma})=p} {\sf Left}_L^q\dot{\sigma}{\sf Right}_L^p
      \]
      We now describe the first function ${\sf t}$ which tags the input with
      look-behind information. Let $\mathcal{L}$ be the set of all
      look-around occurring on the transitions of $T$. Let $Q =
      \biguplus_{L\in\mathcal L} Q_L$. 
      The first function is defined by
      ${\sf t} : \Sigma^\omega \rightarrow (\Sigma\times 2^Q)^\omega$
      which given a word $u\in\Sigma^\omega$ and a position $i$,
      replaces $u[i]$ by $(u[i], P)$ where $P$ are all the states reached
      by the look-around automata over input $u[{:}{i{-}1}]$. In particular, the
      first letter is tagged with the set of initial states of the
      look-around automata.

      The second function, to be composed with ${\sf t}$, is
      realized by a \tDTLAS{} $M$ which simulates
      $\trans$ but only uses look-ahead, thanks to the annotation of its
      inputs. In particular, consider a transition of $\trans$ of the form
      $x\xrightarrow{\sigma,L\mid v,L_{safe}} (y,d)$ where $x,y$ are
      states of $\trans$, $\sigma\in\Sigma$, $L$ is a look-around,
      $L_{safe}$ is a safety constraint, $v\in \Gamma^*$ is an output
      word, and $d$ a direction. For any $P\subseteq Q$, we create in
      $M$ the
      transitions
      $x\xrightarrow{(\sigma,P),{\sf Right}^q,L_{safe}\mid v} (y,d)$
      for all $q\in Q_L$ such that there exists $p\in P$ such that
      $\delta_L(\dot{\sigma},p) = q$. We also have to ensure
      disjointness. Consider two transitions of $M$ the form
      \[
      \begin{array}{llllllll}
        p \xrightarrow{(\sigma,P),\textsf{Right}^{q_1}\mid v_1,L_{\saf_1}} p_1 \\
        p \xrightarrow{(\sigma,P),\textsf{Right}^{q_2}\mid v_2,L_{\saf_2}} p_2 \\
      \end{array}
      \]
      It could be the case that $\textsf{Right}^{q_1}$ and
      $\textsf{Right}^{q_2}$ are not disjoint. In that case, we
      replace those two transitions as follows (taking an arbitrary
      order between $q_1$ and $q_2$):
      \[
      \begin{array}{llllllll}
        p \xrightarrow{(\sigma,P),\textsf{Right}^{q_1}\setminus \textsf{Right}^{q_2}\mid v_1,L_{\saf_1}} p_1 \\
        p \xrightarrow{(\sigma,P),\textsf{Right}^{q_2}\mid v_2,L_{\saf_2}} p_2 \\
      \end{array}
      \]
      We apply this rewriting on all transitions until the
      disjointness condition is met. It is correct for the following
      reason. Assuming $M$ runs on a word annotated with the function
      ${\sf t}$, then since the look-around of $\trans$ are disjoint,
      when $q_1$ and $q_2$ are states of two different look-around
      automata, only one look-ahead $\textsf{Right}^{q_1}$ or
      $\textsf{Right}^{q_2}$ succeeds, otherwise it would contradict
      disjointness. If $q_1$ and $q_2$ are states of the same
      look-around automaton, only one of the two look-ahead can
      succeed by determinism of this look-around automaton (otherwise,
      there would be two different runs of the look-around automaton
      on the same pointed word, one going through $q_1$, the other
      through $q_2$).

    Finally, note that the tagging function ${\sf t}$ can be realized
    by a \oDT{} all states of which are final. So, 
    we have proved that $\trans$ can be defined as a composition of a
    $\tDTLAS$ and such an \oDT{}. Since \tDTLAS = \tDT{} by
    \cref{lem:twowaysafe}, and \tDT{} are closed under
    pre-composition with \oDT{} all states of which are final, by
    Theorem 3.11 of~\cite{carton2022continuous}, we get the desired
    result.
\end{proof}

\subsection{Closure under composition: proof of \texorpdfstring{\cref{thm:closure-comp}}{Theorem~\ref{thm:closure-comp}}}

\begin{proof}[Proof of \cref{thm:closure-comp}]
Let $\trans_1,\trans_2$ denote two \tDT{}.
Our goal is to construct a deterministic two-way transducer \emph{with look-around} that recognizes $\trans_2 \circ \trans_1$.
Then, we use \cref{lem:twowaylarsafe} to obtain a deterministic two-way transducer.

The idea is to do a product construction.
The behavior of $\trans$ is determined by the combined behaviors of $\trans_1$ and $\trans_2$.
At each step, given an input letter $\sigma \in \Sigma$, $\trans$ simulates a matching transition $t_1 \in \delta_1$, then simulates an end-to-end run $\rho_2$ of $\trans_2$ on the production $\lambda_1(t_1) \in \Gamma^*$, and then produces the corresponding output $\outp{(\rho_2)}$.

We assume that $\trans$ is in a state such that the component of $\trans_1$ resp.\ $\trans_2$ is in state $p \in Q_1$ resp.\ $q \in Q_2$.

If $\trans_2$ is moving forward, then we need to advance the run of $\trans_1$.
We can use the determinism of $\trans_1$ to update the component of $\trans_1$ according to $t_1 := (p,\sigma) \mapsto (p',d) \in \delta_1$, and we simulate $\trans_2$ entering $\lambda_1(t_1) \in \Sigma^*$ from the left. 
Hence, $\trans$ contains a transition of the form $((q,p),\sigma) \mapsto ((q',p'),d)$, where $q'$ is the target state of the end-to-end run $\rho_2$ on output $\lambda_1(t_1) \in \Sigma^*$ starting from the left in $q$.
The output of the transition is $\outp{(\rho_2)}$.

If $\trans_2$ is moving backward, then we need to rewind in the run of $\trans_1$.
However, $\trans_1$ is not co-deterministic, so at this point we need to make use of the look-around feature.

Assume that for each transition $t := (p',\sigma) \mapsto (p,d) \in \delta_1$ we have a look-around $L_t$ that holds in a configuration $(p,i)$ of $\trans_1$ if it can be reached from $(p',i')$ via application of the transition $t$.
We explain below how to construct such look-arounds.

Hence, the look-around $L_{t_1}$ is enabled if the previously used transition was some $t_1 := (p',\sigma) \mapsto (p,d) \in \delta_1$.
The \tDT{} $\trans$ simulates $\trans_2$ entering $\lambda_1(t_1) \in \Gamma^*$ from the right.
Thus, $\trans$ contains a transition of the form $((q,p),\sigma) \mapsto ((q',p'),d)$, where $q'$ is the target state of the end-to-end run $\rho_2$ on the production $\lambda_1(t_1) \in \Gamma^*$ starting from the right in $q$.
The output of the transition is $\outp{(\rho_2)}$.

Regarding the acceptance component of $\trans$, it suffices to combine the Büchi conditions of $\trans_1$ and $\trans_2$.

Now we explain how to obtain the desired look-arounds using deterministic two-way automata.
We then use the fact that deterministic two-way automata have the same expressive power as deterministic one-way automata (aka.\ both models recognize regular languages) and the translation is effective \cite{5392614}.

First, recall that a look-around is a regular language of finite pointed words; and when a look-around is evaluated in the computation of a deterministic two-way transducer with look-around, then the pointed letter is located at the position that the two-way transducer currently reads.
Hence, if the two-way transducer $\trans_1$ has reached some configuration $(p,i)$ and we want to determine the the predecessor configuration of $\trans_1$ using a two-way automaton on the input where position $i$ has the pointed letter, then the automaton can simply go to the beginning of the word, then simulate the run of $\trans_1$ until it reaches the pointed letter in state $p$ and remember the transition that was taken to arrive there.

Note that such a two-way automaton only must visit the finite prefix of the input that the two-way transducer has visited so far.
Thus, it is clear that we indeed define a \emph{finite} look-around.
Furthermore, it is easy to see that the look-arounds to be verified when the transducer $\trans_1$ is in a state $p$ at position $i$ (to determine the successor configuration of $(p,i)$) are all pair-wise disjoint, because if not it would imply that two different runs would lead to the same configuration $(p,i)$ which is a contradiction as $\trans_1$ is deterministic.
\end{proof}

%% file: app-proof-mso.tex
\section{Logic-transducer correspondence: proof of
  \texorpdfstring{\cref{thm:main}}{Theorem~\ref{thm:main}}}

\subsection{Structure of the proof}

The proof goes through several intermediate models. We give here an
overview of the chain of (effective) model transformations that is used to prove
the logical characterization. We start from an \GMSOT $\trans_0$. 
\begin{enumerate}
  \item first, $\trans_0$ is converted into a jumping
    \GMSO-transducer $\trans_1$, by Lemma~\ref{lem:gmsot2detreg}.

  \item $\trans_1$ is converted into a walking \GMSO-transducer
    $\trans_2$, by Lemma~\ref{lem:jumpTowalk}

  \item $\trans_2$ is converted into the composition $\trans_3\circ
    \mathcal{U}_1$ of a \tDTLARS{} $\trans_3$ (an extension of \tDT{}
    with look-around and safety constraints, as defined in
    Section~\ref{sec:lookaround}),  and a \oDT{}
    $\mathcal{U}_1$ all states
    of which are final (see Lemma~\ref{lem:walkTodetreg}).

  \item $\trans_3$ is then converted into a  composition
    $\trans_4\circ \mathcal{U}_2$ of a \tDTLAS{} $\trans_4$ and a \oDT{}
    $\mathcal{U}_2$ all state of which are final (see
    \cref{lem:twowaylarsafe}). It shows how to transform look-arounds
    into look-aheads. 

  \item $\trans_4$ is converted into a \tDT{} $\trans_5$, by
    Lemma~\ref{lem:twowaysafe}, whose proof is an extension with safety
    constraints of the look-ahead removal construction of \cref{thm:laremoval}

  \item it yields that $\trans_0$ is equivalent to $\trans_5\circ
    \mathcal{U}_2\circ \mathcal{U}_1$ where $\trans_4$ is a \tDT{} and
    $\mathcal{U}_1, \mathcal{U}_2$ are \oDT{} all states of which are
    final. Theorem 3.11
    of~\cite{carton2022continuous} shows that deterministic regular
    functions are effectively  closed under precomposition with
    functions defined by \oDT{} all states of which are final
    (actually, the result is slightly more general). Therefore,
    $\trans_0$ is equivalent to a \tDT{} $\trans_6$

  \item finally, $\trans_6$ is converted into a \GMSOT $\trans_7$
    equivalent to $\trans_0$, by Lemma~\ref{lem:trans2log}, closing
    the chain of transformations.
\end{enumerate}

\subsection{Proof of \texorpdfstring{\cref{lem:trans2log}}{Proposition~\ref{lem:trans2log}}}

\begin{proof}

    Let $\trans = (\Sigma,\Gamma,Q,q_0,F,\delta,\lambda)$ be a \tDT{}
    recognizing $f$. By Remark~\ref{rem:allstatesfinal}, we can assume
    that $Q = F$. Wlog we also assume that $Q = \{1,\dots,n\}$ for some
    $n\in\mathbb{N}$, and $q_0=1$. We also assume that
    $|\lambda(q,\sigma)|\leq 1$ for all $q\in Q$ and
    $\sigma\in \Sigma$, and $|\lambda(q,\lmark)| = 0$. If $\trans$ produces more than
    one symbol at once, then they can be produced in several steps by
    doing a series of U-turns. Similarly, the second assumption can be
    ensured by forcing the transducer to make some U-turns and
    producing what it was supposed to produce on $\lmark$, but on the
    2nd symbol instead.

    We define an $\GMSOT$ $\mathcal{M} = (k,(\phi_S^{c,d})_{c,d},(\phi_\gamma)_\gamma,\phi_{\fst}^{c_0})$ such that
    $\sem{M} = f$. It takes exactly $n$ copies of the input,
    i.e.\ $k=n$. Intuitively, each copy corresponds to a state of
    $\trans$. The successor relation from a copy $p$ to a copy $q$
    summarizes sequences of non-producing
    transitions starting in state $p$, and followed by exactly one
    transition to $q$, producing some symbol. The label predicates are
    defined by considering precisely this last symbol.

    Formally, let $q = \delta(\lmark,q_0)$. Then, we let $c_0 = q$ and
    $\phi_{\fst}^{c_0}(x) = \first(x)$. Then, given two states
    $p,q$, we let $\text{Run}^{p,q}(x,y)$ the binary predicate which
    expresses the existence of a finite run of $\trans$, which starts at position
    $x$ in state $p$ (it reads the label of $x$), ends in state $q$
    after reading position $y$, produces no output symbols except at
    the very last transition. It is standard to show that this
    predicate is \MSO-definable over finite words, based on the correspondence between
    regular languages of finite words and \MSO. Then, we let 
    $\Phi_S^{p,q}(x,y) = \exists \gv\cdot x\leq \gv\wedge y\leq
    \gv\wedge \text{Run}^{p,q,\gv}(x,y)$ where
    $\text{Run}^{p,q,\gv}(x,y)$ is exactly $\text{Run}^{p,q}(x,y)$
    where all quantifiers are guarded by $\gv$ (so that
    $\text{Run}^{p,q,\gv}(x,y)$ is interpreted over a \emph{finite}
    prefix of the input).

    Finally, the label predicates are defined similarly:
    $\phi_\gamma^p(x)$ expresses that there exists a finite run from
    the initial position in state $q_0$ to position $x$ and state $q$, such that the
    last transition produces the symbol $\gamma$. As before, this can
    be expressed by a guarded formulas. 
\end{proof}

\subsection{Proof of \texorpdfstring{\cref{lem:jumping}}{Lemma~\ref{lem:jumping}}}\label{sec:jumping}

Given a \GMSO-transducer $\trans$, our goal is to construct an equivalent deterministic two-way
transducer with look-around and safety constraints $\trans'$ (an
extension of deterministic two-way transducers (\tDTLARS, defined in \cref{app:twoway}), and
conclude by \cref{lem:twowaylarsafe}, which states that this
extension is equivalent to \tDT. There are two main differences with \tDTLARS: $(i)$ jumping
    \GMSO-transducers jump while \tDTLARS{} walk, and $(ii)$ safety
    constraints in jumping \GMSO-transducers are properties of all
    finite prefixes starting in the first input position and that include the current position, 
    while safety constraints in \tDTLARS{} are properties
    of all finite prefixes starting in the current position.

    We decompose the proof into two lemmas which respectively address
    those two difficulties. \cref{lem:jumpTowalk} below states that any jumping \GMSO-transducer is
equivalent to a \emph{walking} \GMSO-transducer, that is a
\GMSO-transducer such that $\phi_{\mv}(x,y)\Rightarrow S(x,y)\vee
S(y,x)$ holds, for any jumping formula $\phi_{\mv}(x,y)$ occurring on
its transition. Then, \cref{lem:walkTodetreg} below states that any walking
\GMSO-transducer defines a deterministic regular function. The
combination of the two lemmas give the proof of
\cref{lem:jumping}.

\begin{lemma}\label{lem:jumpTowalk}
    Any jumping \GMSO-transducer is equivalent to a walking
    \GMSO-transducer. 
\end{lemma}

\begin{proof}
    Let $\trans$ be a jumping \GMSO-transducer. The transformation of $\trans$
    into a walking transducer $\trans'$ is standard and follows the same line
    as~\cite{engelfriet2001mso,courcelle2012graph}. Thanks to B\"uchi's-Elgot-Trakhtenbrot's
    theorem, any jumping formula $\phi_{\mv}(x,y)$
    of $\trans$ is converted into a DFA $A_{\phi_{\mv}}$ over finite words with two pointed
    positions $x$ and $y$, i.e., words over the alphabet $\Sigma\times
    2^{\{x,y\}}$ in which $x$ and $y$ occurs exactly once. In
    particular, we have $L(A_{\phi_{\mv}}) = \{ (\sigma_1,s_1)\dots
    (\sigma_n,s_n)\mid \exists 1\leq i,j\leq n\cdot
    \sigma_1\dots\sigma_n\models \phi_{\mv}(i,j)\text{ and } \forall
    1\leq k\leq n, i=k\leftrightarrow x\in s_k, j=k\leftrightarrow y\in
    s_k\}$. 
When $\trans$ triggers a transition
\[
    p\xrightarrow{\phi_{\la}(x)\mid
      w,\phi_{\mv}(x,y),\phi_{\saf}(x)} q
\]
    instead $\trans'$ moves to a sub-transducer $\trans_{\phi_{\mv}}$ which simulates
    the move of the DFA $A_{\phi_{\mv}}$ (and produces $\epsilon$
    all the time). We explain the simulation of $A_{\phi_{\mv}}$ in
    the case where $x< y$. $\trans_{\phi_{\mv}}$  starts at current
    position $x$, and uses some look-around (actually a look-behind) to
    determine the state reached by $A_{\phi_{\mv}}$ on the prefix so
    far, with its last position pointed by $x$, i.e., the state
    reached by $A_{\phi_{\mv}}$ on input $(\sigma_1,\varnothing)\dots
    (\sigma_{k-1},\varnothing)(\sigma_{k},\{x\})$, where
    $\sigma_1\dots\sigma_{k}$ is the input prefix up to the reading
    head position. It then starts to
    simulate $A_{\phi_{\mv}}$ in the state determined by the
    look-around. Then, for any subsequent transition, $\trans_{\phi_{\mv}}$ uses finite
    look-ahead to determine whether the current position, call it $y$, satisfies
    $\phi_{\mv}(x,y)$ or whether it should continue the simulation of
    $A_{\phi_{\mv}}$. To do so,  it checks with look-aheads, if there
    exists a finite word $v = (\sigma_1,\{y\})(\sigma_2,\varnothing)\dots
    (\sigma_n,\varnothing)$ such that $\sigma_1\dots \sigma_n$ is a
    prefix of the current infinite suffix (the one starting at the
    current position), such that the run of $A_{\phi_{\mv}}$ can be
    continued to an accepting run on $v$. 
    If the finite look-ahead succeeds, then $\trans_{\phi_{\mv}}$ knows
    that $\phi_{\mv}(x,y)$ holds and that the current position is $y$, and it gives back
    the control to $\trans'$. Note that $\trans_{\phi_{mv}}$ does not need
    to check that $y$ is unique. Indeed, the assumption in the
    definition of jumping \GMSO-transducers guarantees that if at some
    point two positions $y$ satisfy $\phi_{\mv}(x,y)$, then there
    cannot be an infinite run of $\trans$ on the input word, and so no
    infinite run of $\trans'$ as well.
\end{proof}

\begin{lemma}\label{lem:walkTodetreg}
    Any walking \GMSO-transducer $\trans$ defines a deterministic
    regular function. 
\end{lemma}

\begin{proof}
    Let $\trans$ be a jumping \GMSO-transducer. We express $\trans$ as the
    composition of a deterministic regular function $f_{\saf}$ defined by
    a \oDT{} $\trans_{\saf}$ which
    annotates the input with information useful to check safety
    constraints, and a walking
    \GMSO-transducer $\trans'$ such that the safety constraints are
    \MSO-formulas $\psi_{\saf}(x)$
    restricted to positions larger than $x$. So, in addition to
    upper-guard, quantified variables are also restricted to range
    over positions larger than $x$. Note that $\trans'$ is
    nothing but a \tDTLARS{}, whose look-around and safety
    constraints are defined by \GMSO-formulas.

    We now proceed to the definition of $f_{\saf}$ and the construction
    of $\trans'$. Intuitively, all formulas $\phi_{\saf}(x)$ are converted into
    DFA and the function $f_{\saf}$ annotates the input with some states
    reached by those DFA on the prefix so far. This information can then be exploited to
    have safety constraints that talk only about the prefixes starting
    in the current position.

    More precisely, any formula $\phi_{\saf}(x)$ occurring on some
    transition of $\trans$
    is converted into a DFA $A_{\phi_{\saf}}$ over finite pointed words. We assume that all the
    DFA have disjoint sets of states, and let $Q$ be their
    union. Then, the function $f_{\saf} : \Sigma^\omega\parfonc
    (\Sigma\times 2^Q)^\omega$ annotates the input word as follows:
    over an input $u\in\Sigma^\omega$, for any position $i$,
    $f_{\saf}(u)[i] = (u[i], P)$ where $P \subseteq Q$ is the set of
    states reached by all the DFAs over the prefix $u[{:}i]$ in which $u[i]$
    is pointed. So, $f_{\saf}(u)[i]$ contains the states of the DFAs
    right after they have ``selected'' position $i$. Note that each
    set $P$ contains at most one state per DFA $A_{\phi_{\saf}}$,
    because they are deterministic. It is easily seen
    that the function $f_{\saf}$ can be realized by an \oDT{}, which
    simulates all the DFAs in parallel.

    Then, $\trans'$ runs on input words over alphabet $\Sigma\times
    2^Q$, and exploits the information given by $f_{\saf}$ as follows: for each
    transition 
    $t = p\xrightarrow{\phi_{\la}(x)\mid
      w,\phi_{\mv}(x,y),\phi_{\sf}(x)} q$ of $\trans$ and each subset
    $P\subseteq Q$, we create the transition $t_P$ of $\trans'$
    \[
    t_P = p\xrightarrow{\phi^P_{\la}(x)\mid w,\phi^P_{\mv}(x,y),\phi^P_{\saf}(x)} q
    \]
    where:
    \begin{itemize}
      \item $\phi^P_{\la}(x)$ is the conjunction of $\bigvee_{\sigma
          \in \Sigma} (\sigma,P)(x)$ (to check that the second
        input label component is $P$), and of the formula 
        $\phi_{\la}(x)$ in which any
        occurrence of a predicate $\sigma(y)$, for all
        $\sigma\in\Sigma$, is replaced
        by $\bigvee_{R\subseteq Q} (a,R)(y)$.
      \item $\phi^P_{\mv}(x,y)$ is $\phi_{\mv}(x,y)$ where the label
        predicates are modified as before

      \item $\phi ^P_{\saf}(x)$  is $\bot$ if $P$ contains more than
        one state of $A_{\phi_{\saf}}$, or none.  If $P$
        contains exactly one state $s$ of  $A_{\phi_{\saf}}$, then 
        $\phi^P_{\saf}(x)$ expresses the existence of a finite prefix
        on which there exists an accepting run of $A_{\phi_{\saf}}$ starting at position
        $x+1$ in state $s$.
    \end{itemize}
    Note that the safety constraints now only talks about the prefixes
    starting at the current position.

    We also need to make sure that the look-arounds on transitions
    with same source state are disjoint, i.e.\ they are not mutually
    satisfiable. Consider two transitions:
\[
\begin{array}{lllllll}
p\xrightarrow{\phi^{P}_{\la}(x)\mid
  w_1,\phi^P_{\mv}(x,y),\phi^{P}_{\saf}(x)} q \\
p\xrightarrow{\psi^R_{\la}(x)\mid w',\psi^R_{\mv}(x,y),\psi^R_{\saf}(x)} q'
\end{array}    
\]

We know by definition of $\trans'$, that $\phi_\la(x)$ and
$\psi_{\la}(x)$ are disjoint, by definition of $\trans$. This disjointness transfers to the modified formulas 
$\phi^{P}_{\la}(x)$ and $\psi^{R}_{\la}(x)$, for any sets
$P,R\subseteq Q$, because the models of $\phi^{P}_\la(x)$ (resp.\ of
$\psi^{R}_{\la}(x)$), when projected on alphabet $\Sigma$, are exactly
the models of $\phi_\la(x)$ (resp.\ $\psi_\la(x)$).

    Finally, by \cref{coro:twowaylarsafefull}, we know that
    $\trans'$ is effectively equivalent to some \tDT{}, hence it
    defines a deterministic regular function. Note that
    $\trans_{\saf}$ is a \oDT{} all states of which are
    final. Composing such transducers with deterministic regular
    functions is known to yield deterministic regular functions only,
    by Theorem~3.11 of~\cite{carton2022continuous}. 
\end{proof}

\subparagraph{Final arguments for the proof of \cref{lem:jumping}}

The statement of \cref{lem:jumping}
is a direct consequence of \cref{lem:jumpTowalk} which
transforms any jumping \GMSO-transducer into a walking
\GMSO-transducer, of \cref{lem:walkTodetreg}, which proves that
the latter defines deterministic regular functions.


%% file: main.bbl
\begin{thebibliography}{10}

\bibitem{aho1969general}
Alfred~V. Aho, John~E. Hopcroft, and Jeffrey~D. Ullman.
\newblock A general theory of translation.
\newblock {\em Mathematical Systems Theory}, 3(3):193--221, 1969.

\bibitem{alur2010expressiveness}
Rajeev Alur and Pavol Cern{\'{y}}.
\newblock Expressiveness of streaming string transducers.
\newblock In {\em {IARCS} Annual Conference on Foundations of Software
  Technology and Theoretical Computer Science, {FSTTCS} 2010}, volume~8 of {\em
  LIPIcs}, pages 1--12. Schloss Dagstuhl, 2010.

\bibitem{alur2012regular}
Rajeev Alur, Emmanuel Filiot, and Ashutosh Trivedi.
\newblock Regular transformations of infinite strings.
\newblock In {\em Proceedings of the 27th Annual {IEEE} Symposium on Logic in
  Computer Science, {LICS} 2012}, pages 65--74. {IEEE} Computer Society, 2012.

\bibitem{alur2014regular}
Rajeev Alur, Adam Freilich, and Mukund Raghothaman.
\newblock Regular combinators for string transformations.
\newblock In {\em Proceedings of the Joint Meeting of the Twenty-Third EACSL
  Annual Conference on Computer Science Logic (CSL) and the Twenty-Ninth Annual
  ACM/IEEE Symposium on Logic in Computer Science (LICS)}, page~9. ACM, 2014.

\bibitem{baudru2018two}
Nicolas Baudru and Pierre-Alain Reynier.
\newblock From two-way transducers to regular function expressions.
\newblock In {\em International Conference on Developments in Language Theory},
  pages 96--108. Springer, 2018.

\bibitem{bojanczyk2020single}
Miko{\l}aj Boja{\'n}czyk and Rafa{\l} Stefa{\'n}ski.
\newblock Single-use automata and transducers for infinite alphabets.
\newblock In {\em 47th International Colloquium on Automata, Languages, and
  Programming (ICALP 2020)}. Schloss Dagstuhl-Leibniz-Zentrum f{\"u}r
  Informatik, 2020.

\bibitem{bojanczyk2018polyregular}
Mikołaj Bojańczyk.
\newblock Polyregular {Functions}, 2018.
\newblock URL: \url{https://arxiv.org/abs/1810.08760}, \href
  {https://doi.org/10.48550/ARXIV.1810.08760}
  {\path{doi:10.48550/ARXIV.1810.08760}}.

\bibitem{Bu60}
J.~R. B\"uchi.
\newblock Weak second-order arithmetic and finite automata.
\newblock {\em Zeitschrift f\"ur Mathematische Logik und Grundlagen der
  Mathematik}, 6(1--6):66--92, 1960.

\bibitem{carton2022continuous}
Olivier Carton and Ga{\"{e}}tan Dou{\'{e}}neau{-}Tabot.
\newblock Continuous rational functions are deterministic regular.
\newblock In {\em 47th International Symposium on Mathematical Foundations of
  Computer Science, {MFCS} 2022}, 2022.

\bibitem{chytil1977serial}
Michal~P. Chytil and Vojt{\v{e}}ch J{\'a}kl.
\newblock Serial composition of 2-way finite-state transducers and simple
  programs on strings.
\newblock In {\em 4th International Colloquium on Automata, Languages, and
  Programming, {ICALP} 1977}, pages 135--147. Springer, 1977.

\bibitem{colcombet2007combinatorial}
Thomas Colcombet.
\newblock A combinatorial theorem for trees.
\newblock In {\em 34th International Colloquium on Automata, Languages, and
  Programming, {ICALP} 2007}, 2007.

\bibitem{Cou94}
Bruno Courcelle.
\newblock Monadic second-order definable graph transductions: A survey.
\newblock {\em Theor. Comput. Sci.}, 126:53--75, 1994.

\bibitem{courcelle2012graph}
Bruno Courcelle and Joost Engelfriet.
\newblock {\em Graph structure and monadic second-order logic: a
  language-theoretic approach}, volume 138.
\newblock Cambridge University Press, 2012.

\bibitem{dartois2018logics}
Luc Dartois, Emmanuel Filiot, and Nathan Lhote.
\newblock Logics for word transductions with synthesis.
\newblock In {\em Proceedings of the 33rd Annual ACM/IEEE Symposium on Logic in
  Computer Science}, pages 295--304. ACM, 2018.

\bibitem{dartois2017reversible}
Luc Dartois, Paulin Fournier, Isma{\"{e}}l Jecker, and Nathan Lhote.
\newblock On reversible transducers.
\newblock In {\em 44th International Colloquium on Automata, Languages, and
  Programming, {ICALP} 2017}, volume~80 of {\em LIPIcs}, pages 113:1--113:12.
  Schloss Dagstuhl, 2017.

\bibitem{dartois2016aperiodic}
Luc Dartois, Isma{\"{e}}l Jecker, and Pierre{-}Alain Reynier.
\newblock Aperiodic string transducers.
\newblock {\em Int. J. Found. Comput. Sci.}, 29(5):801--824, 2018.

\bibitem{dave2020synthesis}
Vrunda Dave, Emmanuel Filiot, Shankara~Narayanan Krishna, and Nathan Lhote.
\newblock Synthesis of computable regular functions of infinite words.
\newblock In {\em 31st International Conference on Concurrency Theory (CONCUR
  2020)}. Schloss Dagstuhl-Leibniz-Zentrum f{\"u}r Informatik, 2020.

\bibitem{DBLP:journals/lmcs/DaveFKL22}
Vrunda Dave, Emmanuel Filiot, Shankara~Narayanan Krishna, and Nathan Lhote.
\newblock Synthesis of computable regular functions of infinite words.
\newblock {\em Log. Methods Comput. Sci.}, 18(2), 2022.

\bibitem{dave2018regular}
Vrunda Dave, Paul Gastin, and Shankara~Narayanan Krishna.
\newblock Regular transducer expressions for regular transformations.
\newblock In {\em Proceedings of the 33rd Annual ACM/IEEE Symposium on Logic in
  Computer Science}, pages 315--324. ACM, 2018.

\bibitem{Elg61}
C.~C. Elgot.
\newblock Decision problems of finite automata design and related arithmetics.
\newblock {\em In Transactions of the American Mathematical Society},
  98(1):21--51, 1961.

\bibitem{engelfriet2001mso}
Joost Engelfriet and Hendrik~Jan Hoogeboom.
\newblock {MSO} definable string transductions and two-way finite-state
  transducers.
\newblock {\em ACM Transactions on Computational Logic (TOCL)}, 2(2):216--254,
  2001.

\bibitem{filiot2021synthesizing}
Emmanuel Filiot and Sarah Winter.
\newblock Synthesizing computable functions from rational specifications over
  infinite words.
\newblock In {\em 41st {IARCS} Annual Conference on Foundations of Software
  Technology and Theoretical Computer Science, {FSTTCS} 2021, December 15-17,
  2021, Virtual Conference}. Schloss Dagstuhl - Leibniz-Zentrum f{\"{u}}r
  Informatik, 2021.

\bibitem{DBLP:journals/jsyml/Gradel99}
Erich Gr{\"{a}}del.
\newblock On the restraining power of guards.
\newblock {\em J. Symb. Log.}, 64(4):1719--1742, 1999.

\bibitem{gurari1982equivalence}
Eitan~M Gurari.
\newblock The equivalence problem for deterministic two-way sequential
  transducers is decidable.
\newblock {\em SIAM Journal on Computing}, 11(3):448--452, 1982.

\bibitem{perrin2004}
Dominique Perrin and Jean-{\'E}ric Pin.
\newblock {\em Infinite words: automata, semigroups, logic and games}.
\newblock Academic Press, 2004.

\bibitem{5392614}
J.~C. Shepherdson.
\newblock The reduction of two-way automata to one-way automata.
\newblock {\em IBM Journal of Research and Development}, 3(2):198--200, 1959.
\newblock \href {https://doi.org/10.1147/rd.32.0198}
  {\path{doi:10.1147/rd.32.0198}}.

\bibitem{simon1990factorization}
Imre Simon.
\newblock Factorization forests of finite height.
\newblock {\em Theor. Comput. Sci.}, 72(1):65--94, 1990.
\newblock \href {https://doi.org/10.1016/0304-3975(90)90047-L}
  {\path{doi:10.1016/0304-3975(90)90047-L}}.

\bibitem{thomas1997languages}
Wolfgang Thomas.
\newblock Languages, automata, and logic.
\newblock In {\em Handbook of formal languages}, pages 389--455. Springer,
  1997.

\bibitem{tra61short}
Boris~Avraamovich Trakhtenbrot.
\newblock Finite automata and logic of monadic predicates (in {R}ussian).
\newblock {\em Dokl.\ Akad.\ Nauk SSSR}, 140:326--329, 1961.

\end{thebibliography}
